\documentclass[aps, prd, notitlepage, superscriptaddress, amsfonts, amssymb, amsmath, nofootinbib, 
10pt, longbibliography]{revtex4-1}
\usepackage{microtype}
\usepackage[utf8]{inputenc}
\usepackage[english]{babel}
\usepackage{titlesec}
\usepackage{graphicx}   
\usepackage{xcolor}
\usepackage[colorlinks=true, 
	linkcolor=blue, 
	citecolor=blue, 
	urlcolor=blue, 
	linktocpage=true]{hyperref}

\newcommand{\nn}{\nonumber \\}
\newcommand{\eq}[1]{Equation~\eqref{#1}} 
\newcommand{\eqs}[2]{Equations~\eqref{#1} and~\eqref{#2}} 
\newcommand{\fig}[1]{Figure~\ref{#1}}

\newcommand{\be}{\begin{eqnarray}}
\newcommand{\ee}{\end{eqnarray}}
\newcommand{\om}{\ensuremath{\omega}}

\newcommand{\pd}{\ensuremath{\partial}}
\newcommand{\la}{\ensuremath{\lambda}}
\newcommand{\ep}{\ensuremath{\epsilon}}
\newcommand{\lp}{\ensuremath{\left(}}
\newcommand{\rp}{\ensuremath{\right)}}

\newcommand{\abs}[1]{\ensuremath{{\left\lvert #1 \right\rvert}}} 
\newcommand{\dd}{\ensuremath{\mathrm{d}}}
\newcommand{\e}{\ensuremath{\mathrm{e}}}
\newcommand{\ii}{\ensuremath{\mathrm{i}}}
\newcommand{\ord}[1]{\ensuremath{O \left( #1 \right)}} 
\newcommand{\orde}[1]{\ord{\epsilon^{#1}}}
\newcommand{\lb}{\ensuremath{\left\lbrace}}
\newcommand{\rb}{\ensuremath{\right\rbrace}}
\newcommand{\s}{\nobreak\hspace{.08em plus .04em}}

\newcommand{\deta}{\delta\mkern-1.mu \eta}
\newcommand{\Deta}{\Delta\mkern-1.mu \eta}
\newcommand{\w}{\mu} 

\newcommand{\ad}[1]{\thanks{\texttt{\url{#1}}}\;}

\newcommand{\tsetminus}{\, \backslash}

\begin{document}

\title{Gravity waves on modulated flows downstream from an obstacle: 
\texorpdfstring{\\}{} 
The transcritical case} 
\date{\today}

\author{Florent Michel}
\email{florent.c.michel@durham.ac.uk} 
\affiliation{Center for Particle Theory, Durham University, South Road, Durham, DH1 3LE, United Kingdom}
\author{Renaud Parentani}
\email{renaud.parentani@th.u-psud.fr}
\affiliation{Laboratoire de Physique Th\'eorique, CNRS, Univ. Paris-Sud, Universit\'e Paris-Saclay, 91405 Orsay, France}
\author{Scott Robertson}
\email{scott.robertson@th.u-psud.fr} 
\affiliation{Laboratoire de Physique Th\'eorique, CNRS, Univ. Paris-Sud, Universit\'e Paris-Saclay, 91405 Orsay, France}

\begin{abstract}
\defcitealias{Busch:2014hla}{[X.~Busch {\it et al.}, {\it Phys. Rev. D} {\bf 90}, 105005 (2014)]}
\noindent 
Periodic spatial  modulations arise in analogue gravity experiments aimed at detecting the analogue version of the Hawking effect in a white-hole flow. 
Having the same spatial periodicity as low-frequency dispersive modes, they can induce resonances which significantly modify the scattering coefficients. 
This has been shown numerically in a previous work~\cite[X. Busch et al., Phys. Rev. D 90, 105005 (2014),][]{Busch:2014hla}, but the precise dependence of the low-frequency effective temperature on the amplitude and length of the undulation remains elusive. 
In this article, using the Korteweg-de Vries equation, we explicitly compute this dependence in the small-amplitude limit and find three regimes of ``short'', ``intermediate'' and ``long'' undulations showing different scaling laws.  
In the latter, the effective temperature is completely determined by the properties of the undulation, independently of the surface gravity of the analogue white-hole flow. 
These results are extended to a more realistic hydrodynamical model in an appendix. 
\end{abstract}

\maketitle

\section{Introduction}

Undulations, i.e., static spatially periodic modulations of some physical quantity, are ubiquitous in experiments involving a flow over an obstacle. 
They have been studied, for instance, in water currents~\cite{byatt-smith_1971, Lee85, Lawrence87, wu87, johnson_1997, 2004nlin.....12060E, 2005Chaos..15c7102E, chapman_vanden-broeck_2006, Binder2006, Coutant:2012mf, Euve:2015vml} and in Bose-Einstein condensates flowing faster than the Landau critical velocity~\cite{pitaevskii2003bose}. 
Roughly speaking, such modulations can be seen as macroscopic superpositions of low-frequency linear waves. 
As such, small perturbations propagating over them can be strongly affected via $n$-wave mixing, with $n \geq 3$, provided some resonance conditions are satisfied. 

These modulations are particularly relevant for the study of analogue white-hole flows, i.e., flows going from supercritical to subcritical along the direction of the fluid velocity~\cite{Macher:2009tw,Mayoral11,Busch:2014hla}. 
As was shown by William Unruh~\cite{Unruh:1980cg}, there exists a mathematical correspondence between the behavior of sound waves close to the point where the flow velocity crosses the speed of low-frequency waves in the fluid frame and that of scalar fields around the horizon of a black hole. 
This correspondence being independent of the sign of the flow velocity, it therefore also holds for its time-reversed version, known as a white hole. 
The original model of~\cite{Unruh:1980cg} was then extended to a variety of different systems like gravity waves in water~\cite{Schutzhold:2002rf}, cold atoms~\cite{Garay:1999sk}, polaritons~\cite{Gerace:2012a}, light in a nonlinear optical fiber~\cite{Philbin1367}, ``slow'' sound~\cite{Aregan:2015a,Auregan:2015uva}, and magnetohydrodynamics~\cite{Noda:2016pva}, prompting theoretical studies of the link with Hawking radiation (HR) and experimental realizations aimed at detecting its condensed matter analogue~\cite{Weinfurtner:2010nu,Steinhauer:2015saa,Euve:2015vml}. 
An account of the main advances in this field can be found in~\cite{Barcelo:2005fc} and references therein. 
The majority of these works use the assumption that the flow is homogeneous far from the (analogue) horizon, which is necessary for the corresponding space-time to be asymptotically flat. 
However, experimental setups involving white-hole flows generically have undulations extending far from the near-horizon region and whose effect on the scattering of linear waves, used to probe the Hawking effect, remains unclear. 
The main objective of the present work is to tackle this problem and determine, in the case of gravity waves on water, how the presence of an undulation affects the scattering coefficients. 

When considering white-hole flows of ideal fluids, the dispersion relation in the asymptotic downstream (e.g. for surface water waves) or upstream (e.g. for Bose-Einstein condensates) region generically leads to the emission of a wave with zero frequency and nonvanishing wave vector. 
There are then two known mechanisms for such an emission.  
The first one, discussed in~\cite{Whitham_LaNW, Lighthill_WiF} for water waves and in~\cite{pitaevskii2003bose} and Appendix~C of~\cite{Busch:2014hla} for Bose-Einstein condensates, is due to the fact that, given an obstacle and parameters for the upstream flow, the amplitude of the periodic wave is generally nonzero unless these parameters are precisely fine-tuned to suppress it. 
(A procedure to design an obstacle shape reducing the undulation amplitude in water waves was proposed in~\cite{Michel:2014zsa} and applied in~\cite{Euve:2015vml}. 
Another procedure was used in~\cite{Mayoral11} in the context of flowing atomic Bose condensates.) 
The second mechanism is the amplification of incoming low-frequency perturbations by the analogue Hawking effect~\cite{Coutant:2012mf}, exhibiting a close relation between this effect and undulations.   
Moreover, in the experiment~\cite{Euve:2015vml} the undulation seems to play an important role in the wave conversion, which cannot be unambiguously separated from the contribution of the flow gradient localized above the obstacle. 
These observations, calling for a better understanding of the scattering in the presence of an undulation, are the main motivations for the present work. 

The effect of an undulation on the analogue HR was addressed numerically in the context of Bose-Einstein condensates in~\cite{Busch:2014hla}. 
It was shown that a sufficiently long undulation can significantly modify the effective temperature. 
Importantly, it was found to be reduced when the undulation has a phase close to that of a nonlinear solution of the field equations. This study raises two important questions.  
The first one concerns the analytical description of the reduction, which was so far obtained only numerically. 
Second, it is not completely clear from the numerical results whether the effective temperature goes to zero in the limit of a long undulation or saturates to a finite value -- nor, in the latter case, how this value depends on the undulation amplitude. 

We here aim at answering these two questions by computing analytically the effective temperature in the presence of an undulation in a transcritical flow. 
In the main text we use a simple model based on the Korteweg-de Vries (KdV) equation to describe surface waves on a two-dimensional flow of an ideal fluid. 
Although idealized, it has the advantage of showing the main features in a transparent way, keeping technicalities to a minimum.  
In Appendix~\ref{app:UCP}, the same analysis is carried out in a more involved model fully taking into account the dispersion relation of water waves obtained when neglecting viscosity, surface tension, and vorticity. 
We believe that the similarities between the results of our two models and the generality of the arguments used to motivate them indicate they should be only weakly affected when including these three effects. 
In Appendix~\ref{app:sub} we briefly discuss the case of subcritical flows (see also Refs.~\cite{Michel:2014zsa,Robertson:2016ocv,Coutant:2016vsf,Coutant:2017qnz}) to be closer to the experiments of~\cite{Euve:2014aga,Euve:2015vml}. 

This paper is organized as follows. 
Section~\ref{sec:KdV_tuned} is devoted to the computation of the modes over a long undulation. 
We show in which sense it may be thought of as the zero-frequency limit of a three-wave resonance and exhibits the specific features of this limit.   
These results are applied to white-hole-like flows in Section~\ref{sec:WH}, where the implications of a small undulation for the analogue Hawking radiation are determined. 
We conclude in Section~\ref{sec:concl}. 
In Appendix~\ref{app:detuned} we consider the resonant scattering on a ``detuned'' undulation, i.e., a region where some external parameter varies periodically, in the case there is a resonance at finite frequency.  
We take the opportunity to detail two points which are also relevant to the case studied in the main text, namely the Lagrangian description of the KdV equation and the relation between the transfer and scattering matrices. 
Appendix~\ref{app:lingrowmodes} relates the properties of the modes to the nonlinear solutions of the KdV equation. 
Appendix~\ref{app:UCP} generalizes the main results to a more realistic model of water waves, assuming incompressibility, irrotationality, no viscosity, and no surface tension, 
but keeping terms of all orders in the wave vector. 
Finally, in appendix~\ref{app:sub} we briefly comment on the case of a subcritical flow. 

\section{Modes over an infinite undulation}
\label{sec:KdV_tuned}

The aim of this section is to exhibit the general properties of low-frequency modes over an undulation. 
This is a preliminary step to the calculation of the spectrum in analogue white-hole flows done in Section~\ref{sec:WH}. 
In Appendix~\ref{app:detuned}, the interested reader will find a complete treatment of the case of a ``detuned'' undulation, corresponding to an externally imposed modulation of some parameter with an arbitrary period. 
There, we find that the behavior of resonant modes depends on the relative signs of the energies and momenta of the two waves involved in the resonant scattering: the resonant modes are exponentially growing in time (respectively in space) if their energies (respectively energy currents) have opposite signs, and bounded if they have the same sign.   
Here instead, we consider the effects of ``tuned'' undulations which are themselves static solutions of the KdV equation.  
As we shall see, this introduces a qualitative difference: the resonance now involves modes with a vanishing energy, leading to a linear, instead of exponential, growth in space or time. 
In Appendix~\ref{app:lingrowmodes}, we show that this behavior can be directly related to variations of the nonlinear solutions. 

We work with the one-dimensional KdV equation:
\begin{equation} \label{eq:KdV} 
\pd_t \eta + \w \, \pd_x \eta + \pd_x^3 \eta + 6 \, \eta \, \pd_x \eta = 0,
\end{equation}
where $\w > 0$. 
(See appendix~\ref{app:detuned} and Refs.~\cite{Shen1993,Hereman2013}.) 
To study linear perturbations, we write $\eta = \eta^{(0)} + \deta$ where $\eta^{(0)}$ is an exact stationary solution of \eq{eq:KdV} and $\deta$ is a ``small'' perturbation. 
Neglecting terms quadratic in $\deta$, one obtains the linearized KdV equation: 
\begin{equation} \label{eq:KdV-linearized}
\pd_t \deta + \pd_x \lp \lp \w + 6 \, \eta^{(0)} \rp \deta \rp + \pd_x^3 \deta = 0. 
\end{equation}
In the following we first review a few properties of this linear equation which will play an important role in the scattering over an undulation. 
We generalize Equation~\eqref{eq:KdV-linearized}, replacing the factor $\w + 6 \s \eta^{(0)}$ by an arbitrary differentiable function of $x$, so that the same formalism can be applied to ``detuned'' undulations. 
Applying it to \eq{eq:KdV-linearized} allows us to determine the form of the resonant modes as well as the leading terms in the transfer matrix at low frequencies, which will be used in Section~\ref{sec:WH} to determine the corrections to the emission spectrum of a white-hole flow. 
The analysis presented in this section is extended to a more realistic model of water waves in Appendix~\ref{app:UCP}.

\subsection{The linearized KdV equation}
\label{sub:linKdV}

In this subsection we consider the mathematical description of periodic variations of some parameter entering the field equation. 
These might be ``detuned'' (see Appendix~\ref{app:KdV_detuned} for a fuller treatment), corresponding to local variations of some external potential or, in the case of water waves, to the height of the obstacle put at the bottom of the flume. 
However, in the main body of this paper we shall restrict our attention to ``tuned'' undulations, which are themselves static nonlinear solutions of the KdV equation.  

To this end, we consider the following generalized form of Equation~\eqref{eq:KdV-linearized}: 
\begin{equation} \label{eq:linKdV} 
\pd_t \deta(t,x) + \pd_x \lp \bar{\w}(x) \, \deta(t,x) \rp + \pd_x^3 \deta(t,x) = 0, 
\end{equation}
where $\bar{\w} \in C^0 (\mathbb{R})$ is a nonconstant periodic function of $x$.  
The relation between \eq{eq:linKdV} and the forced KdV equation used to model water waves~\cite{Shen1993} is discussed in appendix~\ref{app:lag_lin_KdV}. 
In the case of ``tuned'' undulations considered here, the undulation is a solution $\eta^{(0)}$ of the KdV equation, and enters \eq{eq:linKdV}  through the relation $\bar{\w} = \w + 6 \s \eta^{(0)}$. 

Let us call $\la_\w$ the fundamental period of $\bar{\w}$ and define $k_\w \equiv 2 \pi / \la_\w$. 
For all $n \in \mathbb{Z}$, we define 
\begin{equation} 
\w_n \equiv \frac{1}{\la_\w} \int_0^{\la_\w} \bar{\w}(x) \, \e^{- \ii \s n \s k_\w \s x} \dd x. 
\label{eq:def_w_n}
\end{equation}
Then, for any $x \in \mathbb{R}$,
\begin{equation} \label{eq:expv}
\bar{\w}(x) = \sum_{n \in \mathbb{Z}} \w_n \, \e^{\ii \s n \s k_\w \s x}. 
\end{equation} 
Since $\bar{\w}$ is real valued, $\w_{-n} = \w_n^*$ for all $n \in \mathbb{Z}$. 
To determine the scattering coefficients analytically, we work perturbatively in the variations of $\bar{\w}$. 
More precisely, we define a small parameter $\epsilon > 0$, $\epsilon \ll 1$, and assume the coefficients $\w_n$ scale as
\begin{equation} \label{eq:exp_eps} 
\forall \, n \in \mathbb{Z}, \; \frac{\w_n}{\w_0} = O \big( \ep^\abs{n} \big). 
\end{equation}
The calculation can then be performed to any given order in $\epsilon$ by expanding Equation~\eqref{eq:linKdV} in this parameter. 
(This justifies a posteriori the scaling in \eq{eq:exp_eps}.) 
 
\begin{figure}
\begin{center}
\includegraphics[width = 0.49 \linewidth]{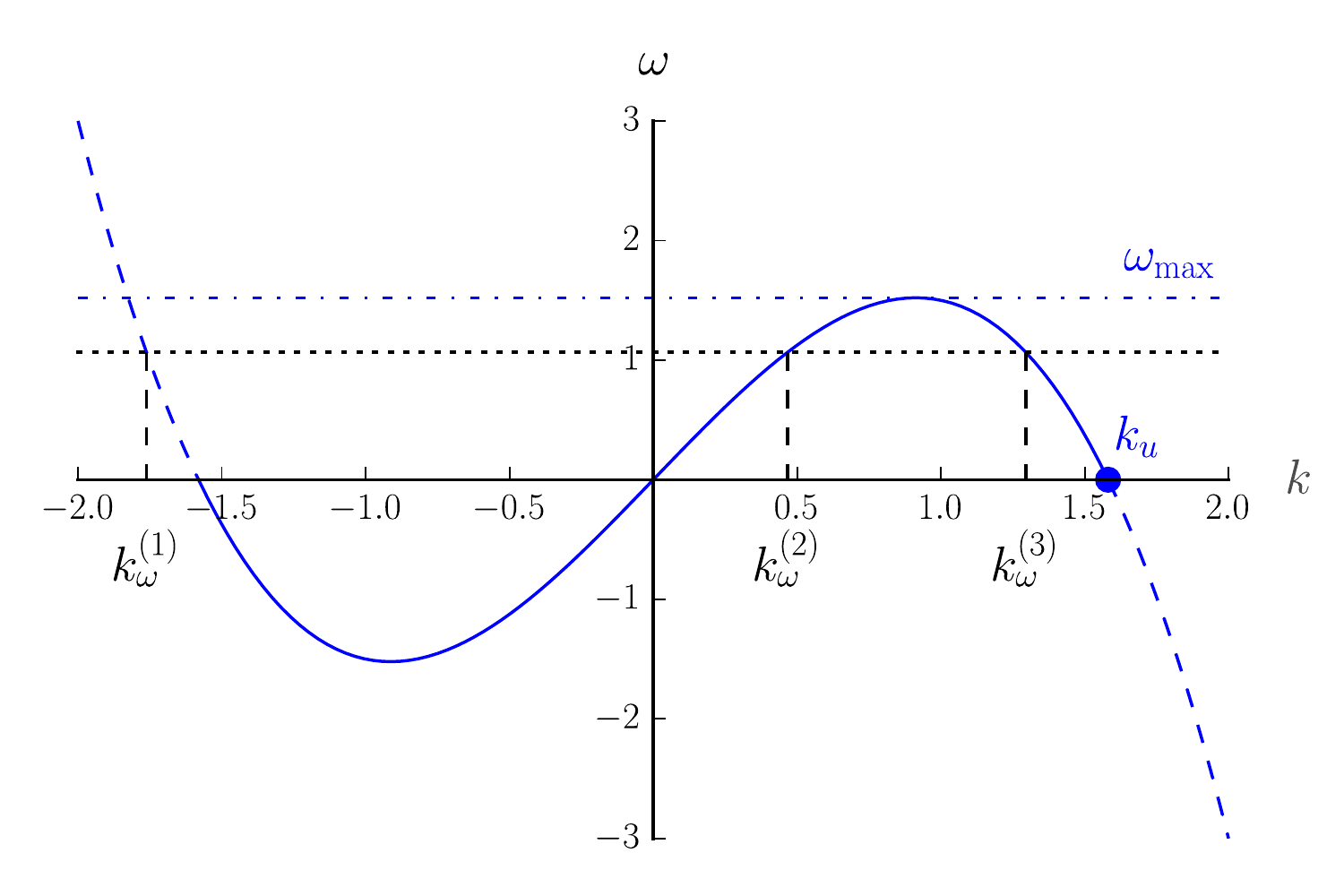} 
\end{center}
\caption{Dispersion relation $\omega$ versus $k$ for the linear KdV equation \eqref{eq:linKdV0} with $\w_0 = 2.5$. 
The (subsonic) flow velocity $v_0 < 0 $ is to the left, and $\w_0 = c_0 + v_0$ where $c_0$ is the low-frequency group velocity of linear waves. 
The style of the line gives the sign of the energy of the corresponding mode: continuous for positive energies and dashed for negative ones. 
The horizontal dot-dashed line shows $\om = \om_\mathrm{max}$ of \eq{eq:ommax}. 
The large dot shows the wave vector $k_u$ of the ``tuned'' undulation. 
The black dotted line materializes a constant value of $\omega$ strictly between $0$ and $\omega_\mathrm{max}$. 
Its intersection points with the blue curve give the three wave vectors $k_{\omega}^{(i)}$ at that frequency. 
} \label{fig:DR_KdV}
\end{figure}

It is of value to first recall some properties of the solutions in the case $\ep = 0$. 
Then, Equation~\eqref{eq:linKdV} becomes 
\begin{equation} \label{eq:linKdV0}
[ \pd_t + \w_0 \, \pd_x + \pd_x^3 ] \deta(t,x) = 0.
\end{equation} 
The constant $\w_0$, chosen here to be positive, is equal to $c_0 + v_0$, where $c_0$ is the low-frequency group velocity in the fluid frame, and $v_0 < 0$ is the flow speed. 
With these choices, Equation~\eqref{eq:linKdV0} describes small-amplitude counterpropagating waves over a (subcritical) flow to the left, i.e., waves whose group velocities in the reference frame of the fluid are all positive. 
As this equation has no explicit space or time dependence, there exists a continuous basis of bounded solutions of the form
\begin{equation} \label{eq:planew}
\deta_{\om, k}: (t,x) \mapsto \e^{\ii \s \lp k \s x - \om \s t \rp},
\end{equation}
where $(\om, k) \in \mathbb{R}^2$. 
Using Equation~\eqref{eq:linKdV0}, one finds Equation~\eqref{eq:planew} is a solution if and only if the dispersion relation 
$\om = \om_0(k)$ is satisfied, where 
\begin{equation} \label{eq:DR} 
\om_0 (k) = \w_0 \, k - k^3.
\end{equation}
It is shown in Figure~\ref{fig:DR_KdV}. 
At fixed value of $\om$, the dispersion relation has three complex roots in $k$. 
For $\om \in \mathbb{R}$, the number of real ones depends on whether $\abs{\om}$ is larger or smaller than a critical value 
\begin{equation}\label{eq:ommax} 
\om_\mathrm{max} = 2 \lp \frac{\w_0}{3} \rp^{3/2}:
\end{equation}
\begin{itemize}
\item If $\abs{\om} < \om_\mathrm{max}$, the three roots are real. 
We denote them as $k_\om^{(i)}$, $i \in \lb 1, 2, 3 \rb$, with $k_\om^{(1)} < k_\om^{(2)} < k_\om^{(3)}$. 
The second one, $k_\om^{(2)}$, corresponds to a right-moving wave, whose group velocity $\lp \dd k_\om^{(2)} / \dd \om \rp^{-1}$ is positive. 
The two other roots correspond to left-moving waves, with negative group velocities. 
For $\om > 0$ , we also have $k_\om^{(1)} < 0$ and $k_\om^{(3)} > k_\om^{(2)} > 0$.
So, the quantity $k_\om^{(i)} \, \dd k_\om^{(i)} / \dd \om$, which will play an important role in the following, is strictly positive for $i \in \lb 1, 2 \rb$ and strictly negative for $i = 3$.  
In the low-frequency limit $\om = 0$, one finds $k_\om^{(2)} = 0$ and $k_\om^{(3)} = - k_\om^{(1)} = k_u$, where $k_u \equiv \sqrt{\w_0}$ is the wave vector of the ``tuned'' undulation $\eta^{(0)}$.
\item If $\abs{\om} > \om_\mathrm{max}$, only one real root remains: $k_\om^{(1)}$ for $\om > 0$ and $k_\om^{(3)}$ for $\om < 0$. The two other roots are complex conjugates with nonvanishing imaginary parts. 
\end{itemize}
As shown in appendix~\ref{app:lag_lin_KdV}, the sign of the energy of a wave is given 
by $\om \, k$. It is indicated in the figure by the style of the curve: continuous for positive-energy modes and dashed for negative-energy ones. 
In the following we concentrate on modes with $0 < \om < \om_\mathrm{max}$. 
Those with $- \om_\mathrm{max} < \om < 0$ are simply their complex conjugates, so there is no need to study them separately. 
The modes with $i \in \lb 2, 3 \rb$ then have positive energies while the mode with $i = 1$ has a negative energy. 
A resonance can occur if there exists $(i,j) \in \lb 1, 2, 3 \rb^2$ and $n_r \in \mathbb{N} \tsetminus \lb 0 \rb$ such that $k_\om^{(i)} - k_\om^{(j)} = n_r \, k_\w$, where $k_\w$ is defined above Equation~\eqref{eq:def_w_n}. 

Scattering on a ``detuned'' undulation is studied in appendix~\ref{app:detuned}. 
The main result is that, when there is a three-wave resonance, i.e., two linear waves interacting resonantly with the undulation, the corresponding modes are either bounded or exponentially increasing in one direction depending on the relative signs of the energy and energy flux of the two interacting waves. 
If they have energies with the same sign, then the mode is bounded in time; otherwise the mode grows or decays exponentially as $t \to \infty$. 
Similarly, it is spatially bounded if their energy fluxes have the same sign, and exponentially increasing or decreasing as $\abs{x} \to \infty$ otherwise. 
As we will see in the following, in the limit of a ``tuned'' undulation, which is itself a static solution of the KdV equation, this exponential behavior is replaced by a linear growth.

\subsection{Periodic static solutions of the (nonlinear) KdV equation} 
\label{sub:persol}

\begin{figure}
\includegraphics[width = 0.49 \linewidth]{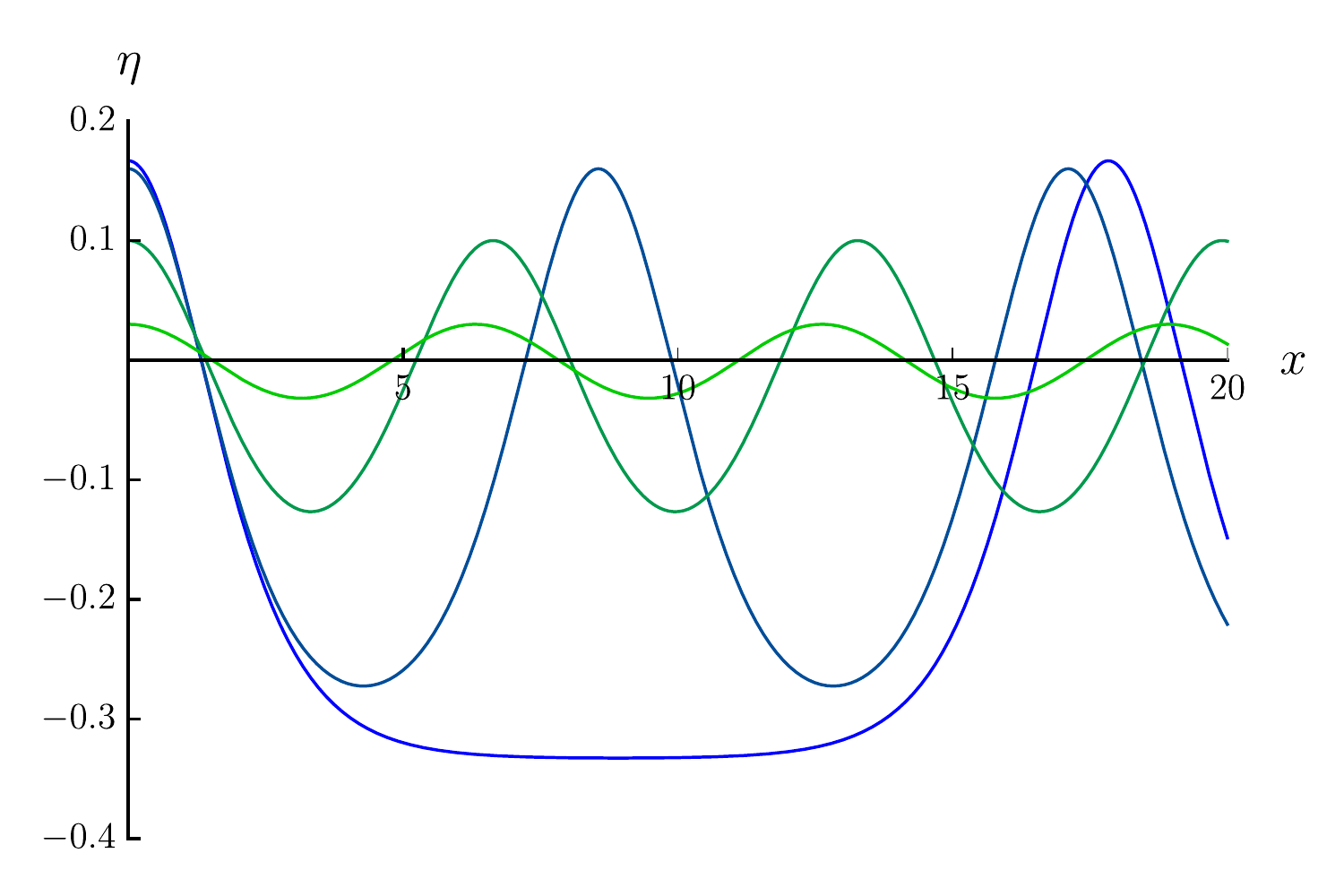}
\includegraphics[width = 0.49 \linewidth]{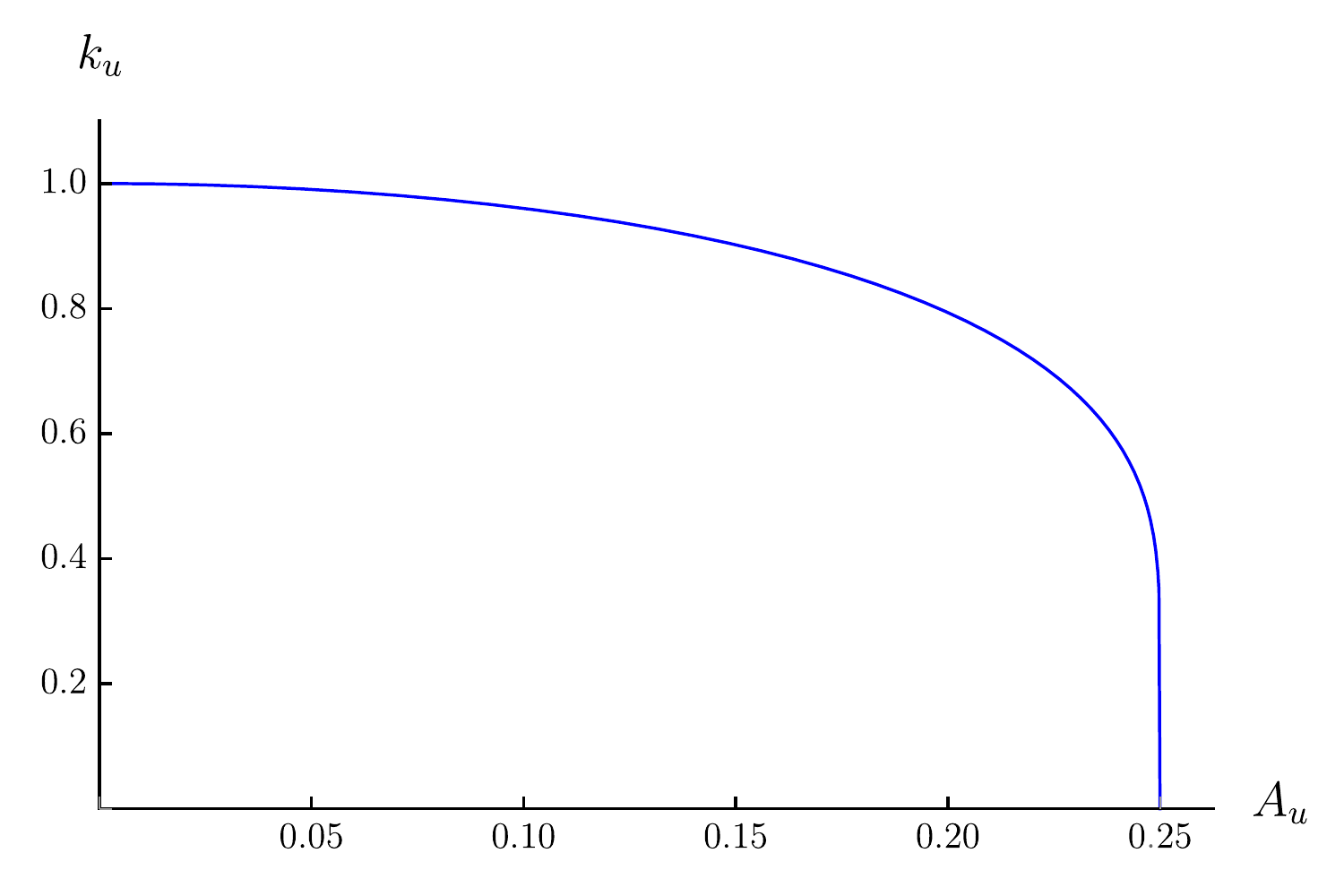}
\caption{\textit{Left panel:} Plots of four different periodic nonlinear solutions of Eq.~(\ref{eq:KdV_stat_int1}) with different amplitudes, for $\w = 1$ and $C = 0$. 
\textit{Right panel:} Wave vector $k_u$ as a function of the amplitude $A_u$, defined as half the difference between the maximum and minimum values reached by $\eta$.} \label{fig:KdV_unduls}
\end{figure}

As our analysis will rely on the structure of the stationary (nonlinear) solutions of \eq{eq:KdV}, we now review their most relevant properties. 
A more detailed account can be found in the textbook~\cite{Kamchatnov}. 
Setting $\pd_t \eta = 0$ in \eq{eq:KdV}, one obtains the stationary KdV equation
\begin{equation}
\pd_x \lp \w \, \eta + \pd_x^2 \eta + 3 \s \eta^2 \rp = 0. 
\end{equation}
Integration over $x$ gives
\begin{equation} \label{eq:KdV_stat_int1}
\w \, \eta + \pd_x^2 \eta + 3 \s \eta^2 = C, 
\end{equation}
where $C$ is an integration constant, see the left panel of Fig.~\ref{fig:KdV_unduls} for four solutions with $\w = 1$ and $C = 0$. 
We assume $\w^2 + 12 \s C > 0$.~\footnote{This condition is equivalent to the existence of real homogeneous solutions, given by $\eta = \lp \pm \sqrt{\w^2 + 12 \s C} - \w \rp \big/ 6$.}
Let $\eta_0 \equiv \lp \sqrt{\w^2 + 12 \, C} - \w \rp \big/ 6$ and $\Deta \equiv \eta - \eta_0$. 
Equation~\eqref{eq:KdV_stat_int1} becomes
\begin{equation}
\lp \w + 6 \s \eta_0 \rp \, \Deta + \pd_x^2 \Deta + 3 \, \Deta^2 = 0.
\end{equation} 
Multiplication by $\pd_x \Deta$ gives
\begin{equation}
\pd_x \lp 
	\frac{\w + 6 \s \eta_0}{2} \, \Deta^2 
	+ \frac{1}{2} \s \lp \pd_x \Deta \rp^2
	+ \Deta^3 
\rp = 0, 
\end{equation}
which can be integrated over $x$, giving 
\begin{equation}
\lp \pd_x \Deta \rp^2 = - 2 \s \lp \Deta - \eta_1 \rp \s \lp \Deta - \eta_2 \rp \s \lp \Deta - \eta_3 \rp,
\end{equation}
where $\lp \eta_1, \eta_2, \eta_3 \rp \in \mathbb{C}^3$ satisfies 
\begin{equation}
\lb
\begin{aligned}
& \eta_1 + \eta_2 + \eta_3 = - \frac{\w + 6 \s \eta_0}{2} \\
& \eta_1 \s \eta_2 + \eta_2 \s \eta_3 + \eta_3 \s \eta_1 = 0
\end{aligned}
\right. ,
\end{equation}
and $\eta_1 \s \eta_2 \s \eta_3$ is another free integration constant. 
Spatially bounded solutions exist if and only if $\lp \eta_1, \eta_2, \eta_3 \rp \in \mathbb{R}^3$. 
Ordering these three numbers as $\eta_1 \leq \eta_2 \leq \eta_3$, the bounded solutions are periodic if $\eta_2 \neq \eta_1$, with $\eta$ oscillating between $\eta_2$ and $\eta_3$. 
The wavelength $\la$ is:
\begin{equation} \label{eq:KdV_la}
\la = \frac{2 \s \sqrt{2}}{\sqrt{\eta_3 - \eta_1}} \, K \lp \sqrt{\frac{\eta_3 - \eta_2}{\eta_3 - \eta_1}} \rp = 4 \int_0^{\pi / 2} \frac{\dd \theta}{\sqrt{\sqrt{\bar{\w}^2 - 3 \, \lp \eta_3 - \eta_2 \rp^2} + \lp \eta_3 - \eta_2 \rp \, \cos \lp 2 \s \theta \rp}},
\end{equation}
where $K$ is the complete elliptic integral of the first kind and $\bar{\w} \equiv \w + 6 \s \eta_0$.
The wavelength $\la$ diverges in the limit $\eta_2 \to \eta_1$. 
In the following, it will be convenient to use the wave vector $k_u \equiv  2 \pi / \la$, whose dependence on the amplitude is shown in \fig{fig:KdV_unduls}, right panel. 

We concentrate on small-amplitude undulations, i.e., with $\eta_2 \approx \eta_3$. 
In this limit, the wave vector $k_u$ tends to $\sqrt{\bar{\w}}$, the strictly positive zero-frequency root of the dispersion relation over the solution $\eta = \eta_0$, see Fig.~\ref{fig:DR_KdV}. 
This is an important point: the wavelength of the undulation is, to lowest order, always equal to that of a zero-frequency solution of the linearized equation. This property, also found in the model of Appendix~\ref{app:UCP}, is the reason why the undulation has an important effect on the scattering at low frequencies. 
To next order in $\eta_3 - \eta_2$, a straightforward calculation using \eq{eq:KdV_la} gives
\begin{equation} \label{eq:KdV_ku}
k_u = \sqrt{\bar{\w}} \, \lp 1 - \frac{15}{4 \bar{\w}^2} A_u^2 + \ord{A_u^3} \rp,
\end{equation}
where $A_u = (\eta_3 - \eta_2) / 2$ is the amplitude of the undulation. 
The solution takes the form
\begin{equation} \label{eq:KdV-undul-sols}
\eta_u (x) =  
	\eta_0 
	+ A_u \cos \lp k_u \s \lp x - x_u \rp \rp
	- \frac{3 A_u^2}{2 \bar{\w}}
	+ \frac{A_u^2}{2 \bar{\w}} \cos \lp 2 \s k_u \s \lp x - x_u \rp \rp
	+ \frac{3 A_u^3}{16 \s \bar{\w}^2} \cos \lp 3 \s k_u \s \lp x - x_u \rp \rp 
	+ \ord{A_u^4},
\end{equation}
where $x_u \in \mathbb{R}$ determines the phase at $x = 0$. 
The main result to keep in mind is that the spatially bounded solutions are labeled by three continuous parameters: $\eta_0$, giving the mean value of $\eta$ to leading order, the amplitude of the oscillations $A_u$, and $x_u$, giving (on multiplication by $k_u$) the phase at the origin $x = 0$. 

\subsection{Modes over the undulation and transfer matrix}  
\label{subsec:tuned_mat}

Let us now determine the low-frequency modes over the undulation from a linear calculation. 
An alternative derivation in the zero-frequency limit is given in Appendix~\ref{app:lingrowmodes}. 
Denoting by $k_\om^{(i)}, \, i \in \lb 1, 2, 3 \rb$ the solutions of the dispersion relation~\eqref{eq:DR} at fixed angular frequency $\om$, ordered as $k_\om^{(1)} < k_\om^{(2)} < k_\om^{(3)}$ for $\abs{\om} < \om_\mathrm{max}$, we have for $\om = 0$:
\begin{equation*}
k_0^{(1)} = - k_u, \; k_0^{(2)} = 0, \; \text{and} \; k_0^{(3)} = k_u.
\end{equation*} 
There are thus two resonances involving two linear waves and the undulation: 
\begin{itemize}
\item a first-order resonance due to $k_0^{(3)} - k_0^{(2)} = k_0^{(2)} - k_0^{(1)} = k_u$, giving a contribution linear in $A_u$,
\item a second-order resonance due to $k_0^{(3)} - k_0^{(1)} = 2 k_u$, whose contribution is of order $A_u^2$. 
\end{itemize}
To obtain the corresponding resonant modes, we solve \eq{eq:KdV-linearized} in a background $\eta^{(0)} = \eta_u$ given by \eq{eq:KdV-undul-sols}.  
Since the coefficients of the differential equation~\eqref{eq:KdV-linearized} are independent of $t$ and periodic in $x$ with period $2 \pi / k_u$, one can look for Bloch wave solutions~\cite{Kittel79}: 
\begin{equation} \label{eq:KdV:etaexp}
\deta: (t,x) \mapsto \e^{- \ii \s \om \s t + \ii \s k \s x} \rho(x),
\end{equation} 
where $\om$ and $k$ are two complex numbers giving respectively the angular frequency and the quasimomentum of the solution $\deta$, and where $\rho \in C^3 \lp \mathbb{R}, \mathbb{C} \rp$ is periodic with period $2 \pi / k_u$. 
Let $(\rho_n)_{n \in \mathbb{Z}}$ be the coefficients of its Fourier expansion, defined so that 
\begin{equation} \label{eq:KdV:rhoexp}
\rho(x) = \sum_{n \in \mathbb{Z}} \rho_n \s \e^{\ii \s n \s k_u \s x}. 
\end{equation}
The mode is normalized with respect to the inner product~\eqref{eq:innerprod} if $\rho_0$ is chosen such that
\begin{equation} 
2 \s \pi \s \abs{\frac{\dd k}{\dd \om}}^{-1} \s \abs{\sum_{n \in \mathbb{Z}} \frac{\abs{\rho_n}^2}{k + n \s k_u}} = 1,
\end{equation}
and the sign of its energy is 
\begin{equation}
\mathrm{sgn} \lp
	 \lp \frac{\dd k}{\dd \om} \rp^{-1} \s 
	 \sum_{n \in \mathbb{Z}} \frac{\abs{\rho_n}^2}{k + n \s k_u}
\rp . 
\end{equation}
Similarly, $\eta_u$ can be expanded as
\begin{equation} \label{eq:KdV:etauexp}
\eta_u (x) = \sum_{n \in \mathbb{Z}} \eta_n \s \e^{\ii \s n \s k_u \s x},
\end{equation} 
where $\lp \eta_n \rp_{n \in \mathbb{Z}} \in \mathbb{C}^\mathbb{Z}$. 
Plugging \eq{eq:KdV:etaexp} in \eq{eq:KdV-linearized} and using Equations~(\ref{eq:KdV:rhoexp}) and~(\ref{eq:KdV:etauexp}) gives the recursion relation:
\begin{equation} \label{eq:KdV:recrel}
\forall \, n \in \mathbb{Z}, \; 
	\lp \om - \om_0 \lp k + n \s k_u \rp \rp \rho_n
	- 6 \s \lp k + n \s k_u \rp \sum_{l \in \mathbb{Z}^*} \eta_l \s \rho_{n-l}
	= 0,
\end{equation}
where 
\begin{equation} \label{eq:KdV:DR_hom_sol} 
\om_0 (k) = \bar{\w} \, k - k^3
\end{equation}
gives the relation between $\om$ and $k$ on a homogeneous solution $\eta = \eta_0$. 
We define a small parameter $\ep$ and solve \eq{eq:KdV:recrel} perturbatively in $\ep$, assuming the following scalings: 
\begin{equation} 
\om = \orde{}; \; \; 
\forall \, n \in \mathbb{Z}, \; \eta_n = O \big( \ep^\abs{n} \big); \; \; \text{and} \; \; 
 \forall \, n \in \mathbb{Z}, \; \rho_n  = O \big( \ep^{\textrm{max}(\abs{n}-1,0)} \big). 
\end{equation} 
As for the previous expansion \eq{eq:exp_eps} in the detuned case, the KdV equation can then be solved order by order in $\ep$. 
As could be expected from the unperturbed dispersion relation, we find three modes: two dispersive ones $\deta_\om^{(\pm)}$ whose group velocities go to $\om_0' \lp k_u \rp$ in the limit $\ep \to 0$, and a hydrodynamic one $\deta_\om^{(h)}$ whose group velocity tends to $\om_0'(0)$. 
The former are given by
\begin{equation} \label{eq:modes_d}
\deta_\om^{(s)}: 
\left\lbrace
\begin{aligned}
& \rho_0 = - \frac{2 \s \om}{k_u^2} \s \rho_c + \orde{2} \\
& \rho_{\pm 1} = \lp
	\frac{\om}{A_u} \mp s \sqrt{\frac{\om^2}{A_u^2} + 4 \s k_u^2 + \orde{3}}
	\rp \rho_c + \orde{} \\
& \rho_{\pm 2} = \lp 1 \pm \frac{\om}{2 \s k_u^3} \rp \frac{A_u}{k_u^2} \s \rho_{\pm 1} 
	+ \orde{3} \\
& \rho_{\pm 3} = \frac{9 \s A_u^2}{16 \s k_u^4} \s \rho_{\pm 1}
	+ \orde{3} 
\end{aligned}
\right. ,
\end{equation}
where $s \in \lb -1, +1 \rb$ and $\rho_c$ is a constant.
The corresponding quasimomenta are
\begin{equation}
k = - \frac{\om}{2 \s k_u^2} + s \s \delta k_d,
\end{equation}
where 
\begin{equation}
\delta k_d \equiv 
	\frac{3}{8 \s k_u^4} \s \mathrm{sgn}(\om)
	\sqrt{\frac{\om^4}{k_u^2}
		+ 4 \s \om^2 \s A_u^2
		+ \orde{5}} .
\end{equation}
The hydrodynamic mode is given by 
\begin{equation} \label{eq:mode_h}
\deta_\om^{(h)} : 
\left\lbrace
\begin{aligned}
& \rho_0 = \frac{\om}{A_u \s k_u} \s \rho_c + \orde{2} \\
& \rho_{\pm 1} = \pm \rho_c + \orde{} \\
& \rho_{\pm 2} = \pm \frac{A_u}{k_u^2} \s \lp 1 \mp \frac{\om}{k_u^2} \rp \rho_c
	+ \orde{3} \\
& \rho_{\pm 3} = \pm \frac{9 \s A_u^2}{16 \s k_u^4} \s \rho_c
	+ \orde{3} 
\end{aligned}
\right. .
\end{equation}
Its quasimomentum satisfies $\om_0(k) = \om + \orde{3}$, i.e., is unchanged to this order. 

The stationary modes obtained by variations of the nonlinear solution of the KdV equation, $\pd_{x_u} \eta_u$, $\pd_{\eta_0} \eta_u$, and $\pd_{A_u} \eta_u$ (see Eqs.~(\ref{KdV_statmode1})--(\ref{KdV_statmode3})), can be obtained as the limit $\om \to 0$ of the three modes of Equations~(\ref{eq:modes_d}) and~(\ref{eq:mode_h}). 
Indeed, a straightforward calculation shows that (up to a global factor) $\deta_\om^{(h)}$ and $\deta_\om^{(\pm)}$ converge uniformly toward $\pd_{x_u} \eta_u$ in the limit $\om \to 0$, while the two other modes are obtained through
\begin{equation} 
\frac{A_u}{4 \s \om \s \rho_c} \s \lp \deta_\om^{(+)} + \deta_\om^{(-)} \rp \mathop{\longrightarrow}_{\om \to 0}
	\pd_{A_u} \eta_u + 2 \s \frac{A_u}{\w} \s \lp 1 - \frac{9 \s A_u^2}{\w^2} \rp \pd_{\eta_0} \eta_u 
\end{equation}
and
\begin{equation} 
\frac{A_u \s k_u}{\om \s \rho_c} \s \deta_\om^{(h)} + \frac{A_u}{4 \s \om \s \rho_c} \s \lp \deta_\om^{(+)} - \deta_\om^{(-)} \rp
	\mathop{\longrightarrow}_{\om \to 0} \lp 1 - \frac{9 \s A_u^2}{\w^2} \rp \s \pd_{\eta_0} \eta_u.
\end{equation}
(In these expressions, the convergence is uniform on any bounded domain of $\mathbb{R}^2$.) 

\begin{figure}
\begin{center}
\includegraphics[width = 0.49 \linewidth]{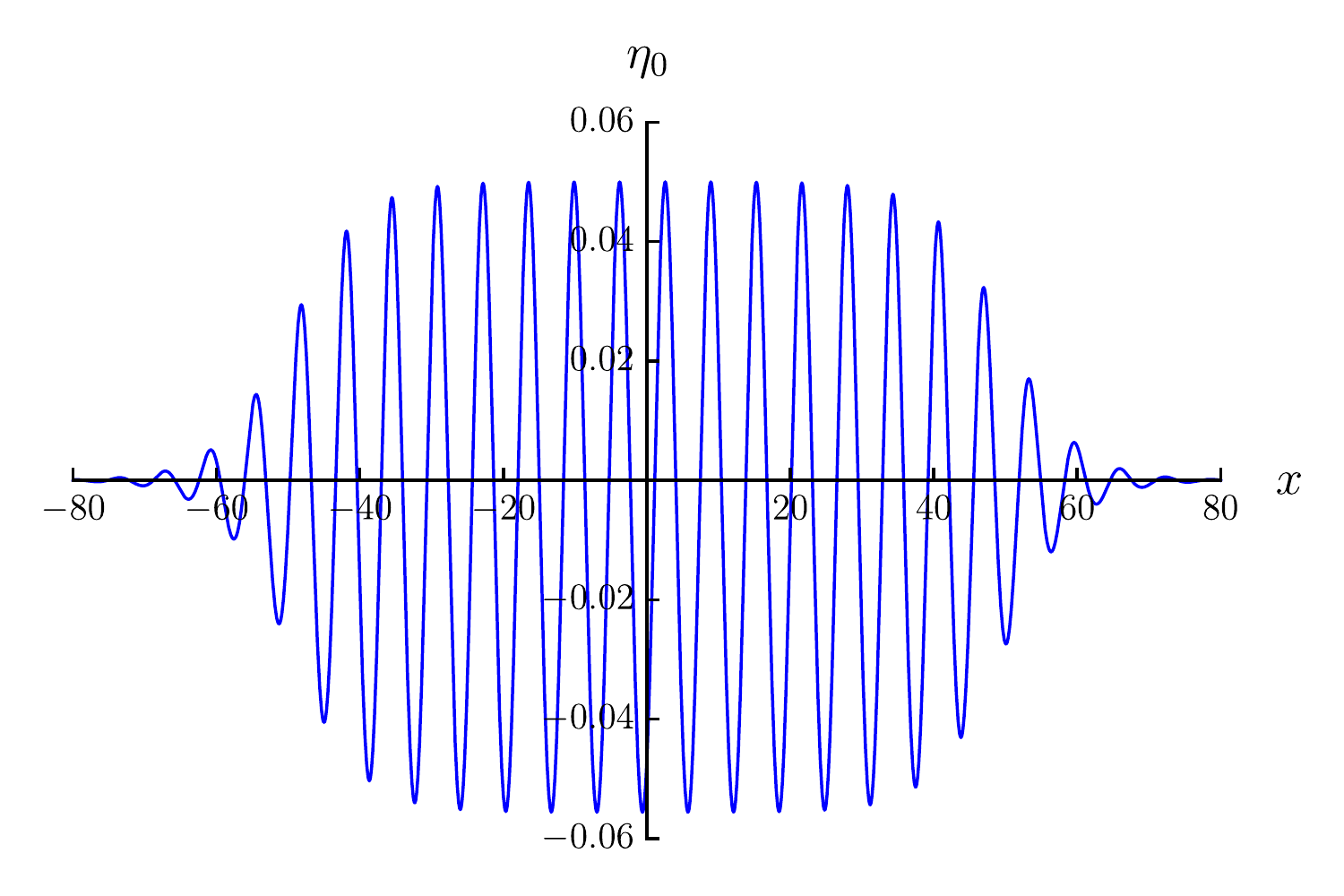}
\end{center}
\caption{Asymptotically turned-off undulation (see~\eq{eq:offundul}) for $L_u = 100$, $A_u \approx 0.1$, and $\sigma_u = 0.1$.} \label{fig:offundul}
\end{figure}

\begin{figure} 
\begin{center}
\includegraphics[width = 0.49 \linewidth]{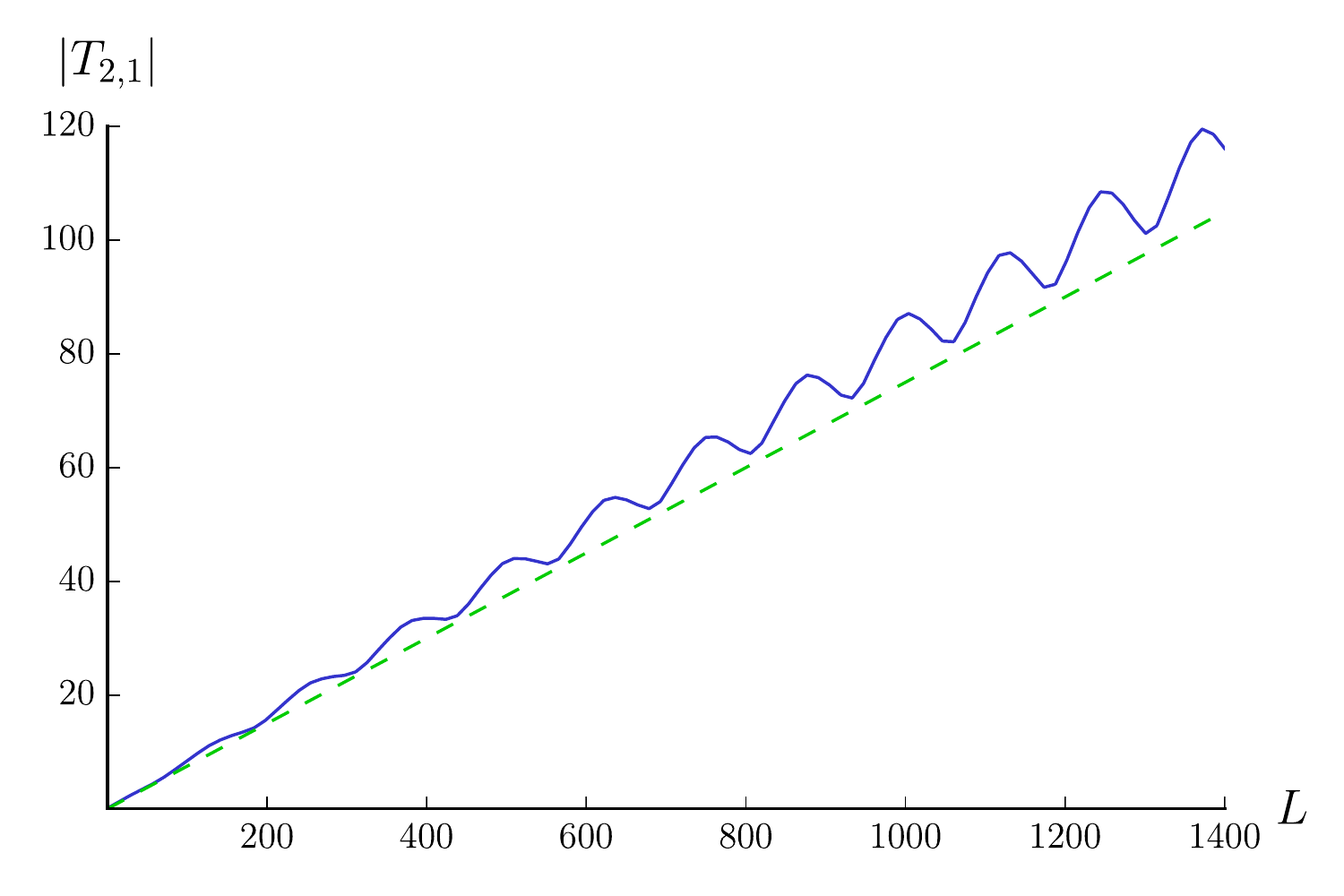}
\includegraphics[width = 0.49 \linewidth]{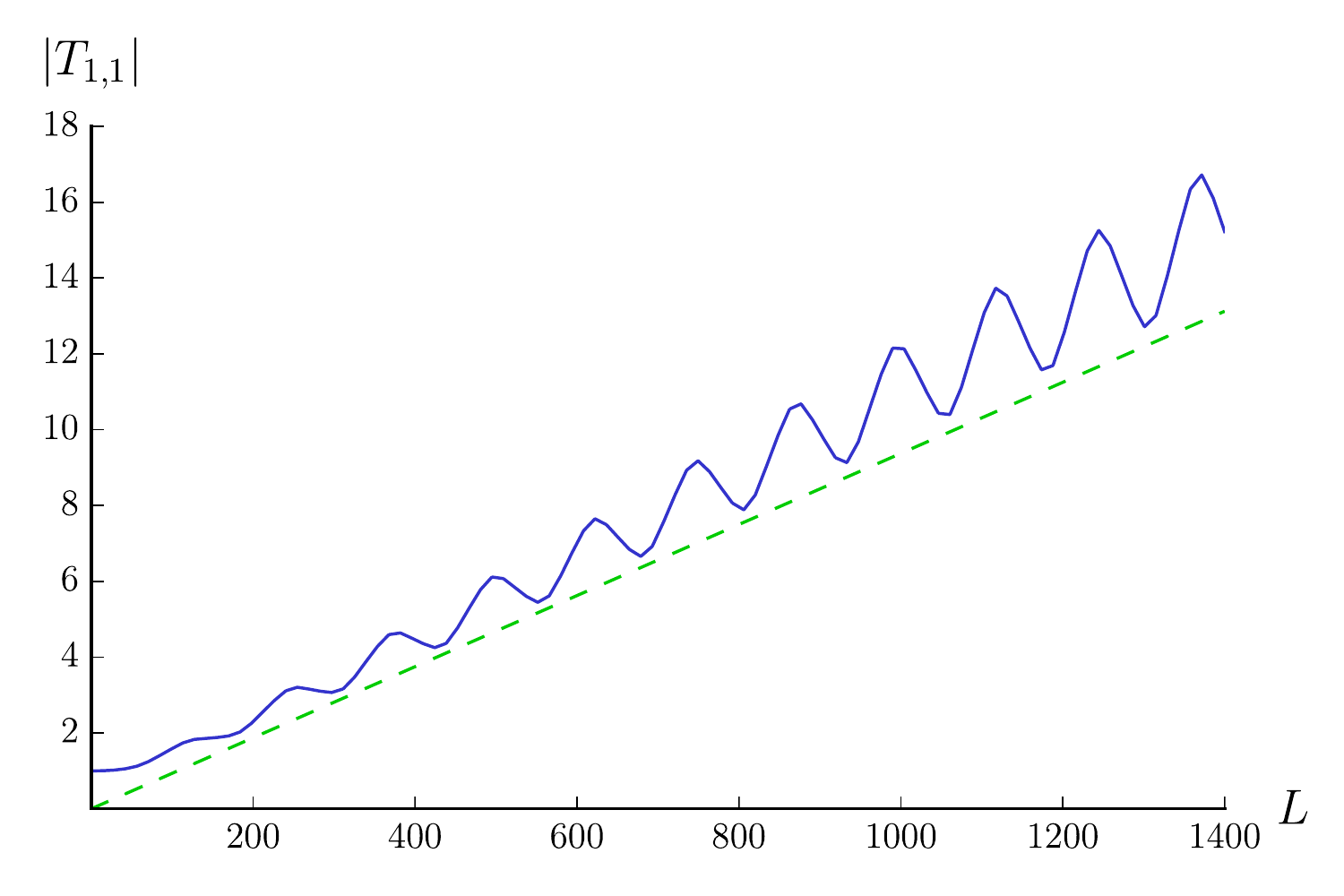}
\includegraphics[width = 0.49 \linewidth]{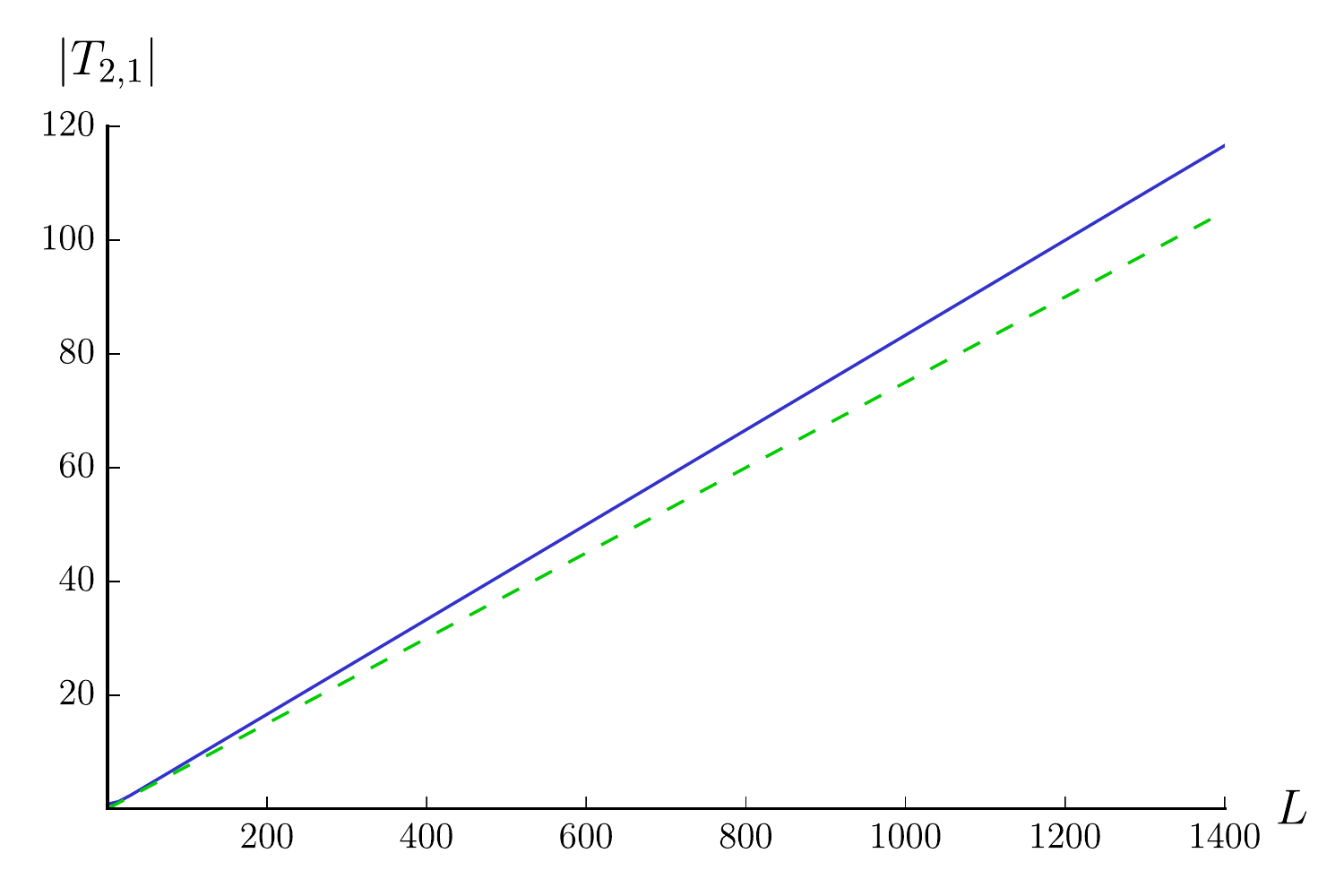}
\includegraphics[width = 0.49 \linewidth]{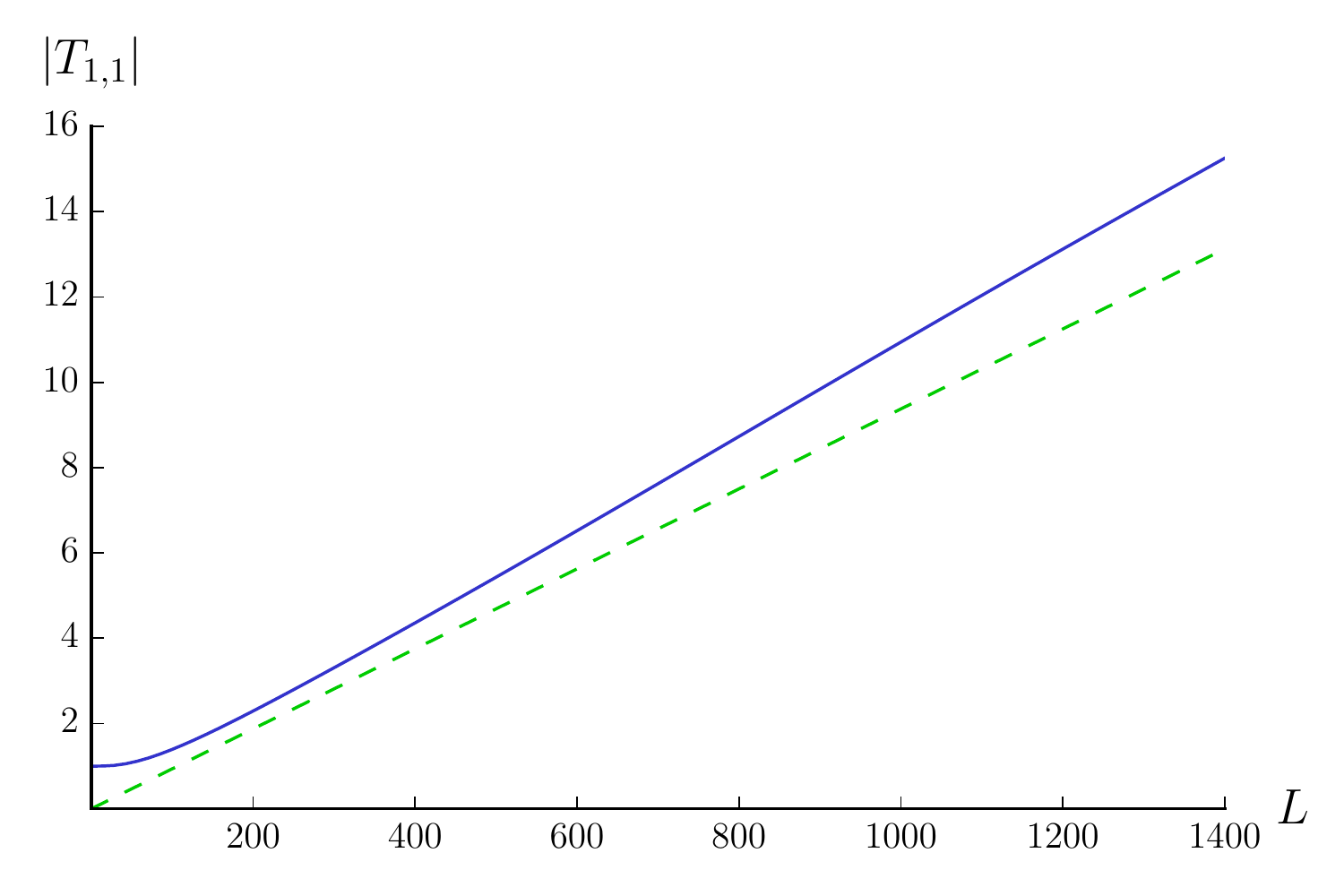}
\end{center}
\caption{Plots of the absolute values of the coefficients $T_{2,1}$ (left panels) and $T_{1,1}$ (right panels) as functions of the length $L$ of an undulation of the form~\eqref{eq:offundul}, for $A_u \approx 0.05$, $\w = 1$, and for $\sigma_u = 1$ (top panels) and $\sigma_u = 0.1$ (bottom panels).  
Blue, continuous lines show numerical results, while the green, dashed lines show the analytical approximation~\eqref{eq:KdV_T_main}. 
The difference between the average slopes is due to higher-order effects, and the oscillations visible on the top plots seem due to the sharp variation of the amplitude at both ends of the undulation.} \label{fig:KdV_T}
\end{figure}

One can compute the transfer matrix $T$ at zero frequency over a damped undulation of the form
\begin{equation} \label{eq:offundul}
\eta_{u,d} (x) = \frac{\eta_u(x)}{2} \s \left[ 1 - \tanh \lp \sigma_u \s \lp x - L_u / 2 \rp \rp \tanh \lp \sigma_u \s \lp x + L_u / 2 \rp \rp \right], 
\end{equation}
where $\sigma_u$ and $L_u$ are two strictly positive numbers. 
One example is shown in \fig{fig:offundul}.  
Following the notation adopted in Fig.~\ref{fig:DR_KdV}, let us denote with an index $2$ the hydrodynamic wave with vanishing wave vector while indices $1$ and $3$ denote the dispersive waves with wave vectors going to $\pm \sqrt{\bar{\w}}$ for $\om \to 0$. 
To simplify the expressions, we here give the results for modes normalized to have a single plane wave with unit amplitude on the left or on the right of the undulation.
Taking the double limit $A_u \to 0$, $L_u \to \infty$ at fixed $A_u \s L$, 
the results become independent of $\sigma_u$. 
Retaining only the leading terms gives
\begin{equation} \label{eq:KdV_T_main} 
\begin{aligned}
& \abs{T_{2,1}} \sim \abs{T_{2,3}} \sim \frac{3 \, A_u}{2 \, \sqrt{\w}} \, L_u, \\
& \abs{T_{1,1} - 1} \sim \abs{T_{1,3}} \sim \abs{T_{3,1}} \sim \abs{T_{3,3} - 1} \sim \frac{15 \, A_u^2}{4 \, \w^{3/2}} \, L_u .
\end{aligned}
\end{equation}
The transfer coefficients $T_{2,2}-1$, $T_{1,2}$, and $T_{3,2}$ go to zero faster than $A_u$ in this limit. 
The behavior of the coefficients $T_{2,1}$ and $T_{1,1}$ is compared with results from a numerical resolution of the KdV equation in \fig{fig:KdV_T}. 
The important point is that imposing a hydrodynamic wave with an amplitude of order $1$ on one side will generate dispersive waves with amplitudes linear in $A_u \s L_u$ on the other side, while the coefficients relating two dispersive waves are in $A_u^2 \s L_u$. 
As will be shown in Section~\ref{sec:WH}, this is crucial to understand the effect of the undulation on the scattering in a white-hole-like flow. 
Moreover, this scaling is preserved in the more realistic model of appendix~\ref{app:UCP_tuned}, showing that the results obtained with the KdV equation qualitatively extend to that model.

In this section we have thus obtained the low-frequency modes over a small-amplitude undulation and determined the corresponding transfer matrix. 
To study the scattering of water waves in the presence of undulation, there remains to determine which linear combination of these modes enters the incoming mode in a given configuration. 
This is done in Section~\ref{sec:WH} for a ``white-hole'' flow.

\section{White-hole-like flows} 
\label{sec:WH}

\subsection{Setup and general idea}

As mentioned in the introduction, one motivation for studying water waves in the context of analogue gravity is the possibility to realize ``white hole'' flows with an analogue horizon. 
For definiteness, we work with flows oriented to the left. 
To fix the ideas, we here define a white-hole flow by the five properties: 
\begin{enumerate}
\item The flow velocity $v$ and speed of long-wavelength perturbations in the fluid frame $c$ are piecewise-continuous functions of $x$ such that $c$ is positive and $v$ does not change sign;
\item \label{ite:1} $v(x)$ and $c(x)$ are asymptotically uniform in the limits $x \to \pm \infty$;
\item \label{ite:2} The flow is subcritical in the downstream region: 
\begin{equation}
\lim_{x \to - \infty} \abs{\frac{v(x)}{c(x)}} < 1; 
\end{equation}
\item It is supercritical in the upstream region: 
\begin{equation}
\lim_{x \to + \infty} \abs{\frac{v(x)}{c(x)}} > 1; 
\end{equation}
\item \label{ite:4} There exists only one point $x_h \in \mathbb{R}$ such that 
\begin{equation}
\lim_{x \to x_h^-} \abs{\frac{v(x)}{c(x)}} \leq 1 \; \; \text{and} \; \; \lim_{x \to x_h^+} \abs{\frac{v(x)}{c(x)}} \geq 1. 
\end{equation}
The point $x_h$ is the analogue horizon.~\footnote{Assuming $v / c$ is continuous, this condition simplifies to: there exists only one $x_h \in \mathbb{R}$ such that $\abs{v(x_h) / c(x_h)} = 1$.} 
\end{enumerate}  
Such setups have been extensively studied, see for instance Refs.~\cite{Macher:2009tw,Coutant:2011in,Coutant:2012mf,Coutant:2014cwa,Michel:2014zsa,Robertson:2016ocv}. 
One important result, related to the Hawking effect in astrophysical black holes, is the divergence of the scattering coefficient $\beta_\om$ relating the incoming counterpropagating mode (corresponding to a wave sent from the downstream region upward) to the outgoing negative-energy wave. 
(An analytical proof is given in~\cite{Coutant:2017qnz}.) 
More precisely, $\beta_\om$ gives the amplitude of the normalized (in the sense of \eq{eq:normom}, see Appendix~\ref{app:lag_lin_KdV} for details) negative-energy wave obtained when sending from the left a normalized counterpropagating incident wave with angular frequency $\om$. 
The crucial point is that $\abs{\om \s \beta_\om^2}$ has a finite limit as $\om \to 0$, which is interpreted as the effective temperature $T_\mathrm{eff}$ of the analogue horizon.~\footnote{In quantum fluids,
choosing units in which the Boltzmann and Planck constants are equal to $1$, $T_\mathrm{eff}$ is the low-frequency temperature of the ensemble of quasiparticles spontaneously produced by the analogue Hawking effect. 
In general, the effective temperatures defined using the various incoming modes will differ due to the coupling between co- and counterpropagating modes in the rest frame of the fluid~\cite{Macher:2009tw}. 
However, when using the KdV equation there are only two independent incoming modes over a transcritical flow, and conservation of the inner product~\eqref{eq:innerprod} ensures their effective temperatures are equal. 
}

The effective temperature can be determined from the zero-frequency limits of the amplitudes of the plane waves in the asymptotic downstream region. 
Let us denote by $A_\mathrm{in}$ that of the incident wave and $A_\mathrm{neg}$ that of the negative-energy one. 
Using \eq{eq:norm_homo} to relate them to normalized waves and the dispersion relation~\eqref{eq:DR} to relate the wave vectors to the angular frequency $\om$, one finds
\begin{equation} \label{eq:effTeff}
T_\mathrm{eff} = 2 \s \bar{\w}(-\infty)^{3/2} \s \abs{\frac{A_\mathrm{neg}}{A_\mathrm{in}}}^2.
\end{equation}
For sufficiently smooth white-hole flows, one can show~\cite{Coutant:2011in} that $T_\mathrm{eff} \approx \kappa / (2 \s \pi)$, where $\kappa \equiv \abs{\pd_x \lp v + c \rp_{x = x_h}}$ is the analogue of the surface gravity.  

However, as also mentioned in the Introduction, flows realized in experiments generally have an undulation in the downstream region. 
To avoid ambiguity, we shall refer to them as ``white-hole-like'' flows, for which the conditions~\ref{ite:1} and~\ref{ite:2} above are replaced, respectively, by
\begin{enumerate}
\item[2'.] $v(x)$ and $c(x)$ are asymptotically uniform in the limit $x \to + \infty$ and periodic at $x \to - \infty$;
\item[3'.] The flow is subcritical in the downstream region:
\begin{equation}
\exists \, x_M \in \mathbb{R}, \; \forall \s x \in \mathbb{R}, \; x < x_M \Rightarrow \abs{v(x) / c(x)} < 1.
\end{equation}
\end{enumerate}
Our aim in this section is to understand the effects of a small-amplitude undulation on $T_\mathrm{eff}$. 
To this end, we determine the wave content of the relevant mode as a function of $x$, 
which may be thought of as the position of a detector measuring the local wave amplitudes using a WKB approximation (or, equivalently, as the total length of the undulation if the measurement is performed in the flat asymptotic downstream region). 
As we shall see, because of the resonance discussed in Section~\ref{sec:KdV_tuned}, their amplitudes change significantly even when going far away from the horizon, which affects the effective temperature. 
When decreasing $x$ from $x_h$ toward $-\infty$, one obtains two limit regimes:
\begin{itemize}
\item Close to the horizon (more precisely, for small values of $A_u \s \lp x_h - x \rp$), in the limit $A_u \ll 1$, the contribution from the undulation is negligible and one recovers the temperature $T_\mathrm{eff}^{(0)}$ of a white-hole flow without undulation;
\item For $x_h - x \gg A_u^{-2}$, the wave content is dominated by the linearly growing terms over the undulation: $\beta_\om$ then becomes proportional to $A_u$ and the effective temperature scales like $A_u^2$.
\end{itemize}
For this reason, and as will be shown more precisely below, the two limits $A_u \to 0$ (small undulation) and $x \to -\infty$ (long undulation) do not commute: sending first $A_u$ to $0$ and then $x$ to $-\infty$ gives $T_\mathrm{eff} = T_\mathrm{eff}^{(0)}$, while taking first $x \to -\infty$ and then $A_u \ll 1$ gives a different effective temperature $T_\mathrm{eff}^{(u)}$, determined by the behavior of the modes of Section~\ref{sec:KdV_tuned} over the undulation. 
In Section~\ref{KdV_WH:2} we consider a toy model in which $T_\mathrm{eff}^{(0)}$ and $T_\mathrm{eff}^{(u)}$ can be computed explicitly as well as the interpolation between them when decreasing $x$. 
In Section~\ref{KdV_WH:3} we generalize the calculation of $T_\mathrm{eff}^{(u)}$ using a forced KdV equation. 

\subsection{Low-frequency effective temperature: An explicit calculation}
\label{KdV_WH:2}

Let us consider the KdV equation with variable coefficient:
\begin{equation} \label{eq:varKdV}
\pd_t \eta + \pd_x \lp \w \s \eta \rp + \pd_x^3 \eta + 6 \s \eta \s \pd_x \eta = 0,
\end{equation}
where $\w$ is a given function of $x$. 
We focus on time-independent solutions. 
Integrating \eq{eq:varKdV} over $x$ gives
\begin{equation}
\w \s \eta + \pd_x^2 \eta + 3 \s \eta^2 = C,
\end{equation}
where $C$ is a real constant. 
To be specific, let us assume that $\w$ has a steplike profile:
\begin{equation} \label{eq:step}
\w (x) = 
\lb 
\begin{aligned}
& \w_- & & x < 0 \\
& \w_+ & & x > 0
\end{aligned}
\right. 
\end{equation}
for some real numbers $\w_- > 0$ and $\w_+ < 0$. 
The trivial solution $\eta = 0$ then corresponds to a white-hole flow with negative velocity. 
Stationary perturbations in the downstream region $x < 0$ are given by Eqs.~(\ref{eq:KdV_ku},\ref{eq:KdV-undul-sols}) with $\bar{\w}$ replaced by $\sqrt{\w_-^2 + 12 \s C}$ and with $\eta_0 = (k_u^2 - \w_-)/6$. 
A similar procedure can be used to find the stationary solutions in the upstream region $x > 0$. 
At fixed $C$, they are described by a single parameter $A_d \in \mathbb{R}$:
\begin{equation} \label{eq:dec_sol} 
\eta_d (x) = 
	- \frac{\w_+ + k_d^2}{6}
	+ A_d \s \e^{- k_d \s x} 
	- \frac{A_d^2}{k_d^2} \s \e^{- 2 \s k_d \s x}
	+ \frac{3 \s A_d^3}{4 \s k_d^4} \e^{- 3 \s k_d \s x}
	+ \ord{A_d^4},
\end{equation}
where $k_d \equiv \lp \w_+^2 + 12 \s C \rp^{1/4}$ is the decay rate of the perturbation to linear order. 
The general global solution is then given by \eq{eq:dec_sol} for $x > 0$ and by \eq{eq:KdV-undul-sols} for $x < 0$, with amplitude $A_u$ and phase $k_u \s x_u$ related to $A_d$ by the requirement that $\eta$ be continuous and differentiable at $x = 0$. 
To leading order in $A_d$, these two requirements become
\begin{equation}
\sin \lp k_u \s x_u \rp \approx - \frac{k_d \s A_d}{k_u \s A_u}
\end{equation}
and
\begin{equation} \label{eq:AuAd}
A_u \approx \sqrt{\lp \frac{k_d^2 + k_u^2 + \w_+ - \w_-}{6} - A_d \rp^2 + \frac{k_d^2}{k_u^2} \s A_d^2}.
\end{equation}

We now wish to compute the incoming, counterpropagating mode $\phi^{(\mathrm{in})}_\om$ and its derivative $\deta^{(\mathrm{in})}_\om \equiv \pd_x \phi_\om^{(\mathrm{in})}$ at low frequency.~\footnote{The field $\phi$ is defined as the integral of $\deta$ over $x$. As is shown in appendix~\ref{app:lag_lin_KdV}, $\phi$ is the field appearing in the Lagrangian formulation of the KdV equation, and the natural quantity to define normalized modes.} 
We consider two cases: 
\begin{enumerate}
\item As a warm-up exercise, we compute it over the homogeneous solution $\eta^{(0)} = 0$ and determine the corresponding effective temperature $T_\mathrm{eff}^{(0)}$;
\item We then turn to the case $A_d \neq 0$, and thus $A_u \neq 0$, to see the effects of the undulation.  
\end{enumerate}

\noindent \textbf{\\ Low-frequency effective temperature on the homogeneous solution $\eta = 0$: \\ \phantom{0}} 

\noindent 
To compute the effective temperature, one needs the structure of the incoming counterpropagating mode
$\deta_\om^{(\mathrm{in})}$, which corresponds to sending a wave from the left, in the low-frequency limit. 
We first notice that the waves with real wave vectors in the region $x > 0$ are incoming, 
with group velocities $v \pm c < 0$, and are thus absent in the mode we are interested in.~\footnote{In the derivation of the KdV equation, the limit $\abs{v / c} \to 1$ is taken. As a result, $v-c$ is not well defined and copropagating waves are uniform. Their group velocity is thus undefined. However, when using a more refined model such as that of Appendix~\ref{app:UCP_undul}, one finds that their group velocity is negative. 
The interested reader may also notice that this degeneracy is related to the constraint mentioned in footnote~\ref{foot:cons}.} 
This mode thus must be exponentially decreasing for $x \to \infty$. 
Second, we note that at zero frequency any stationary mode is an infinitesimal difference between two nonlinear stationary solutions. 
(See Section~\ref{subsec:tuned_mat} and appendix~\ref{app:lingrowmodes}.) 

In our case, the relevant stationary solution depends on the two parameters $C$ and $A_d$, giving two possible asymptotically bounded modes. 
However, as can be seen from \eq{eq:dec_sol}, $\pd_{A_d} \eta_d$ is exponentially decreasing as $x \to + \infty$ while $\pd_C \eta_d$ goes to a finite constant $-1 / \w_+$. 
In the low-frequency limit, we thus have
\begin{equation}
\deta^{(\mathrm{in})}_\om \propto \lp \pd_{A_d} \eta^{(0)} \rp_{A_d = C = 0}.
\end{equation}
Evaluating the derivative and using the matching conditions at $x = 0$ gives
\begin{equation}
\deta^{(\mathrm{in})}_\om \propto
\lb 
\begin{aligned}
	& \e^{- k_d \s x} & & x > 0 \\
	& \cos \lp k_u \s x \rp - \frac{k_d}{k_u} \sin \lp k_u \s x \rp & & x < 0
\end{aligned}
\right. .
\end{equation}
Integrating over $x$, one obtains:
\begin{equation} 
\phi^{(\mathrm{in})}_\om \propto
\lb 
\begin{aligned}
	& \e^{- k_d \s x} & & x > 0 \\
	& 1 + \frac{k_d^2}{k_u^2} - \frac{k_d}{k_u} \sin \lp k_u \s x \rp - \frac{k_d^2}{k_u^2} \cos \lp k_u \s x \rp & & x < 0
\end{aligned}
\right. .
\end{equation}
The effective temperature can then be determined using \eq{eq:effTeff}: the amplitude of the incoming wave, which is uniform in the limit $\om \to 0$, goes to $1 + k_d^2 / k_u^2$ while that of the negative-energy wave, with wave vector $- k_u$ for $\om = 0^+$, goes to $(k_d / (2 \s k_u)) \s \sqrt{1 + k_d^2 / k_u^2}$, giving
\begin{equation}\label{eq:Teff0}
T_\mathrm{eff}^{(0)} = \frac{\w_-^{3/2}}{2} \s \lp 1 + \abs{\frac{\w_-}{\w_+}} \rp^{-1}. 
\end{equation}
Recalling from Equation~\eqref{eq:ommax}  that $\omega_{\rm max} \propto \w_-^{3/2}$, this expression is of the same form as those found in the steplike regime of Bogoliubov-de Gennes~\cite{Finazzi:2012iu} and the quartic dispersion relation considered in~\cite{Robertson:2012ku}. 

\noindent \textbf{\\ Low-frequency effective temperature on a white-hole-like flow with undulation: \\ \phantom{0}} 

\begin{figure} 
\centering
\includegraphics[width=0.49\linewidth]{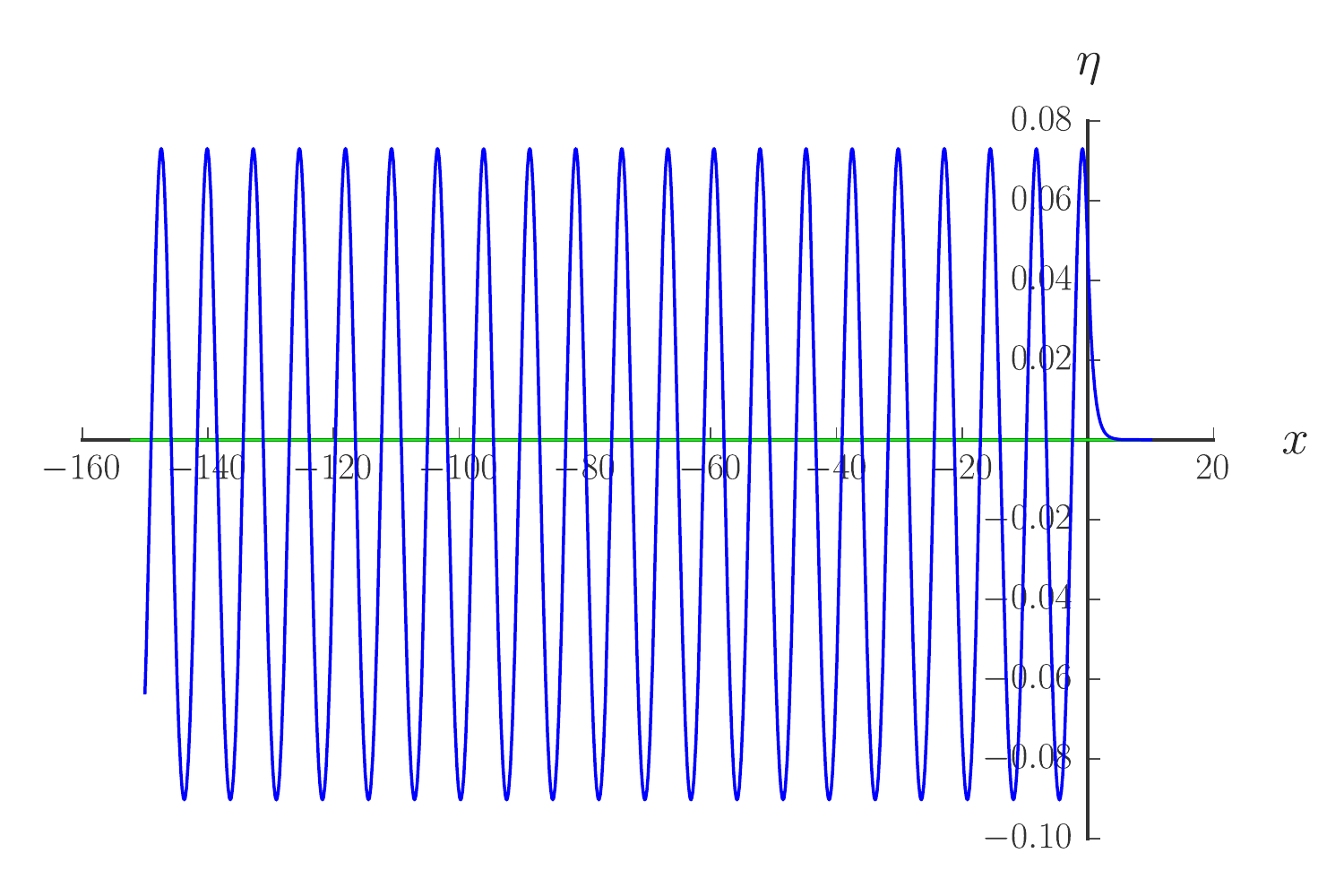}
\includegraphics[width=0.49\linewidth]{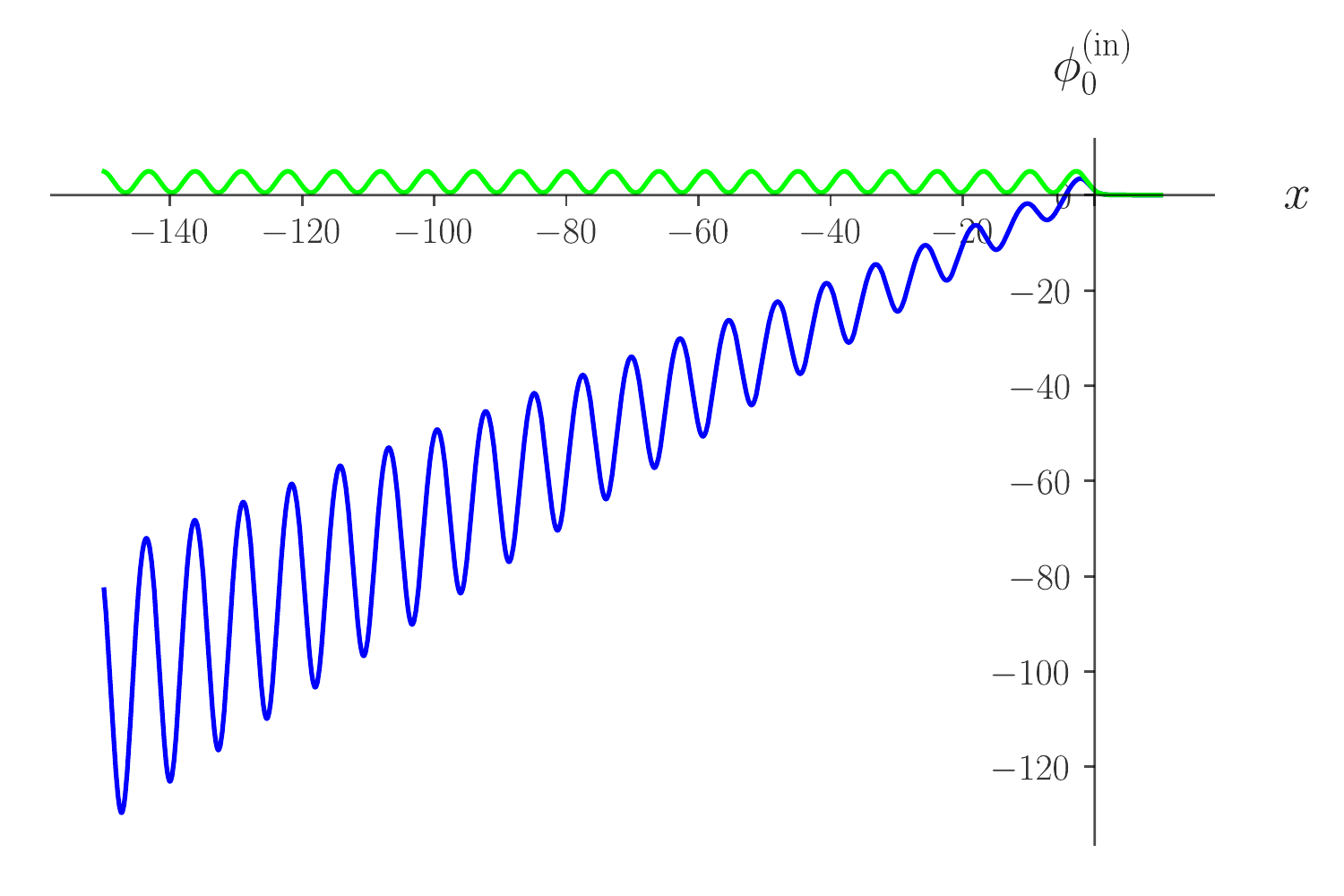} 
\caption{\textit{Left panel:} Plot of the homogeneous solution $\eta = 0$ (green) and of a modulated solution (blue) of the KdV equation with variable coefficient~\eqref{eq:varKdV}.  
The function $\w$ has a hyperbolic tangent profile $\w(x) = A + B \tanh (\sigma \s x)$ with $\sigma = 100$ and $(A,B) \in \mathbb{R}^2$ chosen so that $\w(-\infty) = 0.8$ and $\w(+\infty) = -1.5$. 
\textit{Right panel :} Incoming, counterpropagating mode over the modulated solution (blue) and over the homogeneous solution $\eta = 0$ (green). 
(The normalization is arbitrary.)} \label{fig:KdV_backgs_ccmodes}
\end{figure}

\noindent
We now consider the case $A_d \neq 0$, and thus $A_u \neq 0$, keeping for simplicity $C = 0$ for the background flow. 
(This hypothesis will be relaxed below.) 
The idea of the calculation is the same as above. 
Since the background solution goes to zero exponentially in the limit $x \to + \infty$, the structure of the modes in this region is unaffected and $\phi_0^{(\mathrm{in})}$ must still be exponentially decreasing, which implies $\eta^{(\mathrm{in})}_0 \propto \pd_{A_d} \eta^{(0)}$. 
However, the additional terms in Equation~\eqref{eq:KdV-undul-sols} give linearly growing ones in $\phi_0^{(\mathrm{in})}$ in the limit $x \to - \infty$, represented in \fig{fig:KdV_backgs_ccmodes}.
For sufficiently large values of $-x$ and small amplitudes $A_u$, the relative variations of the amplitudes of the waves over a distance $2 \s \pi / k_u$ becomes negligible. 
Using a WKB approximation, the amplitudes of the incident and negative-energy waves for $\om = 0^+$ can thus be locally evaluated unambiguously. 

An interesting point is that, when focusing on the limit $x \to - \infty$, one does not need to know the relationship between $(A_u, x_u)$ and $A_d$. 
Indeed, assuming $\pd_{A_d} A_u \neq 0$ (which can be verified explicitly using \eq{eq:AuAd}), and keeping only the constant term (which gives a linearly growing contribution to $\phi_0^{(\mathrm{in})}$ after integration over $x$) and the linearly growing one, we obtain
\begin{equation}
\deta_0^{(\mathrm{in})} \propto \lp - \frac{3 \s A_u}{k_u^2} + \frac{15 \s A_u^2}{2 \s \w_-^{3/2}} \s x \sin  \lp k_u \s \lp x - x_u \rp \rp \rp \pd_{A_d} A_u + \cdots,
\end{equation}
where the neglected terms give finite contributions to $\phi_0^{(\mathrm{in})}$ in the limit $x \to - \infty$ or grow linearly with coefficients of higher order in $A_u$. 
Integrating over $x$ gives
\begin{equation}
\phi_0^{(\mathrm{in})} \propto - \lp \frac{3 \s A_u}{k_u^2} + \frac{15 \s A_u^2}{2 \s \w_-^2} \cos  \lp k_u \s \lp x - x_u \rp \rp \rp x \s \pd_{A_d} A_u + \cdots
\end{equation}
Using again \eq{eq:effTeff}, one obtains the effective temperature:
\begin{equation}\label{eq:Teff1}
T_\mathrm{eff}^{(u)} \approx \frac{25 \s A_u^2}{8 \s \sqrt{\w_-}}.
\end{equation}
Equation~\eqref{eq:Teff1}  (along with its generalization given later in~\eqref{Teffinftygen}) is the main result of this work. 
They show that an undulation with an arbitrarily small amplitude will, if extending over a sufficiently long domain (this condition is made more precise below), efficiently suppress the scattering, replacing the effective temperature $T_\mathrm{eff}^{(0)}$ of \eq{eq:Teff0} by $T_\mathrm{eff}^{(u)}$ of \eq{eq:Teff1}. 
Moreover, this new effective temperature goes to zero like $A_u^2$ for $A_u \to 0$. 
Although the above calculation is done in a very specific case, it is clear from the derivation that this scaling only depends on the properties of the periodic, stationary solutions for $x \to - \infty$ and thus applies much more generally. 
In particular, Equation~\eqref{eq:Teff1}, whose derivation involves only the properties of the flow in the asymptotic regions $x \to \pm \infty$, remains valid when replacing the steplike profile for $\mu$ with a smooth one. 
Below we consider two generalizations: in Section~\ref{KdV_WH:3} we use white-hole-like flows of the forced KdV equation and  show that \eq{eq:Teff1} is recovered up to adding a constant to $\w_-$ accounting for the different mean background flow. 
In Appendix~\ref{app:UCP_tuned}, we use a more realistic model of water waves and find the same scaling between the effective temperature computed far from the horizon and the amplitude of the undulation. 

\begin{figure} 
\centering
\includegraphics[width=0.47\linewidth]{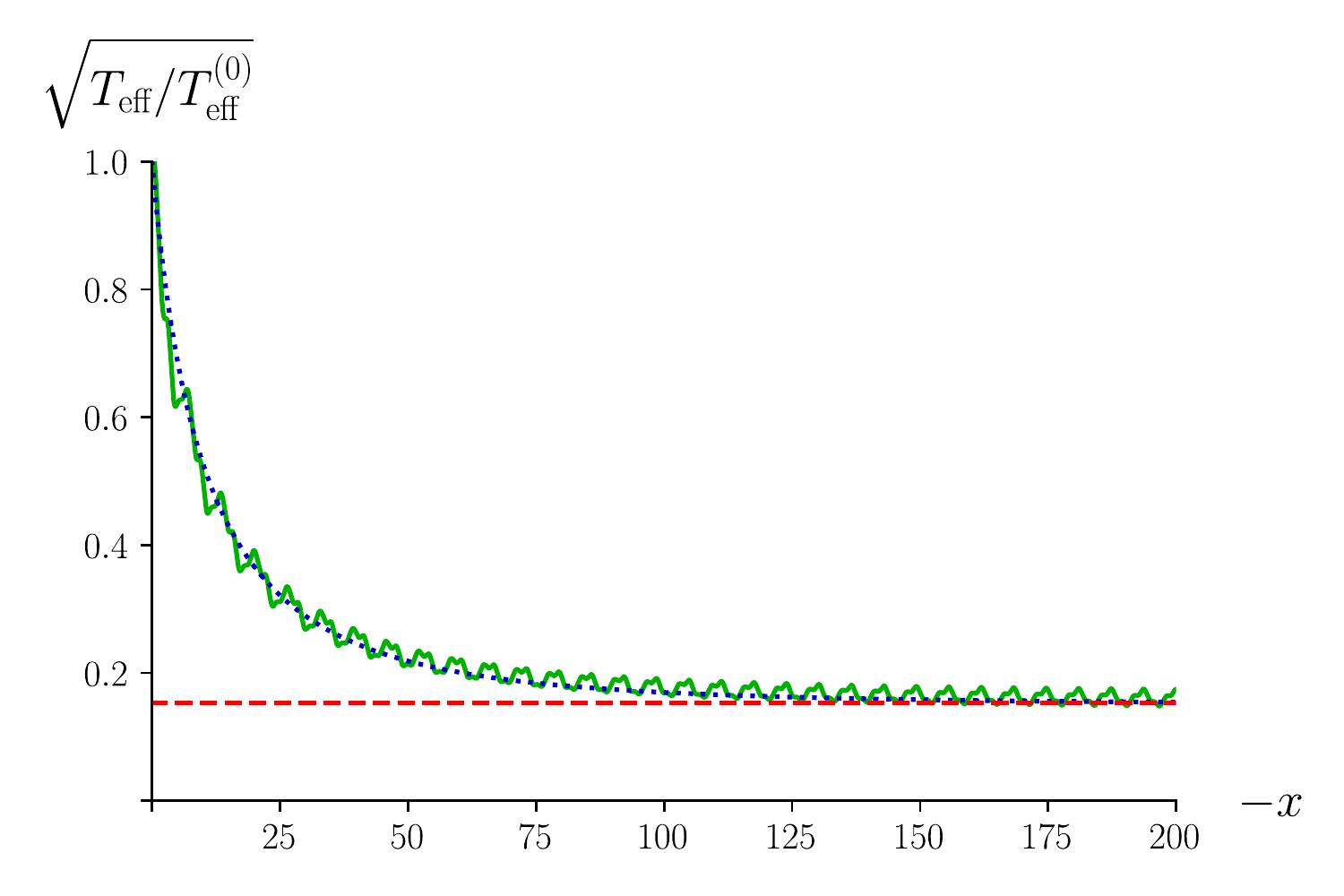}
\caption{Square root of the effective temperature computed numerically over a finite-length undulation with sharp cutoff (which causes the rapid oscillations with small amplitude) with $A_d = -0.03$, $\w_+ = 1$, and $\w_- = -1$, as a function of the cutoff position $x$. 
The temperature is adimensionalized by $T_\mathrm{eff}^{(0)}$ of \eq{eq:Teff0}. 
The green curve shows the numerical result. The blue, dotted one shows the approximation from \eq{eq:Teff_x} and the red, dashed line its limit $T_\mathrm{eff}^{(u)}$ of Equation~\eqref{eq:Teff1} for $x \to - \infty$. 
} \label{fig:KdV_T_x}
\end{figure}

Before that, it is useful to consider the effective temperature $T_\mathrm{eff}(x)$ obtained when doing the measurement at a finite (negative) value of $x$. 
Redoing the above calculation while keeping the leading terms of constant amplitude in $\phi_0^{(\mathrm{in})}$ gives~\footnote{
If $A_d > 0$, \eq{eq:Teff_x} has a divergence for a finite negative value of $x$. 
This does not seem to indicate anything dramatic: it is only the consequence of a local cancellation between the wave with low wave vector present in the absence of undulation and the linearly growing part due to the latter, so that the amplitude of the incoming mode as measured there would vanish.}
\begin{equation} \label{eq:Teff_x}
T_\mathrm{eff}(x) \approx \frac{\w_-^{3/2}}{2} \s 
	\frac{1 + \frac{225 \s A_u^4}{4 \s \w_-^3} \s x^2}
		{\lp \sqrt{1 - \frac{\w_-}{\w_+}} + \mathrm{sgn}(A_d) \frac{3 \s \abs{A_u}}{\sqrt{\w_-}} \s x \rp^2}. 
\end{equation}
This is represented in \fig{fig:KdV_T_x}. 
Three regimes can be distinguished:
\begin{itemize}
\item For $0 < - x \ll \sqrt{\w_- - \w_-^2 / \w_+} / (3 \s \abs{A_u})$, the terms linear in $x$ are negligible and one recovers $T_\mathrm{eff}(x) \approx T_\mathrm{eff}^{(0)}$;
\item For $- x \gg 2 \w_-^{3/2} / (15 \s A_u^2)$, they become dominant and the effective temperature becomes $T_\mathrm{eff}(x) \approx T_\mathrm{eff}^{(u)}$;
\item For $\sqrt{\w_- - \w_-^2 / \w_+} / (3 \s \abs{A_u}) \ll - x \ll 2 \w_-^{3/2} / (15 \s A_u^2)$, one finds an intermediate regime with
\begin{equation}
T_\mathrm{eff}(x) \approx \frac{\w_-^{5/2}}{18 \s A_u^2 \s x^2}. 
\end{equation}
This regime is interesting as it shows how the temperature goes from the value $T^{(0)}_\mathrm{eff}$ to the (under our hypotheses) much smaller value $T^{(u)}_\mathrm{eff}$ when increasing the length of the undulation.
In particular, one sees that the main parameter describing this transition is the product $A_u \s x$ of the amplitude of the undulation and the length $x$ separating the point where the measurement is made from the analogue horizon. 
\end{itemize}

As a consistency check, we compare in the right panel of the figure results from \eq{eq:Teff_x} to those from a numerical calculation of the effective temperature with a cutoff undulation. 
This finite undulation is defined by multiplying the background solution $\eta^{(0)}$ by $\theta(x - x_0)$, where $\theta$ is Heaviside's step function and $x_0 < 0$, so that the flow is asymptotically homogeneous and the scattering coefficients can be computed in the usual way~\cite{Macher:2009tw,Michel:2014zsa}. 
Apart from small oscillations coming from higher-order effects, whose relative amplitude seems linear in $A_u$, we observe a good agreement between the numerical and analytical results.

Before moving on, let us pause to make a qualitative summary. 
As mentioned above, there are two resonances at zero frequency: a first-order one involving the ``hydrodynamic'' root $k = 0$ and one dispersive root, and a second-order one involving the two dispersive roots. 
The first resonance is responsible for the term linear in $x$.~\footnote{More precisely, it gives a finite constant to $\deta$, which after integration yields a linear term in $\phi$.}
The second-order one yields the term in $x \cos (k_u \s x)$. 
When moving away from $x = 0$, in the negative-$x$ direction, both the hydrodynamic and dispersive waves thus have amplitudes growing linearly in $\abs{x}$, with coefficients of orders $A_u$ and $A_u^2$, respectively. 
For sufficiently long undulations, these linearly growing terms dominate over the constant ones coming from the scattering in the region $x \approx 0$, giving values of $\beta_{\om \approx 0}$ proportional to $A_u$, and thus a temperature proportional to $A_u^2$. 
We briefly investigated numerically the case of a finite slope of $u$ near the sonic horizon and found that: first, the qualitative behavior of $T_{\mathrm{eff}}(x)$ is similar to that in Figure~\ref{fig:KdV_T_x}; and second, the asymptotic value of $T_{\mathrm{eff}}(x)$ as $x \to -\infty$ is still given by Equation~(\ref{eq:Teff1}), as expected from the above analysis. 

\noindent \textbf{\\ Effect of a damped undulation: \\ \phantom{0}} 

\begin{figure}
\centering
\includegraphics[width=0.49\linewidth]{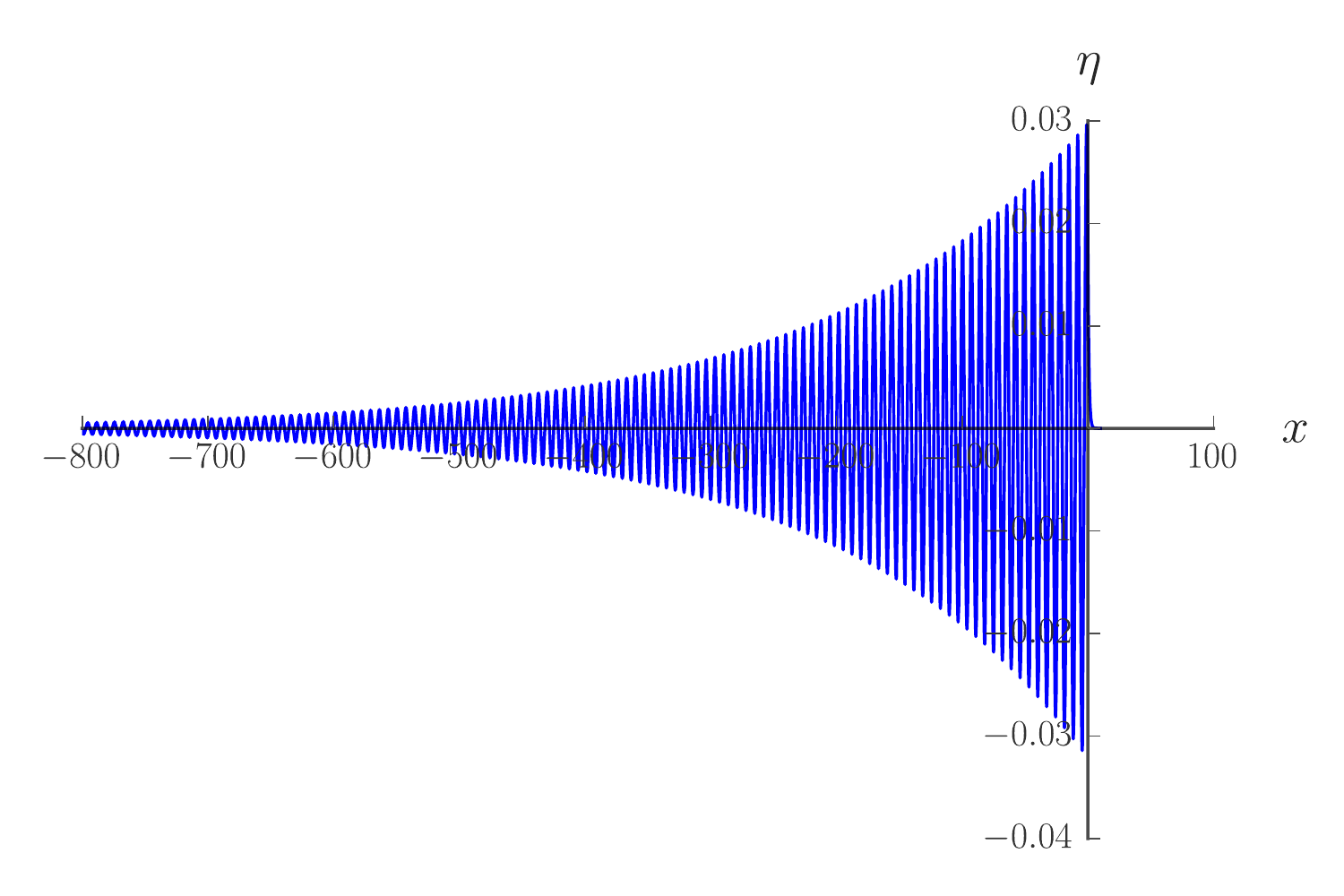}
\includegraphics[width=0.49\linewidth]{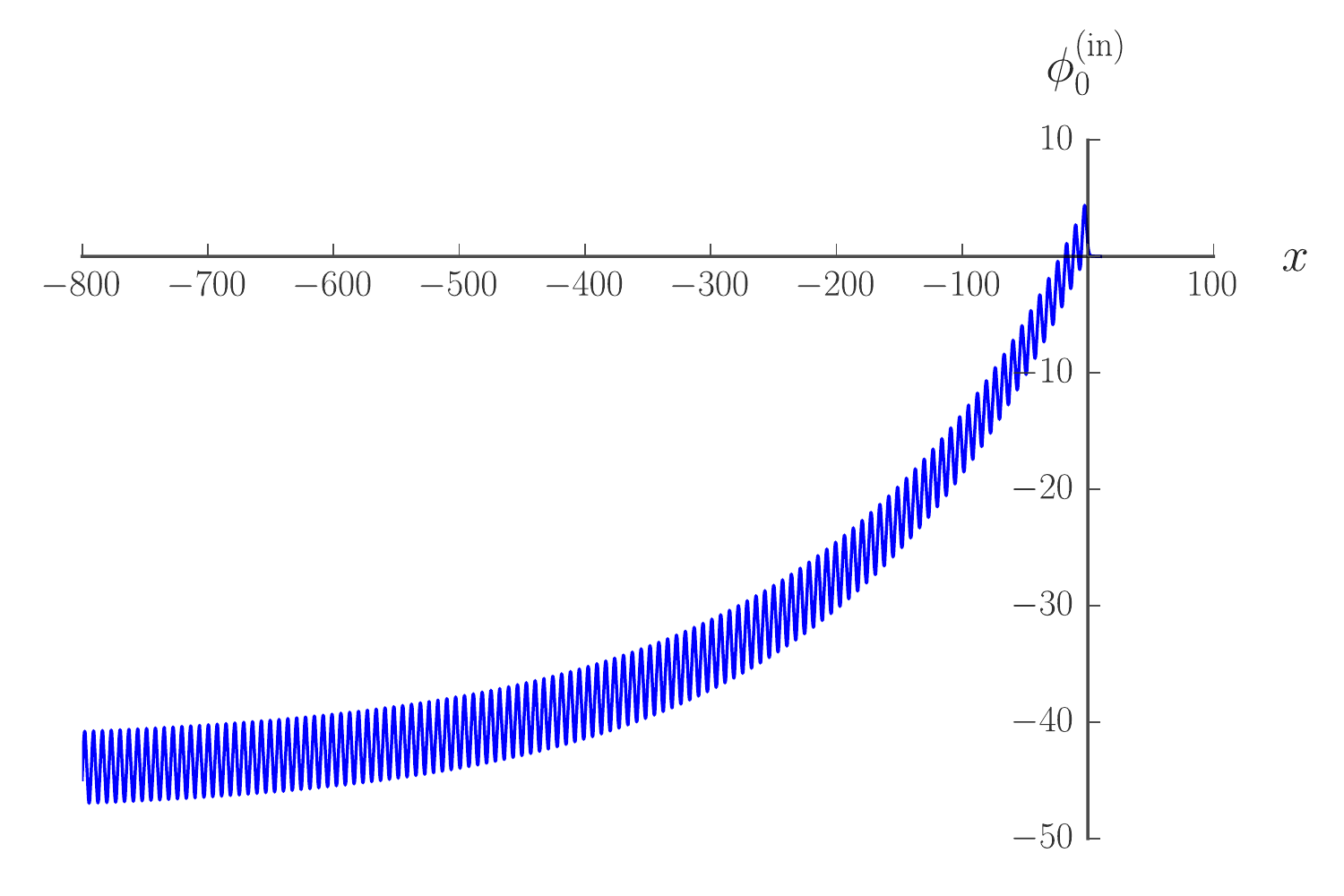} 
\includegraphics[width=0.5\linewidth]{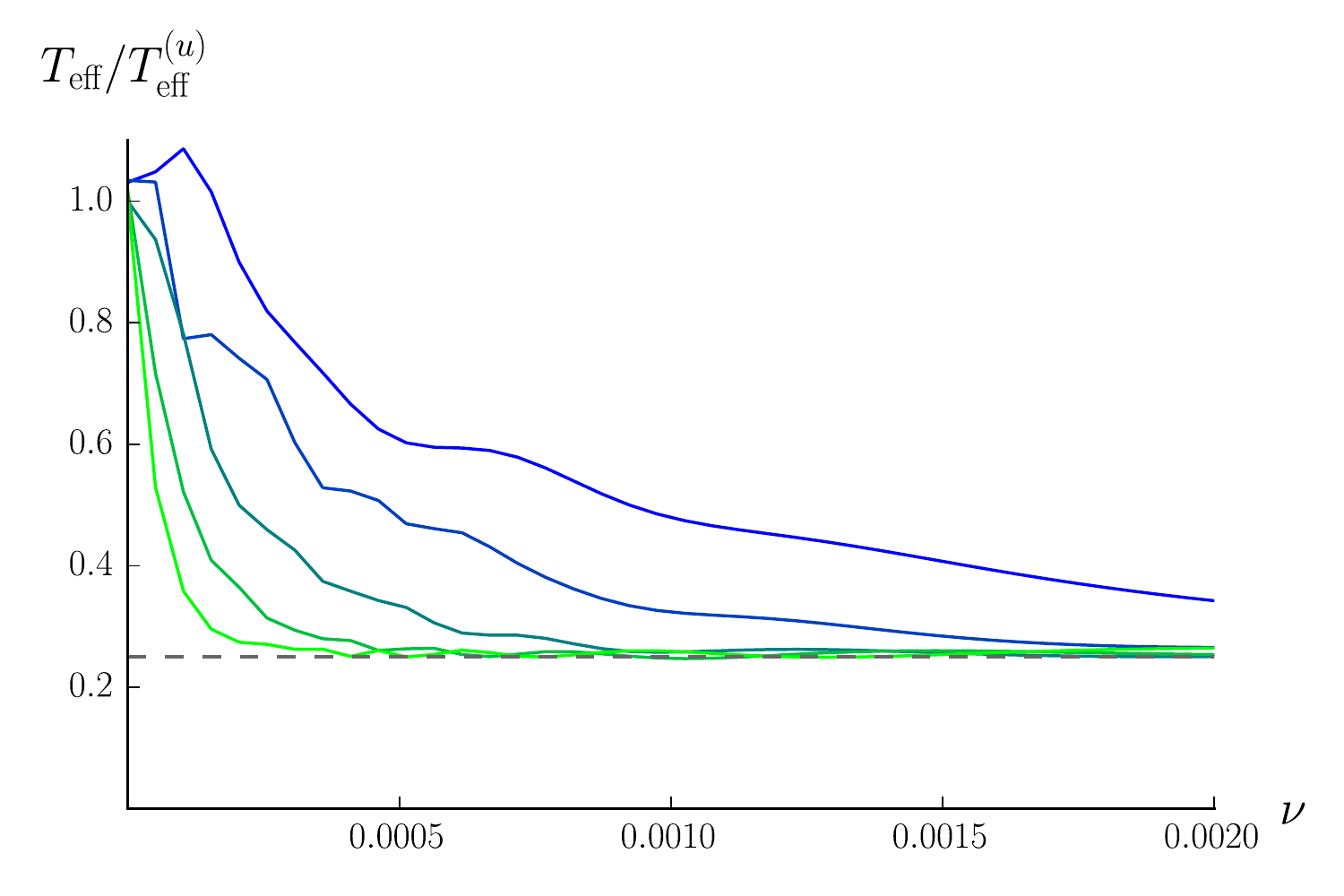}
\caption{\textit{Top:} A damped undulation solution of \eq{eq:KdV_dis} (left) and the incoming counterpropagating mode over this solution (right). The parameters are the same as in the right panel of \fig{fig:KdV_backgs_ccmodes}, and $\nu$ is set to $0.01$.
\textit{Bottom:} Variation of the effective temperature with $\nu$, for relatively small values of this parameter. 
The value of $-x$, is multiplied by $2$ between each curve, from blue to green, with minimum value $-x_{\mathrm{max}} = 2000$. The dashed line materializes the value $1/4$.} \label{fig:varphi5}
\end{figure}

\noindent
A more physical way to send the amplitude of the undulation to zero at spatial infinity is to add a 
small dissipative term to the KdV equation. 
Focusing on stationary solutions, we work with the equation
\begin{equation} \label{eq:KdV_dis}
\pd_x^2 \eta + \w \s \eta + 3 \s \eta^2 - \nu \s \pd_x \eta = 0,
\end{equation}
where $\nu \ll k_d, k_u$ is a small positive number. 
This equation may be written as
\begin{equation}
\pd_x \lp 
	\frac{1}{2} \lp \pd_x \eta \rp^2
	+ \frac{\w}{2} \s \eta^2
	+ \eta^3
\rp
 = \nu \s \lp \pd_x \eta \rp^2.
\end{equation}
To leading order, the evolution of the amplitude $A_u$ of the undulation is thus given by
\begin{equation}
\pd_x \lp A_u^2 \rp \approx 2 \s \nu \sin^2 \lp k_u \s x \rp A_u^2.
\end{equation}
Since $\nu \ll k_u$, one can average this equation over a few wavelengths, leading to 
\begin{equation}
A_u \propto \e^{\nu \s x / 2} 
\end{equation}
for $x < 0$. 
The effects of this damped undulation on the scattering can be estimated as follows. 
Locally, if $\nu$ is sufficiently small, the modes of Equation~\eqref{eq:KdV-linearized} have the same form as above, with $A_u$ now slowly varying with $x$. 
In particular, over an interval 
of length $\delta x$ centered on $x$ and such that $\nu \s \abs{\delta x} \ll 1$, the nonoscillating part of $\phi^{(\mathrm{in})}_0$ will grow by a quantity proportional to $A_u(x) \s \delta x$ and the oscillating part by a quantity proportional to $A_u(x)^2 \s \delta x$. 
When computing their ratio for $x \to - \infty$ , $A_u$ should thus be replaced by
\begin{equation}
\frac{\int_{-\infty}^0 A_u(x)^2 \s \dd x}{\int_{-\infty}^0 A_u(x) \s \dd x} = \frac{A_u(0)}{2}. 
\end{equation}
For a given value of $A_u(0)$, one thus expects that the effective temperature $T_\mathrm{eff}^{(\mathrm{d})}$ measured at $x \to - \infty$ over a slowly damped undulation is $4$ times smaller than the result obtained without damping:
\begin{equation}\label{eq:Td}
T_\mathrm{eff}^{(\mathrm{d})} \approx \frac{25 \s A_u(0)^2}{32 \s \sqrt{\w_-}}. 
\end{equation}
To illustrate this result, we show in \fig{fig:varphi5} the evolution of the effective temperature with $\nu$, computed numerically for different values of $x$. 
As expected from the above calculation, the two limits $x \to - \infty$ and $\nu \to 0^+$ do not commute:
\begin{itemize}
\item When sending $\nu$ to $0^+$ at fixed $x$ and then $x$ to $- \infty$, one recovers the effective temperature $T_\mathrm{eff}^{(u)}$ computed for $\nu = 0$;
\item When sending first $x$ to $- \infty$ and then $\nu$ to $0^+$, the effective temperature we obtain is $4$ times smaller, in accordance with \eq{eq:Td}. 
\end{itemize}

\subsection{Low-frequency effective temperature: Generalization}
\label{KdV_WH:3}

We now wish to generalize the main result of the above analysis, i.e., Equation~\eqref{eq:Teff1}, to a wider class of white-hole-like flows. 
Let us consider the forced KdV equation
\begin{equation} \label{eq:fKdV}
\pd_t \eta + \pd_x \lp \w \s \eta \rp + \pd_x^3 \eta + 6 \s \eta \s \pd_x \eta = f,
\end{equation}
where $\w$ and $f$ are two smooth functions of $x$. 
We assume $\w$ has finite limits $\w_\pm$ as $x \to \pm \infty$ and that $f$ is integrable over $\mathbb{R}$. 
We denote as $h$ a primitive of $f$, and by $h_\pm$ its limits $x \to \pm \infty$. 
When looking for stationary solutions, \eq{eq:fKdV} can be integrated over $x$, giving
\begin{equation}
\w \s \eta + \pd_x^2 \eta + 3 \s \eta^2 = h + C,
\end{equation}
where $C$ is an integration constant. 
Locally, the properties of the solutions are identical to those obtained for $f = 0$, with $C$ shifted by $h$. 
In particular, if $\w$ and $h$ are homogeneous we have two uniform solutions $\eta(x) = (- \w \pm \sqrt{\w^2 + 12 \s (C + h)}) / 6$, the one with the $+$ sign being subcritical while that with the $-$ sign is supercritical. 

We assume that $\w$ and $f$ are such that there exists a value $C_0$ of $C$ for which there is a white-hole solution, i.e., a solution going to $(- \w_\pm \mp \sqrt{\w_\pm^2 + 12 \s (C_0 + h_\pm)}) / 6$ as $x \to \pm \infty$.~\footnote{This hypothesis should be satisfied without the need to fine-tune the functions $\w$ and $f$, as explained in~\cite{Michel:2014zsa}. 
Indeed, since the KdV equation is of order $3$, the general solution has three parameters. 
The condition that it is asymptotically uniform in the subcritical region gives two constraints to linear order while the same condition in the supercritical region gives only one constraint since one of the waves is exponentially decreasing. 
The number of constraints is thus equal to that of degrees of freedom, so one generally expects to find a discrete set of solutions.}
Stationary solutions close to this white-hole flow are given 
\begin{itemize}
\item in the limit $x \to + \infty$, by \eq{eq:dec_sol} with $C$ replaced by $C + h_+$,
\item in the limit $x \to - \infty$, by Eqs.~(\ref{eq:KdV_ku},\ref{eq:KdV-undul-sols}) with $\w$ replaced by $\sqrt{\w_-^2 + 12 \s (C+h_-)}$ and with $\eta_0 = (k_u^2 - \w_-)/6$. 
\end{itemize}
As in the previous case, they depend on the two parameters $A_d$ and $C$, the quantities $A_u$ and $x_u$ being fixed by solving \eq{eq:fKdV}. 
As above also, the incoming counterpropagating mode must go to zero as $x \to +\infty$, which tells us that $\deta_0^{(\mathrm{in})} \propto \pd_{A_d} \eta$ at fixed $C$. 
Using Eqs.~(\ref{eq:KdV_ku},\ref{eq:KdV-undul-sols}) to determine the behavior of $\deta_0^{(\mathrm{in})}$ for $x \to - \infty$, integrating over $x$ to obtain $\phi_0^{(\mathrm{in})}$, and extracting the effective temperature using Equation~\eqref{eq:effTeff} gives, to leading order in $A_u$ and assuming $A_u \neq 0$,
\begin{equation}\label{Teffinftygen}
T_\mathrm{eff}^{(u)} \approx \frac{25 \s A_u^2}{8 \s \lp \w_-^2 + 12 \s \lp C + h_- \rp \rp^{1/4}}.
\end{equation}

The other results of the previous subsection also extend qualitatively to the present case. 
In particular, one still finds the three regimes: assuming the origin of $x$ is chosen close to the analogue horizon, and up to coefficients depending on the profiles of $\w$ and $f$, 
\begin{itemize}
\item for $-x \ll A_u^{-1}$, $T_\mathrm{eff}(x)$ is close to that computed without the undulation;
\item for $-x \gg A_u^{-2}$, $T_\mathrm{eff}(x) \approx T_\mathrm{eff}^{(u)}$;
\item in the intermediate range $A_u^{-1} \ll -x \ll A_u^{-2}$, $T_\mathrm{eff}(x)$ scales like $1 / (A_u^2 x^2)$.
\end{itemize}
The reasoning leading to \eq{eq:Td} also remains valid up to minor modifications, showing that adding a small dissipative term will reduce the effective temperature as measured at infinity by a factor $4$ with respect to \eq{Teffinftygen}.

\section{Conclusions}
\label{sec:concl}

We have studied the propagation of linear waves in spatially modulated water flows. 
In the main text we used a simple model based on the KdV equation, which shows the generic features in a relatively transparent way. 
The same problem is addressed in a more realistic model of water waves in Appendix~\ref{app:UCP}. 
In Appendix~\ref{app:detuned} we focus on the simpler case of a ``detuned'' undulation, whose wavelength is different from that of the low-frequency dispersive waves and which may be understood as a modulation of some external parameter such as, for instance, the height of an obstacle. 
In that case the resonance occurs when the wave number of the modulation, or an integer multiple of it, is equal to the difference between two roots of the dispersion relation at the same frequency. 

We considered waves propagating over stationary inhomogeneous solutions of the nonlinear KdV equation and found a low-frequency resonance with peculiar features. 
In particular, two modes grow linearly in space, which can be understood as the limit of the usual exponential behavior when the energy carried by each wave, and thus the growth rate, goes to zero. 
We show in appendix~\ref{app:lingrowmodes} that the linear modes are closely related to the structure of the nearby nonlinear solutions. 
Applying these results to white-hole-like flows with the analogue of a Killing horizon and a modulated free surface, we found that the latter can drastically modify the analogue Hawking emission provided it extends far enough from the horizon. 
For a sufficiently small amplitude (such that $T_{\rm eff}^{u} < T_{\rm eff}^{(0)}$), we analytically determined three regimes, depending on the length $L$ and amplitude $A$ of the undulation: a regime of ``short'' undulation where the latter hardly affects the scattering, one of ``long'' undulation where it completely fixes the effective temperature $T_\mathrm{eff} \propto A^2$, and an intermediate one where $T_\mathrm{eff} \propto 1/(A^2 L^2)$.  
These scaling laws constitute the main result of our work. 
Their derivation shows that they arise directly from the structure of the solutions close to asymptotically homogeneous white-hole ones, and we thus expect them to be fairly general. 
The analysis of Appendix~\ref{app:UCP} confirms that they also hold in the case of a more refined model of water waves. 
Finally, we considered the effect of a long undulation in the presence of a dissipative term sending its amplitude to zero at infinity. 
We found that, in the limit of small dissipation, the effective temperature is divided by $4$ with respect to the nondissipative result. 
These results were verified numerically by solving the KdV equation.  

These findings raise two important questions which we hope to address in future works. 
The first and, in principle, most straightforward one is to consider the same problem in other analogue gravity systems such as cold atoms or quantum fluids of light. 
While the similarities between the nonlinear solutions of the Gross-Pitaevskii and KdV equations seem to indicate that similar results will hold, it would be interesting to verify this explicitly and to determine the coefficients of the linearly growing terms in different systems. 
The second question concerns their application to experiments. 
Considering for instance the realization of~\cite{Euve:2015vml}, two elements prevent our results from being directly applicable. 
First, the experimental setup was slightly different from the one considered here, consisting of a subcritical water flow over a localized obstacle (briefly discussed in Appendix~\ref{app:sub}). 
Second, the undulation amplitude varied significantly over a few wavelengths. 
It is thus not clear to what extent the present analysis, based on the assumption of a slowly varying amplitude, can be trusted in this regime. 
We expect, however, that low-frequency resonances still play an important role. 
It would be of interest to obtain experimental data with a transcritical flow to be able to compare them with our results: the former could point to physical effects not included in the above analysis while the latter could help disentangle the contribution from the scattering on the undulation and from the analogue Hawking effect. 
As they stand, the main results of this article can already be used for estimating the parameter range in which the Hawking effect should dominate, which we hope will help guide the design of future analogue gravity experiments.

\vspace*{2em}
{\centering \bfseries \fontsize{12pt}{14pt}\selectfont  Acknowledgments \par} 
\vspace*{0.5em}

\noindent 
We thank Antonin Coutant, Silke Weinfurtner, and Vincent Pagneux for interesting discussions while preparing this work. 
We also thank Simone Speziale and our anonymous referee for bringing our attention to the Dirac procedure when handling the Hamiltonian formulation of the KdV equation and Roger Grimshaw for pointing out that some of our results may be recovered using Whitham’s modulation theory. 
We are also grateful to Antonin Coutant for his useful remarks on a previous version of this paper. 
F.~M. thanks the LPT Orsay, where part of this work was done, for its hospitality.
This work was supported by the French National Research Agency through the Grant No. ANR-15-CE30-0017-04 associated with the project HARALAB. 
F.~M. is supported by the Leverhulme grant RPG-2016-233. This research was supported in part by Perimeter Institute for Theoretical Physics. Research at Perimeter Institute is supported by the Government of Canada through the Department of Innovation, Science and Economic Development and by the Province of Ontario through the Ministry of Research and Innovation.

\titleformat{\section}{\large\bfseries}{\appendixname~\thesection .}{0.5em}{}
\appendix 

\section{Detuned undulation}
\label{app:detuned}

\subsection{Lagrangian description of the linear KdV equation}  
\label{app:lag_lin_KdV} 

We first look for a simple linear equation describing dispersive, small-amplitude water waves in the presence of a bottom topography inducing a modulated background flow. 
We focus on counterpropagating waves and assume that the wave vector of the modulation differs from the zero-frequency wave number $k_u$. 
For definiteness, we consider a flow from right to left. 
The waves we shall consider then all propagate to the right in the frame of the fluid. 
Yet, because of the flow, some of them are dragged and propagate to the left in the laboratory frame. 

We start from the (nonlinear) forced KdV equation~\cite{Shen1993} 
\begin{equation} \label{eq:forcedKdV}
[ \pd_t + \w_0 \, \pd_x + \pd_x^3  + 6 \s \eta(t,x) \, \pd_x ] \eta(t,x) = \pd_x h(x), 
\end{equation}
where $\eta$ gives the water height measured with respect to a uniform reference solution of the fluid equations for $\pd_x h = 0$, and where $\w_0$ is a real constant. 
Denoting as $v_0$ the fluid velocity and $c_0$ the speed of long-wavelength perturbations 
of the background solution $\eta = 0$ in the fluid frame, $\w_0$ is equal to the group velocity $v_0 + c_0$ of long-wavelength linear perturbations in the laboratory frame. 
We assume the flow is subcritical, i.e., $\abs{v_0} < c_0$, so that $\w_0 > 0$. 
The function $h$ describes the stationary bottom topography. We use a unit system in which the uniform water depth of the background flow is $\sqrt{3/2}$ and the gravitational acceleration is $16 \s \sqrt{2/3}$. 
(These values are chosen to set the coefficients of the dispersive and nonlinear terms in \eq{eq:forcedKdV} to simple values, see for instance~\cite{Hereman2013}.) 
Let us assume we know a time-independent solution $\eta_0$. 
We look for $C^3$ perturbations $\deta$ of the form $\eta (t,x) = \eta_0(x) + \deta(t,x)$. 

To interpret and generalize the results of the following subsections, it is useful to adopt a Lagrangian description. 
To this end, we define the auxiliary field $\phi$, akin to a velocity potential, by
\begin{equation}
\phi (t,x) = \int_0^x \deta(t,y) \, \dd y. 
\end{equation}
Linearizing the forced KdV equation~\eqref{eq:forcedKdV} gives
\begin{equation} \label{eq:KdV_phi}
\pd_t \pd_x \phi(t,x) 
+ \pd_x \lp \w(x) \, \pd_x \phi(t,x) \rp 
+ \pd_x^4 \phi(t,x) 
= 0.
\end{equation}
Notice that, using $\pd_x \phi = \deta$, one recovers Equation~\eqref{eq:linKdV}. 
Equation~\eqref{eq:KdV_phi} is the Euler-Lagrange equation obtained from extremization of the quadratic action $S_q = \int_{\mathbb{R}^2} \dd t \, \dd x \, \mathcal{L}_q$ with
\begin{equation} \label{eq:KdV_Lq_real}  
\mathcal{L}_q = - [ \lp \pd_t \phi \rp \s \lp \pd_x \phi \rp
	+ \w \s \lp \pd_x \phi \rp^2
	- \lp \pd_x^2 \phi \rp^2 
	] . 
\end{equation}
In practical calculations it is convenient to work with complex solutions of \eq{eq:KdV_phi}. The Lagrangian density is then 
\begin{equation} \label{eq:KdV_Lq} 
\mathcal{L}_q = -
[ \mathrm{Re} \lp \lp \pd_t \phi^* \rp \, \lp \pd_x \phi \rp \rp
	+ \w \, \abs{\pd_x \phi}^2
	- \abs{\pd_x^2 \phi}^2 
	] .
\end{equation}
A straightforward calculation shows that the inner product between two arbitrary square integrable solutions $\phi_1, \phi_2$ of \eq{eq:KdV_phi}, defined by 
\begin{equation} \label{eq:innerprod} 
\lp \phi_1 \vert \phi_2 \rp = \frac{\ii}{2} 
\int_{-\infty}^{+\infty} \lp \phi_2(t,x) \, \pd_x \phi_1^*(t,x) - \phi_1^*(t,x) \, \pd_x \phi_2(t,x) \rp \, \dd x, 
\end{equation} 
is conserved by the time evolution. 
As usual, this definition can be extended from $L^2$ solutions to plane waves in the sense of distributions. 
If $\phi_i: (t,x) \mapsto A_i \, \e^{\ii \lp k_i \s x - \om_0(k_i) \s t \rp}$ for $i \in \lb 1, 2 \rb$ with $(A_1, A_2, k_1, k_2) \in \mathbb{C}^2 \times \mathbb{R}^2$, we have 
\begin{equation}
\lp \phi_1 \vert \phi_2 \rp = 2 \pi A_1^* A_2 k_1 \delta \lp k_1 - k_2 \rp. 
\end{equation}
In this work we focus on stationary, inhomogeneous configurations. 
It is then convenient to work with the normalized waves $\phi_{\om,i}^\mathcal{N}$, where $\om$ denotes the angular frequency and $i$ is a discrete parameter distinguishing the different solutions with the same frequency, satisfying
\begin{equation} \label{eq:normom}
\abs{\lp \phi_{\om,i}^\mathcal{N} \vert \phi_{\om',j}^\mathcal{N} \rp} = \delta \lp \om - \om' \rp \delta_{i,j}. 
\end{equation}
In a homogeneous region, these normalized waves are given by
\begin{equation} \label{eq:norm_homo} 
\phi_{\om,i}^\mathcal{N} (t,x) = \frac{\e^{\ii \s \lp k_i \s x - \om \s t \rp}}
	{\sqrt{2 \s \pi \s \abs{k_i \s \frac{\dd \om}{\dd k_i}}}},
\end{equation}
where $k_i$ denotes the $i$th solution of the dispersion relation for the angular frequency $\om$. 

This inner product is  related to the wave energy in the following way. 
The 
(naive~\footnote{The Lagrangian density~\eqref{eq:KdV_Lq_real} is linear 
in $\pd_t \phi$, so that the system is constrained and the Dirac formalism should be used to obtain the Hamiltonian formulation. A precise analysis is done in~\cite{doi:10.1063/1.526395} using a two-field model and in~\cite{Restuccia} for more general systems of KdV equations. In the present case, it amounts to adding to the Hamiltonian density $\la \left( p- \pd_x \phi \right)$, where $p$ is the momentum conjugate to $\phi$ and $\la$ is a Lagrange multiplier. A straightforward calculation shows that the Poisson bracket of $p - \pd_x \phi$ and the Hamiltonian vanishes for $\la = w \s \pd_x \phi + \pd_x^3 \phi$, so that there is no secondary constraint. Hamilton's principle then gives back the linearized KdV equation~\eqref{eq:KdV_phi}.
Notice also that the additional term in the Hamiltonian vanishes for solutions of the field equation, so that it neither affects Eqs.~(\ref{eq:innerprod})-(\ref{eq:Hq_om}) nor the discussion of the wave energy in the main text. \label{foot:cons}}) 
Hamiltonian associated with \eq{eq:KdV_Lq} is
\begin{equation} \label{eq:nH}
H_q [\phi] = \int_{-\infty}^{+\infty} \lp \w \s \abs{\pd_x \phi}^2 - \abs{\pd_x^2 \phi}^2 \rp \dd x. 
\end{equation}
Assuming $\phi$, its space derivatives up to the third, and $\pd_t \phi$ are square integrable, a straightforward calculation shows that
\begin{equation}
H_q [\phi] = \lp \phi \vert \ii \s \pd_t \phi \rp. 
\end{equation}
This equation is valid for both complex and real solutions of \eq{eq:KdV_phi}. 
If $\phi_\om$ is a solution with fixed angular frequency $\om \in \mathbb{R}$, we have
\begin{equation} \label{eq:Hq_om}
H_q[\phi] = \om \lp \phi_\om \vert \phi_\om \rp.
\end{equation}
If $\phi_\om$ is a plane wave with wave vector $k$, its energy thus has the same sign as $\om \s k$. 
The energy flux can be obtained by multiplication by the group velocity. 

\subsection{Modes over a detuned undulation}
\label{app:KdV_detuned}

We now determine solutions of Equation~\eqref{eq:linKdV} in the presence of a detuned undulation. 
We use the notations of Section~\ref{sub:linKdV} and work to leading nontrivial order in $\ep$.  
The calculation can be done to higher orders following the same lines. 
Our goal is to show using a simple example the generic features due to the undulation, which will reappear in other contexts such as a refined description of water waves (see Appendix~\ref{app:UCP_detuned}), and which will guide the calculation in the case of a tuned undulation, done in Section~\ref{sec:KdV_tuned} and in Appendix~\ref{app:UCP_tuned}.  
We assume the resonance condition is satisfied with $n_r = 1$, i.e.,  there exists $\om_r \in \left] 0,  \om_\mathrm{max} \right[$ and $(i_r,j_r) \in \lb 1,2,3 \rb^2$ with $i_r > j_r$ such that $k_{\om_r}^{(i_r)} - k_{\om_r}^{(j_r)} = k_\w$. 
We work with values of the angular frequency $\om$ close to $\om_r$. 

Since $\w$ is independent of $t$ and periodic in $x$ with period $\la_\w$, we can look for solutions of Equation~\eqref{eq:linKdV} of the form
\begin{equation}
\deta (t,x) = \e^{- \ii \s \om \s t + \ii \s k \s x} \xi(x),
\end{equation}
where 
$\om \in \mathbb{C}$ is the angular frequency, $k \in \mathbb{C}$ is the quasimomentum, and $\xi \in C^3 \lp \mathbb{R}, \mathbb{C} \rp$ is periodic with period $\la_\w$. 
The function $\xi$ can be expanded as 
\begin{equation} \label{eq:expxi}
\xi: x \mapsto \sum_{n \in \mathbb{Z}} \xi_n \, \e^{\ii \s n \s k_\w \s x},
\end{equation}
where, for all $n \in \mathbb{Z}$, 
\begin{equation}
\xi_n = \frac{1}{\la_\w} \int_0^{\la_\w} \e^{- \ii \s n \s k_\w \s x} \xi(x) \, \dd x. 
\end{equation}
Plugging this form into Equation~\eqref{eq:linKdV} gives
\begin{equation} 
\forall \, x \in \left[ 0, \la_\w \right[, \; 
\sum_{n \in \mathbb{Z}} \lp - \om + \w(x) \, \lp k + n \, k_\w \rp - \ii \, \w'(x) - \lp k + n \, k_\w \rp^3 \rp \, \xi_n \, \e^{\ii \s n \s k_\w \s x} = 0 . 
\end{equation}
Expanding $\w$ using Equation~\eqref{eq:expv} and taking the Fourier transform of the result gives after a few lines of algebra (assuming $\xi_n$ decreases sufficiently fast when $\abs{n} \to \infty$ to be able to exchange the sum and integral):
\begin{equation} \label{eq:KdV_recrel}
\forall \, m \in \mathbb{Z}, \; \sum_{n \in \mathbb{Z}} M_{m,n} \, \xi_n = 0,
\end{equation}
where $\lp M_{m,n} \rp_{(m,n) \in \mathbb{Z}^2}$ is a two-dimensional sequence defined by
\begin{equation}
M_{m,n} = \lb
\begin{aligned}
& \om - \om_0 \lp k + n \, k_\w \rp \; & \;  n = m \\
& - \lp k + m \, k_\w \rp \, \w_{m-n} \; & \;  n \neq m 
\end{aligned}
\right. .
\end{equation}
In the absence of resonance, the system~\eqref{eq:KdV_recrel} has perturbed plane wave solutions with $\xi_0 = 1$, $\xi_n = O \big( \ep^\abs{n} \big)$ for $n \in \mathbb{Z} \tsetminus \lb 0 \rb$, and $k$ such that $\om_0(k) \approx \om$. 
The modulation thus has little effect on the solutions for small values of $\ep$. 
When a resonance is present, however, such solutions in general do not exist: as we now show, there is a strong coupling between two waves, lifting the degeneracy between the wave vectors $k$ and $k + k_u$. 

The solutions exhibiting the resonance correspond to $k \approx k_{\om_r}^{(i_r)}$ and $k \approx k_{\om_r}^{(j_r)}$. 
The two strongly coupled terms will then be those with $n = 0$ and $n = -1$ or $n = +1$, respectively. 
Without loss of generality (up to shifting $n$ by one unit in Equation~\eqref{eq:expxi}), we can assume $k \approx k_{\om_r}^{(j_r)}$. 
To simplify the notation, we will write $k_r \equiv k_{\om_r}^{(j_r)}$. 
It satisfies
\begin{equation}
\om_0 \lp k_r + k_\w \rp = \om_0 \lp k_r \rp. 
\end{equation}
One can then look for solutions with
\begin{equation} \label{eq:KdV_order_xi}
\forall \, n \in \mathbb{Z}, \; \xi_n = O \big( \ep^{\abs{n - 1/2}-1/2} \big),
\end{equation} 
i.e., with $\xi_0$ and $\xi_1$ in $\ord{1}$, $\xi_{-1}$ and $\xi_2$ in $\orde{}$,  
$\xi_{-2}$ and $\xi_3$ in $\orde{2}$, and so on.~\footnote{The terms $- 1/2$ in the exponent of \eq{eq:KdV_order_xi} allow for the coefficients of the waves with wave vectors $k$ and $k + k_\w$ to both have amplitudes of order $1$, as required close to the resonance.} 
Notice that this scaling is consistent with the recursion relation~\eqref{eq:KdV_recrel}, 
which gives for any $n \geq 1$ 
\begin{equation}
\xi_{n+1} = \ord{\ep \s \xi_n} \text{ and } \xi_{-n} = \ord{\ep \s \xi_{-(n-1)}}.
\end{equation}
Keeping only terms of order $\epsilon$ in this relation, choosing $m = 2$ and $m = -1$, and using that $\w_{-1} = \w_1^*$ since the function $\w$ is real, gives 
\begin{equation}\label{eq:KdV_xi2}
\xi_2 = \frac{(k + 2 \, k_\w) \, \w_1}{\om - \om_0 \lp k + 2 \, k_\w \rp} \, \xi_1 + \orde{2}
\end{equation}
and
\begin{equation}\label{eq:KdV_xim1}
\xi_{-1} = \frac{\lp k - \, k_\w \rp \, \w_1^*}{\om - \om_0 \lp k - \, k_\w \rp} \, \xi_0 + \orde{2}. 
\end{equation}
Choosing $m = 0$ and $m = 1$ gives 
\begin{equation} \label{eq:KdV_xi1_xi0} 
\begin{pmatrix}
\om - \om_0 \lp k + k_\w \rp & - \lp k + k_\w \rp  \, \w_1 \\
- k \, \w_1^* & \om - \om_0 (k)
\end{pmatrix}
\begin{pmatrix}
\xi_1 \\
\xi_0
\end{pmatrix}
 = \orde{2}. 
\end{equation}

From this matricial equation, two complementary points of view can be obtained by working at fixed angular frequency $\om$ or at fixed wave vector $k$. 
Working at fixed $k \in \mathbb{R}$, and assuming $k \s \lp k + k_\w \rp \s \w_1 \neq 0$, one obtains after a few algebraic manipulations:
\begin{equation} \label{eq:KdV_detuned_om} 
\om = \frac{1}{2} \lp \om_0(k) + \om_0 \lp k + k_\w \rp \pm 
	\sqrt{4 \s k \s \lp k + k_\w \rp \s \abs{\w_1}^2 + \lp \om_0(k) - \om_0 \lp k + k_\w \rp \rp^2} 
	\lp 1 + \orde{} \rp \rp. 
\end{equation}
The evolution of the wave in time thus crucially depends on the sign of $k_r \s \lp k_r + k_\w \rp$, i.e., whether $j_r = 1$ or $j_r = 2$:
\begin{itemize}
\item If $j_r = 2$, the only possibility for $i_r$ is $i_r = 3$. 
So, $k_r$ and $k_r + k_\w$ are both positive and $\om$ remains real around the resonance. 
This can be understood by noting that the sign of the energy of the wave is the same as that of the wave vector (see appendix~\ref{app:lag_lin_KdV}, \eq{eq:Hq_om}). If $k \, (k + k_r) > 0$, energy conservation thus prevents any amplification as the two waves involved have energies of the same sign. 
\item If $j_r = 1$, $k_r < 0$ while $k_r + k_\w > 0$ (since the dispersion relation has only one negative root for $\om_0 > 0$).  
So, in a finite range of values of $k$ containing $k_r$, the argument of the square root in Equation~\eqref{eq:KdV_detuned_om} is negative. 
The homogeneous solution $\eta = 0$ is thus dynamically unstable~\footnote{The appearance of complex frequencies due to the presence of the undulation is reminiscent of the modulation instability (see Sec.~5.1 of~\cite{Agrawal} for an analysis in nonlinear optics), also known as the Benjamin-Feir instability~\cite{Benjamin-Feir} in the context of water waves.  The modulation instability occurs due to a resonance between the sidebands $k_{c} \pm \delta k$ of a carrier wave, which resonate with the carrier through the relation $\left(k_{c} + \delta k\right) + \left(k_{c} - \delta k\right) = 2 k_{c}$.  It is thus of second order in the amplitude of the carrier wave.}, some perturbations growing exponentially in time with a maximum growth rate close to $\sqrt{\abs{k_r \s \lp k_r + k_\w \rp}} \abs{\w_1}$. 
The reason is that the two waves entering the resonance now have opposite energies. They can thus be both amplified while conserving the total energy of the system. 
\end{itemize}

To determine the transfer matrix of stationary modes, it is convenient to work at fixed $\om \in \mathbb{R}$. 
A few algebraic manipulations from \eq{eq:KdV_xi1_xi0} give, assuming $\delta \om \equiv \om - \om_r = \orde{}$:
\begin{equation} \label{eq:KdV_detuned_k} 
\begin{aligned}
2 \, \om_0' \lp k_r \rp \s \om_0' \lp k_r + k_\w \rp \s \delta k = & \:
	\lp \om_0' \lp k_r \rp + \om_0' \lp k_r + k_\w \rp \rp \s \delta \om \\
&	\pm \sqrt{
		\lp \om_0' \lp k_r \rp - \om_0' \lp k_r + k_\w \rp \rp^2 \, \delta \om^2
		+ 4 \, \om_0' \lp k_r \rp \, \om_0' \lp k_r + k_\w \rp \, k_r \, \lp k_r + k_\w \rp \, \abs{\w_1}^2
		+ \orde{3}
	} , 
\end{aligned} 
\end{equation}
where $\delta k \equiv k - k_r$, and $\om_0' \equiv \partial_k \om_0$.  
Just as the frequency of \eq{eq:KdV_detuned_om}, the wave vector given by \eq{eq:KdV_detuned_k} can be real or complex depending on which waves are involved in the resonance. 
We first notice that in the limit $\ep \to 0$, i.e., in the absence of a resonant undulation, we recover 
\begin{equation*}
\delta k \approx \frac{\delta \om}{\om_0' \lp k_r \rp}
 \text{ or }  
\delta k \approx \frac{\delta \om}{\om_0' \lp k_r + k_\w \rp},
\end{equation*}
as expected since the two modes we are computing then have wave vectors close to $k_r$ and $k_r + k_\w$. 
On the other hand, at resonance, i.e., for $\delta \om = 0$, we obtain
\begin{equation*}
\delta k = \pm \sqrt{\frac{k_r}{\om_0' \lp k_r \rp} \frac{k_r + k_\w}{\om_0' \lp k_r + k_\w \rp} \abs{\w_1}^2
	+ \orde{3}}. 
\end{equation*}
The behavior of the modes at or near the resonance thus depends on the relative signs of $\om_0'(k) \, k$ evaluated at $k = k_r$ and $k = k_r + k_\w$:
\begin{itemize}
\item If they have opposite signs (i.e., if $i_r = 3$), $\delta k$ is purely imaginary at resonance. 
The two modes are thus exponentially increasing and decreasing in $x$, with a growth/decay rate equal to
\begin{equation}\label{KdV_growth_rate}
\mathrm{Im} \lp k \rp = \sqrt{\abs{\frac{k_r}{\om_0' \lp k_r \rp} \frac{k_r + k_\w}{\om_0' \lp k_r + k_\w \rp}}} \abs{\w_1} + \orde{2}
\end{equation}
for $\delta \om = 0$. 
More generally, $k$ has a nonvanishing imaginary part provided $\abs{\delta \om} < \delta \om_\textrm{c}$, where 
\begin{equation} \label{eq:KdV_om_c}
\delta \om_\textrm{c} = 2 \abs{\w_1} \frac{\sqrt{\abs{\om_0' \lp k_r \rp \, \om_0' \lp k_r + k_\w \rp \, k_r \, \lp k_r + k_\w \rp}}}{\abs{\om_0' \lp k_r \rp - \om_0' \lp k_r + k_\w \rp}} 
	+ \orde{2}. 
\end{equation}
\item If they have the same sign, $k$ remains real and the two modes are bounded. 
\end{itemize} 
Since the wave vector gives the sign of the energy of a wave, multiplying it by the group velocity $\om_0'$ gives the sign of the energy flux.  
These results can thus also be interpreted in terms of energy conservation: when the energy fluxes of the two waves involved in the resonance have the same sign, the sum of their squared amplitude (after proper normalization) must be uniform in a stationary solution. When they have opposite signs, however, they can both grow without bound while maintaining a uniform energy flux if the growth of the wave with a negative flux exactly compensates that of the wave with positive flux. 

\begin{figure} 
\includegraphics[width = 0.49 \linewidth]{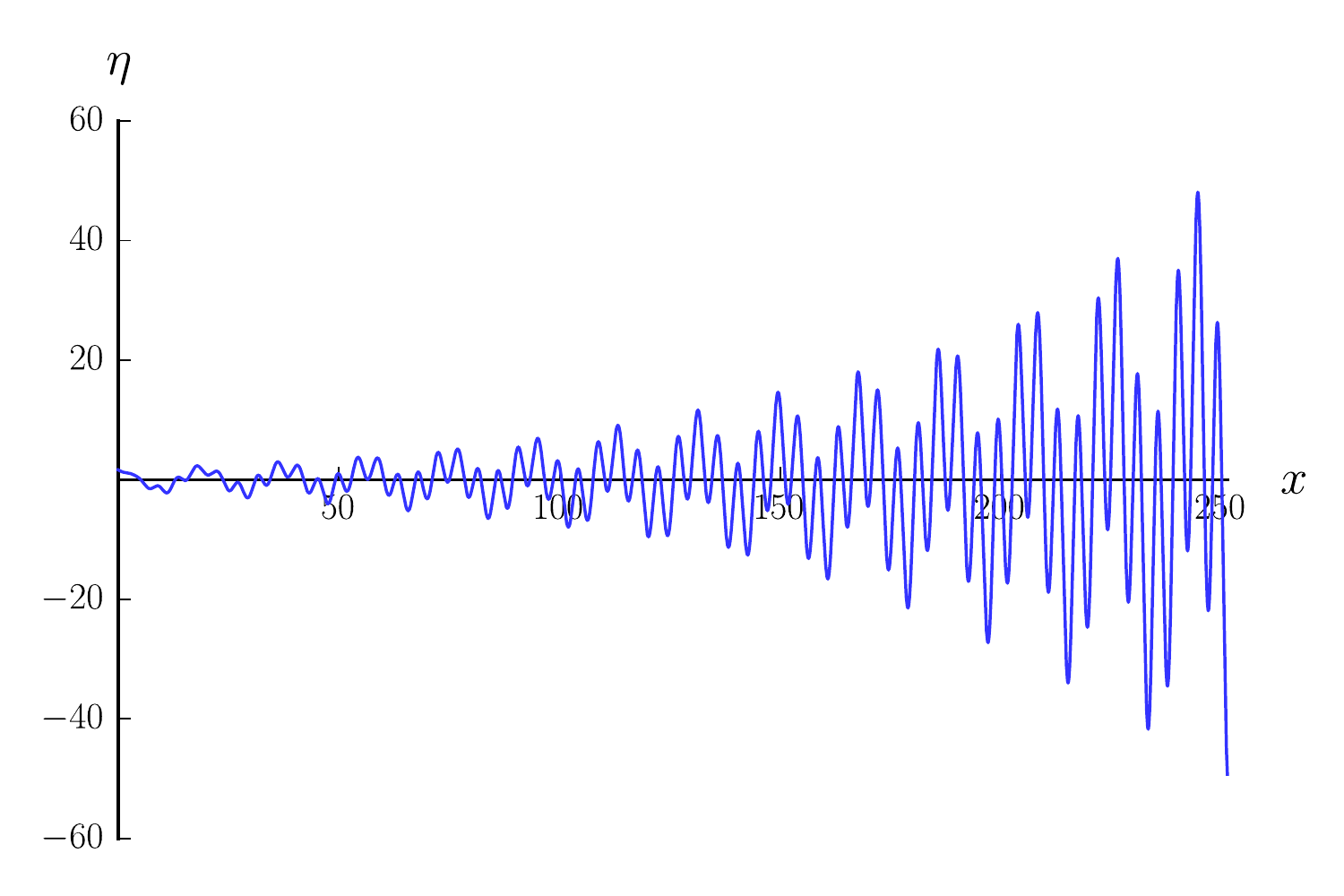}
\includegraphics[width = 0.49 \linewidth]{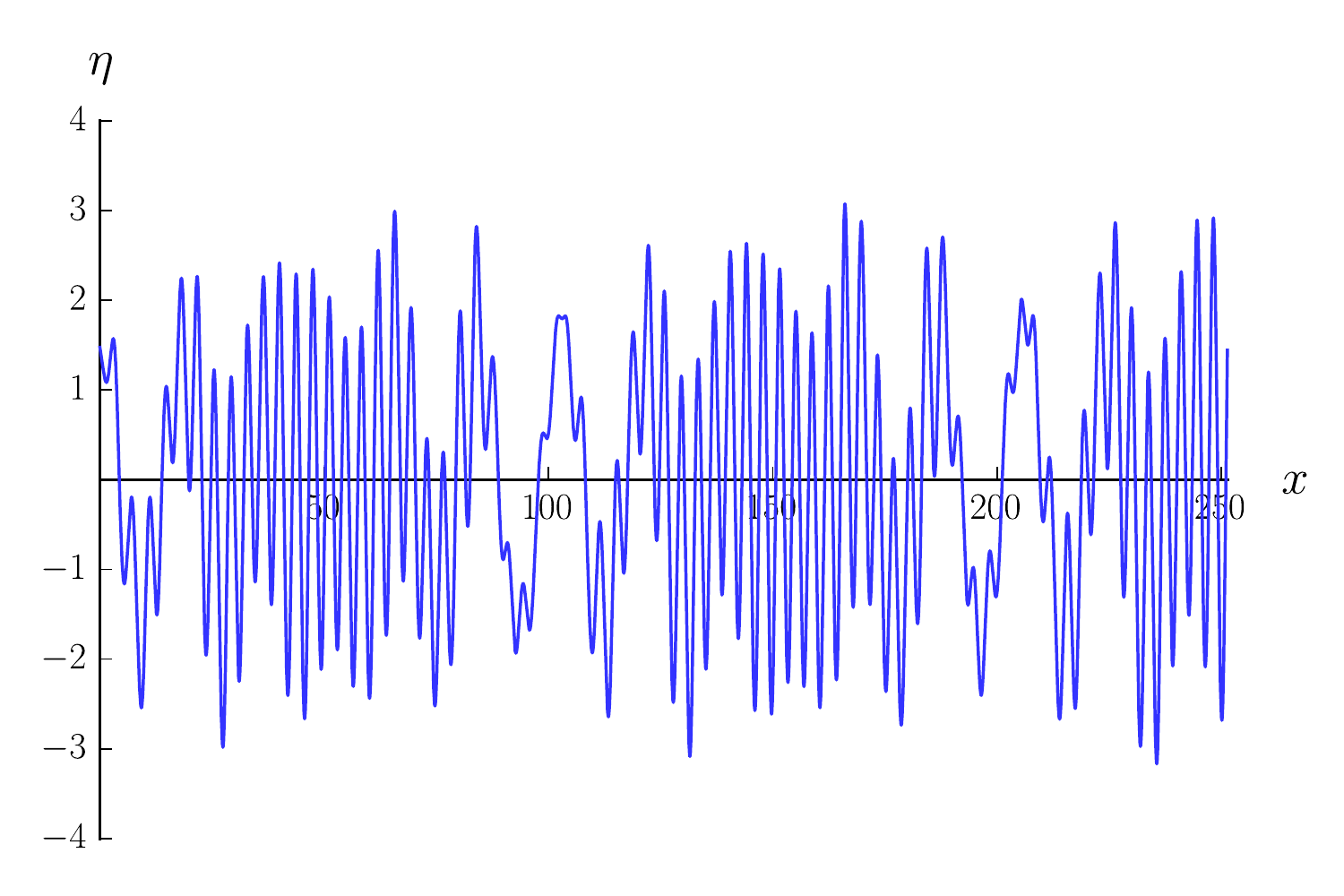}
\caption{Illustrations of the two possible resonant mode behaviors over a ``detuned'' undulation. 
Only the real part is shown. 
The imaginary part is similar, with a dephasing of approximately $\pi / 2$. 
In the left plot, we show the exponentially growing case, with an undulation of wave vector $k_{\om_r}^{(3)} - k_{\om_r}^{(2)}$, thereby mixing modes with opposite energy fluxes (as explained in the main text). 
In the right plot, the undulation wave vector is instead equal to $k_{\om_r}^{(2)} - k_{\om_r}^{(1)}$, and mixes modes with the same energy flux. 
For these two plots, the angular frequency at resonance is $\omega = 0.8$. 
The two undulations have the form \eqref{eq:undul_shape_KdV} with $\w_0 = 2.5$, and their amplitudes are $\abs{\w_1} = 0.05$ for the left panel, and $0.15$ for the right panel.} \label{fig:KdV_detuned_sols}
\end{figure}

These two behaviors are shown in \fig{fig:KdV_detuned_sols} in the case $\w_1 \in \ii \s \mathbb{R}_*^+$ and $\w_n = 0$ for $\abs{n} \geq 2$. 
The undulation thus has the form 
\begin{equation} \label{eq:undul_shape_KdV}
\w(x) = \w_0 + 2 \abs{\w_1} \, \sin \lp k_\w \, x \rp. 
\end{equation}
These plots are obtained by solving the linear KdV equation~\eqref{eq:linKdV} numerically using a finite difference method. 
The modes are shown at resonance $\om = \om_r$, in the cases $i_r = j_r + 1 = 3$ (left panel) and $i_r = j_r + 1 = 2$ (right panel). 
The first case thus corresponds to a dynamically stable modulation of $\w$ with an exponentially growing mode in $x$, shown in the figure. 
One can verify that the growth rate agrees with \eq{KdV_growth_rate}. 
The second case corresponds to a dynamically unstable modulation with spatially bounded modes for $\om \in \mathbb{R}$.

To summarize, the main results of this subsection are: 
\begin{itemize}
\item when working at fixed real quasimomentum, a dynamical instability (presence of exponentially growing modes in time) is present if the waves coupled by the undulation have energies of opposite signs;
\item when working at fixed real frequency, exponentially growing mode in space are present if the waves coupled by the undulation have energy fluxes of opposite signs.
\end{itemize}
A corollary of these observations is that the presence of exponentially growing modes in space is equivalent to the existence of dynamical instabilities if and only if the waves at play have group velocities with the same sign. 

\subsection{Transfer matrix}
\label{app:transfer}

In this subsection, we work at fixed $\om \in \left] 0, \om_\mathrm{max} \right[$ and compute the transfer matrix over a detuned, localized undulation. 
We thus consider undulations with space-dependent amplitudes going to zero as $x \to \pm \infty$, sufficiently fast for asymptotic modes to be well defined. 
The transfer matrix then relates the amplitudes of each wave on the left and on the right of the undulation. 
It encodes the effect of the undulation on the solutions of the linearized KdV equation~\eqref{eq:linKdV}, and can be used to determine the scattering in more complicated setups.  
For instance, in the problem of an obstacle followed by an undulation, in the case where there is a neat separation of scales between the two (e.g., if the obstacle is narrow but high while the undulation has a small amplitude but extends over a long region), the total transfer matrix is simply the product of those of the obstacle and undulation. 
The scattering matrix, more often used in analogue gravity studies, can then be obtained straightforwardly, see appendix~\ref{app:TtoS}. 

The first step is to compute the modes with fixed angular frequency over the undulation. 
There are three of them. 
Two are obtained from Eqs.~(\ref{eq:KdV_xi2},\ref{eq:KdV_xim1},\ref{eq:KdV_xi1_xi0}). 
The possible values of their quasimomentum $k$ are $k_r + \delta k$, where $\delta k$ takes the two values of \eq{eq:KdV_detuned_k}. 
In the following, to simplify the notations we denote by $\delta k_s$ the half sum of the two solutions \eqref{eq:KdV_detuned_k} and by $\delta k_d$ their half difference.  
The first one, $\delta k_s$, is thus always real, while $\delta k_d$ becomes imaginary close to resonance if the energy fluxes of the two waves have opposite signs. 
Explicitly, these two modes read
\begin{align} \label{eq:KdV_detuned_res_modes} 
\eta_\pm: (t,x) \mapsto \e^{\ii \s \lp k_r + \delta k_\pm \rp \s x - \ii \s \om \s t}
\left[ 
	\frac{\lp k_r - k_\w \rp \s \w_1}{\om - (k_r - k_\w) \s \w_0 + (k_r - k_\w)^3} \s \e^{- \ii \s k_\w \s x} +
	1 +
	\lp \frac{\delta \om - \om_0'(k_r) \s \delta k_\pm + \ord{\delta k^2}}{k_r \s \w_1^*} \rp \e^{\ii \s k_\w \s x} +
	\right. \nn \left. 
	+ \frac{\delta \om - \om_0'(k) \s \delta k_\pm}{\om - \om_0 \lp k_r + 2 k_\w \rp} \s \lp 1 + 2 \frac{k_\w}{k_r} \rp \s \frac{\w_1}{\w_1^*}  \s \e^{2 \s \ii \s k_\w \s x}
	+ \orde{2}
\right] ,
\end{align}
where $\delta k_\pm \equiv \delta k_s \pm \delta k_d$ are the two solutions \eqref{eq:KdV_detuned_k}. 
The third, nonresonant mode $\eta_\textrm{nr}$ can be obtained in a similar way. 
The main difference is that there is only one term of order $0$ in $\ep$, with a wave vector equal to $k_\om^{(6 - i_r - j_r)}$. 
We obtain
\begin{equation} \label{eq:KdV_detuned_nonres_modes} 
\begin{aligned}
\eta_\textrm{nr}: (t,x) \mapsto
	\e^{\ii \s k \s \lp 1 + \orde{2} \rp \s x - \ii \s \om \s t} \s 
	& \lp 
		1 +
		\frac{(k + k_\w) \s \w_1}{\om - (k + k_\w) + (k + k_\w)^3} \s \e^{+ \ii \s k_\w \s x} 
		\right. \\ & \phantom{(1+} \left. + 
		\frac{(k - k_\w) \s \w_1^*}{\om - (k - k_\w) + (k - k_\w)^3} \s \e^{- \ii \s k_\w \s x}
		+ \orde{2}
	\rp ,
\end{aligned}
\end{equation}
evaluated at $k = k_\om^{(6 - i_r - j_r)}$. 

When considering undulations whose amplitudes go to zero sufficiently fast as $x \to \pm \infty$, one can define two bases of global modes $\lp \eta_{i,\om}^{(L)} \rp_{i \in \lb 1, 2, 3 \rb}$ and $\lp \eta_{i,\om}^{(R)} \rp_{i \in \lb 1, 2, 3 \rb}$, defined respectively by 
\begin{equation} \label{eq:L} 
\forall \, i \in \lb 1,2,3 \rb, \; \forall \, t \in \mathbb{R}, \; 
	\eta_{i,\om}^{(L)}(t,x) \mathop{\sim}_{x \to - \infty} \sqrt{\abs{\frac{k_\om^{(i)}}{2 \s \pi \s \om_0' \lp k_\om^{(i)} \rp}}} \, \e^{\ii \s \lp k_\om^{(i)} \s x - \om \s t \rp}
\end{equation}
and
\begin{equation} \label{eq:R} 
\forall \, i \in \lb 1,2,3 \rb, \; \forall \, t \in \mathbb{R}, \; 
	\eta_{i,\om}^{(R)}(t,x) \mathop{\sim}_{x \to + \infty} \sqrt{\abs{\frac{k_\om^{(i)}}{2 \s \pi \s \om_0' \lp k_\om^{(i)} \rp}}} \, \e^{\ii \s \lp k_\om^{(i)} \s x - \om \s t \rp}.
\end{equation}
The prefactors in these expressions are chosen so that the modes are normalized, in the following sense. 
Defining the canonical field $\phi$ for each value of $i$ and $\om$ by 
\begin{equation} 
\phi_{i,\om}^{(L/R)} (t,x) = \int_0^x \eta_{i, \om}^{(L/R)}(t,y) \, \dd y, 
\end{equation} 
we impose that it satisfies 
\begin{equation} 
\forall (i,j) \in [\![ 1,3 ]\!]^2, \, \forall \lp \om, \om' \rp \in \mathbb{R}_+^2, \, \abs{\lp \phi_{i,\om}^{(L/R)} \Big\vert \phi_{j,\om'}^{(L/R)} \rp} = \delta_{i,j} \, \delta \lp \om - \om' \rp,
\end{equation}
where $\lp \cdot \vert \cdot \rp$ is the inner product conserved by the linear KdV equation, see appendix~\ref{app:lag_lin_KdV}. 
There it is also shown that, for positive values of $\om$, the inner product of $\phi$ with itself gives the sign of the energy of the wave. 

The transfer matrix at fixed frequency, $T^{(\om)}$, is the $3$ by $3$ complex matrix defined by
\begin{equation} \label{eq:def_Tom} 
\begin{pmatrix}
\eta_{1,\om}^{(L)} \\
\eta_{2,\om}^{(L)} \\
\eta_{3,\om}^{(L)} 
\end{pmatrix}
 = T^{(\om)} \, 
\begin{pmatrix}
\eta_{1,\om}^{(R)} \\
\eta_{2,\om}^{(R)} \\
\eta_{3,\om}^{(R)} 
\end{pmatrix} . 
\end{equation} 
The $i$th line of $T^{(\om)}$ thus contains the coefficients of the expansion of $\eta_{i,\om}^{(L)}$ in the basis $\lp \eta_{j,\om}^{(R)} \rp_{j \in \lb 1, 2, 3 \rb}$. 
We now determine the leading contributions to $T^{(\om)}$ in two opposite limits of an undulation with sharp and slowly varying amplitudes. 
As it relates two bases of normalized modes, two of them having positive energy fluxes and one with a negative energy flux, $T^{(\om)}$ is an element of $\mathrm{SU}(2,1)$. 
It can be related to the scattering matrix $S_\om$ by identifying the incoming and outgoing parts of each mode, see Appendix~\ref{app:TtoS}. 

\subsubsection{Undulation with steplike amplitude}
\label{subsubsec:KdV_step} 

\begin{figure} 
\includegraphics[width = 0.49 \linewidth]{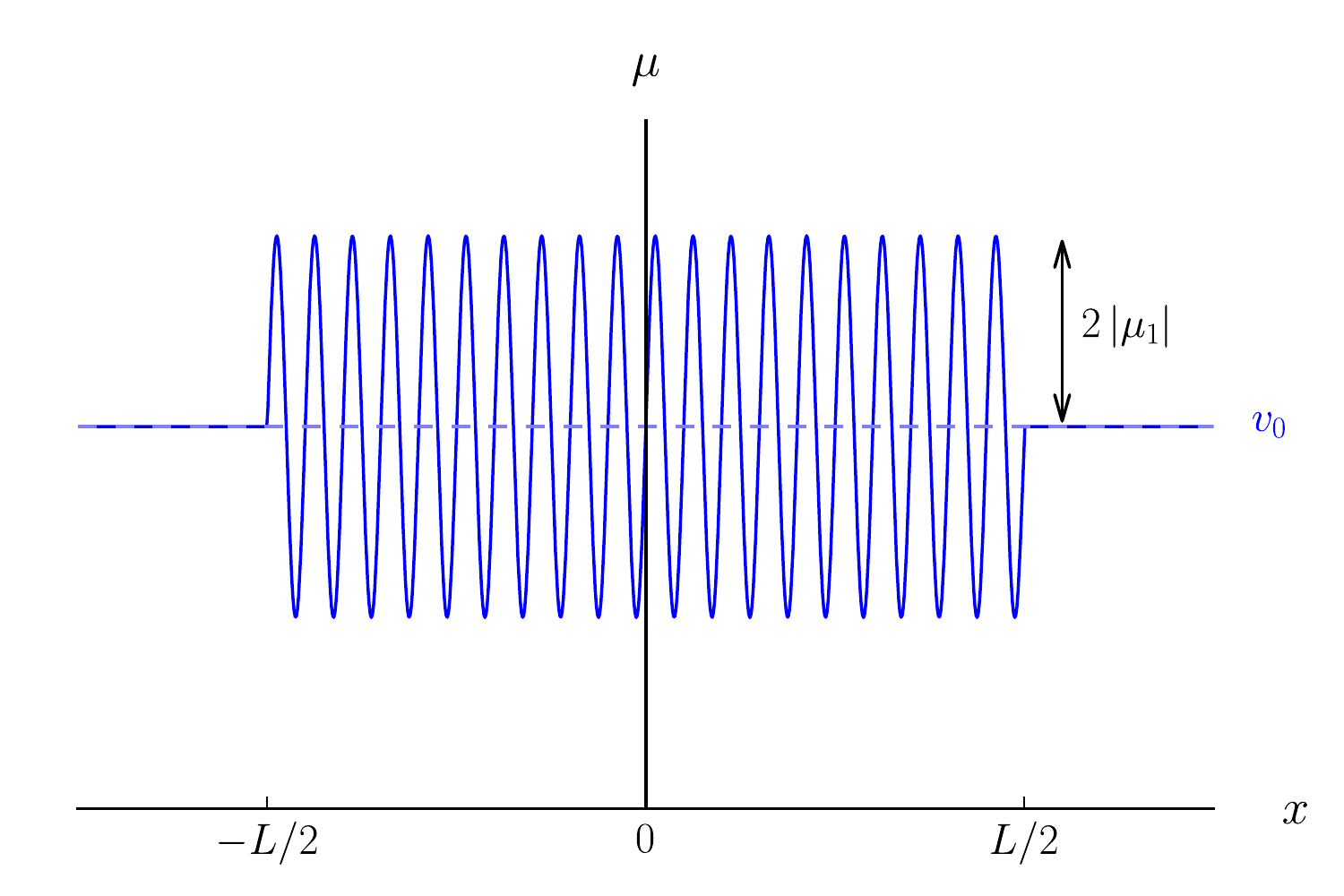} 
\includegraphics[width = 0.49 \linewidth]{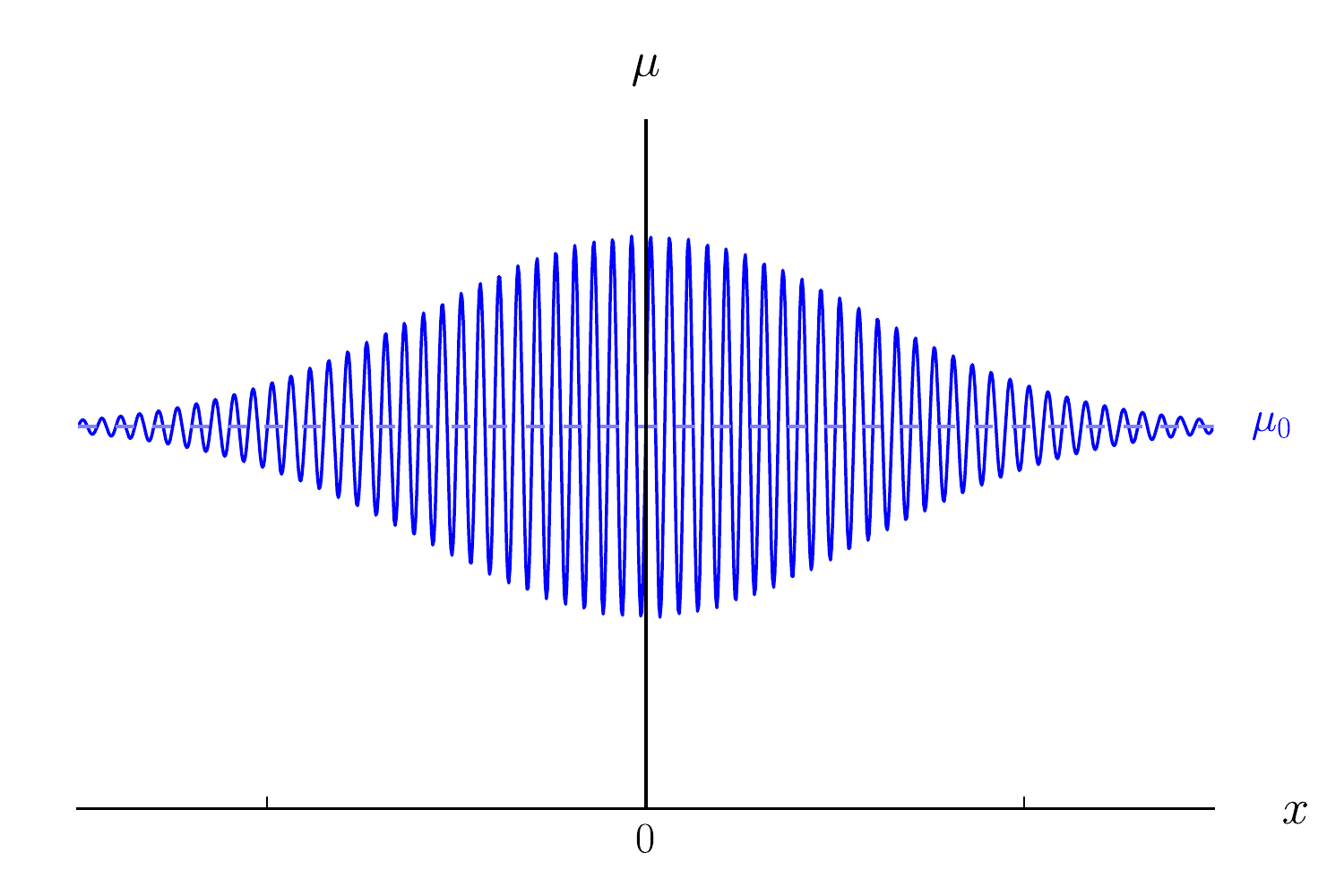}
\caption{Schematic drawings of undulations with ``steplike'' (left panel) and slowly varying (right panel) amplitudes, used in the computation of the transfer matrix.
The dashed line shows the asymptotic value $\w_0$ of $\w$.} 
\label{fig:KdV_undul_det}
\end{figure}

We first consider an undulation with an amplitude vanishing outside a finite interval $\left[ -L/2, +L/2 \right]$ for some positive number $L$ and constant inside it. 
The parameter $\w$ takes the form 
\begin{equation} \label{eq:sharp_undul_KdV}
\w(x) = \w_0 
+ 2 \abs{\w_1} \, \cos \lp k_\w \s \lp x - x_\w \rp \rp \, \theta \lp x - L/2 \rp \, \theta \lp x + L/2 \rp  + \orde{2}, 
\end{equation}
where $x_\w \in \mathbb{R}$. 
It is shown in the left panel of \fig{fig:KdV_undul_det}. 
Notice that it has discontinuities at $x = \pm L / 2$ unless $k_\w \, \lp \pm L/2 - x_\w \rp = \pi / 2 \mod \pi$. 
However, if $k_\w \, L$ is sufficiently large the results should be only marginally affected by a local change of $\w$ smoothly sending the amplitude of the oscillation to $0$ near $x = \pm L$. 
The relation between the two aforementioned bases can be computed straightforwardly by using the matching conditions at $x = \pm L/2$, i.e., continuity of $\eta$, $\pd_x \eta$, and $\w \s \eta + \pd_x^2 \eta$. 

Let us first focus on the resonant case $\delta \om = 0$, which gives simple expressions. 
For definiteness, we also assume $i_r = j_r + 1 = 3$. 
(The expressions for other values of $i_r$ and $j_r$ are obtained by exchanging the corresponding lines and columns of $T^{(\om)}$.) 
We obtain
\begin{equation}
T^{(\om)} = 
\begin{pmatrix}
1 & 0 & 0 \\
0 & \cos \lp \delta k_d \, L \rp & - \ii \, \e^{- \ii \s k_\w \s x_\w} \sin \lp \delta k_d \, L \rp \\
0 & - \ii \, \e^{\ii \s k_\w \s x_\w} \sin \lp \delta k_d \, L \rp & \cos \lp \delta k_d \, L \rp
\end{pmatrix}
+ \orde{}. 
\end{equation}
(The next order in $\ep$ can also be obtained from \eqs{eq:KdV_detuned_res_modes}{eq:KdV_detuned_nonres_modes}.) 
Since $\delta k_d \in \ii \mathbb{R}$, the four coefficients relating two resonant waves grow exponentially with $L$. 
This is the most important effect of the undulation: even if $\abs{\w_1} \ll 1$, the amplification of the waves across the undulation can be large provided the latter is sufficiently long. 
More precisely, it remains finite in the limit $\ep \to 0$ provided $L$ scales like $\ep^{-1}$. 
If $i_r = 2$ or $j_r = 1$, $\delta k_d \in \mathbb{R}$. 
Then $T^{(\om)}$ is a rotation matrix up to a phase, in the sense that one can find a diagonal, real matrix $D_\om$ such that $\e^{\ii \s D_\om} \, T^{(\om)} \, \e^{-\ii \s D_\om} \in \mathrm{SO}(3)$. 

\begin{figure}
\begin{center}
\includegraphics[width = 0.49 \linewidth]{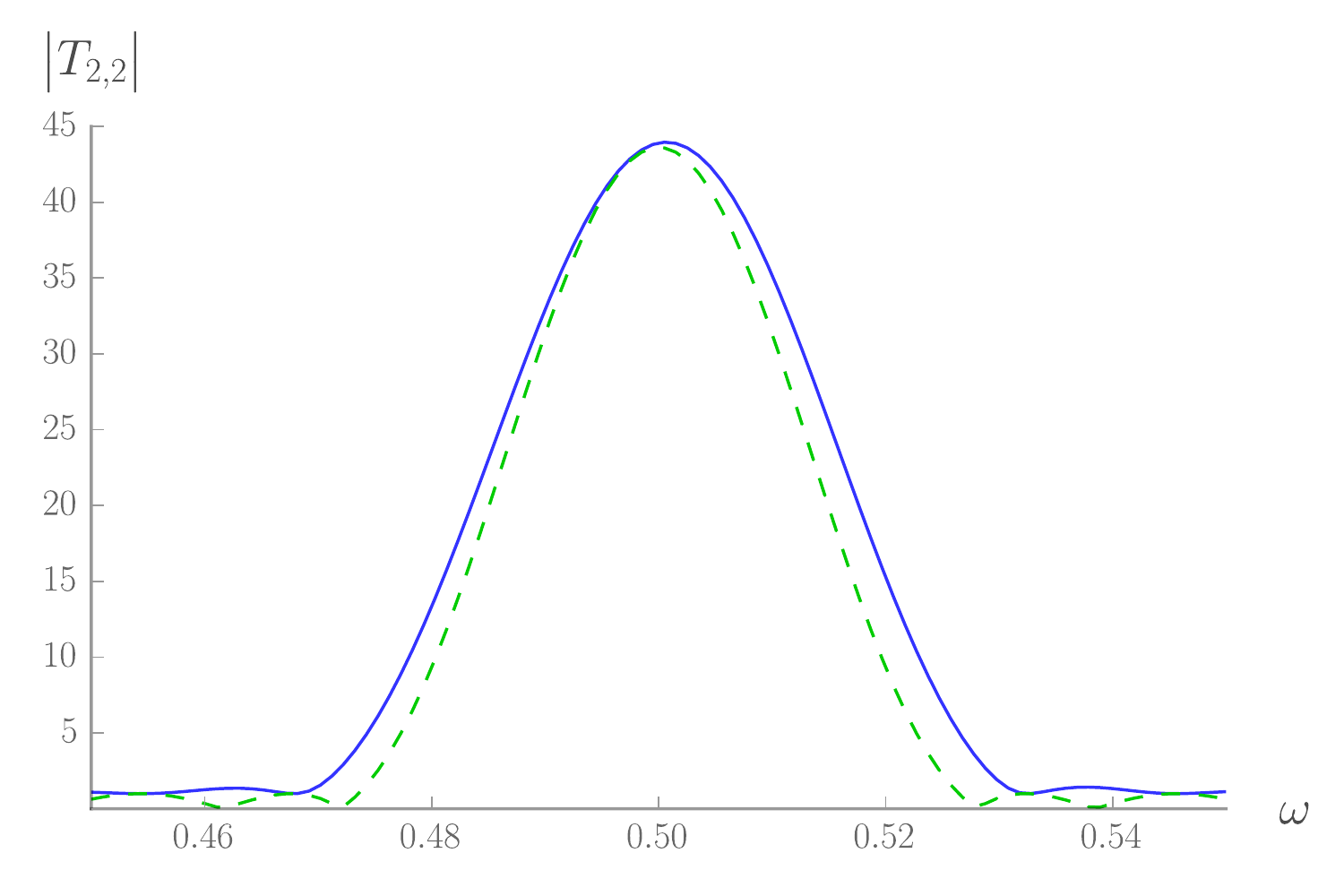} 
\end{center}
\caption{Plots of the transfer coefficient $T^{(\om)}_{2,2}$ relating the coefficients of the plane wave of wave vector $k_\om^{(2)}$ for an undulation of the form \eqref{eq:sharp_undul_KdV}, where $k_\w = k_\om^{(3)} - k_\om^{(2)}$ for $\om = 0.5$, $\w_0 = 2.5$, $\abs{\w_1} = 0.05$, and $L \approx 503$.
The continuous, blue line shows results from the numerical integration of the linear KdV equation. 
The dashed, green line shows the analytical result from Equation~\eqref{eq:transf_sharp_indul_KdV}.
} 
\label{fig:KdV_transfer_coeffs}
\end{figure}

When going slightly off resonance, the transfer matrix can be read from the following relations between the global modes:
\begin{equation} \label{eq:transf_sharp_indul_KdV} 
\begin{aligned}
\eta_{1,\om}^{(L)} & = \eta_{1,\om}^{(R)} + \orde{} , \\
\eta_{2,\om}^{(L)} & = 
	\e^{+\ii \s \delta k_s \s L} \s \lp \cos \lp \delta k_d \s L \rp 
		- \ii \frac{\delta k_s}{\delta k_d} \sin \lp \delta k_d \s L \rp \rp \s \eta_{2,\om}^{(R)} 
	- \ii \s \frac{\w_1}{\delta k_d} \s \sqrt{\abs{\frac{k_\om^{(2)} \s k_\om^{(3)}}{\om_0' \lp k_\om^{(2)} \rp \s \om_0' \lp k_\om^{(3)} \rp}}} \s \e^{-\ii \s \delta k_s \s L} \s \sin \lp \delta k_d \s L \rp \s \eta_{3,\om}^{(R)}
	+ \orde{} , \\
\eta_{3,\om}^{(L)} & = 
	-\ii \s \frac{\w_1^*}{\delta k_d} \s \sqrt{\abs{\frac{k_\om^{(2)} \s k_\om^{(3)}}{\om_0' \lp k_\om^{(2)} \rp \s \om_0' \lp k_\om^{(3)} \rp}}} \s \e^{+\ii \s \delta k_s \s L} \s \sin \lp \delta k_d \s L \rp \s \eta_{2,\om}^{(R)} + 
	\e^{-\ii \s \delta k_s \s L} \s \lp \cos \lp \delta k_d \s L \rp 
		+ \ii \frac{\delta k_s}{\delta k_d} \sin \lp \delta k_d \s L \rp \rp \s \eta_{3,\om}^{(R)}
	+ \orde{} .
\end{aligned}
\end{equation}
The corresponding values for the $(2,2)$ coefficient of the transfer matrix are compared with results from a numerical resolution of \eq{eq:linKdV} in \fig{fig:KdV_transfer_coeffs}. 
(The plots are similar for the three other coefficients involving $\eta_2^{(L)}$ and/or $\eta_3^{(L)}$.) 
We observe the same exponential growth in $\delta k_d \, L$ when $\delta k_d \in \ii \mathbb{R}$. 
From \eq{eq:KdV_detuned_k}, one sees that $\delta \om$ tends to reduce $\abs{\mathrm{Im} \lp \delta k_d \rp}$ when moving away from the resonance. 
The growth rate vanishes for $\abs{\delta \om}$ larger than a critical value. 

\subsubsection{Undulation with slowly varying amplitude}

We now consider the case where the amplitude $\w_1$ of the undulation varies slowly with $x$, shown schematically on the right panel of \fig{fig:KdV_undul_det}. 
We work in the limit $\abs{\pd_x \w_1} / (k_\w \, \w_1) \ll 1$. 
The transfer matrix over some interval $[x_1, x_2]$ with $x_2 \gg 1 / k_\w, \, -x_1 \gg 1 / k_\w$ can then be computed by dividing $[x_1, x_2]$ into $N \gg 1$ subintervals over which $\w_1$ can be approximated by a constant, multiplying the transfer matrices over each subinterval, and taking the limit $N \to \infty$. 
Restricting attention to the subspace spanned by the two waves of wave vectors $k_\om^{(i_r)}$ and $k_\om^{(j_r)}$, one obtains the transfer matrix
\begin{equation}
T^{(\om)} \lp x_1, x_2 \rp \approx 
	\begin{pmatrix}
	\e^{+\ii \s \delta k_d \s x_f} & 0 \\
	 0 & \e^{-\ii \s \delta k_d \s x_f}
	\end{pmatrix} \,
	\mathrm{OE} \lp
		\int_{x_1}^{x_2} - \ii 
		\begin{pmatrix}
		\delta k_s(x) & \Gamma \, \w_1^* \\
		\Gamma \, \w_1 & - \delta k_s(x)
		\end{pmatrix}
		\dd x
	\rp \,
	\begin{pmatrix}
	\e^{-\ii \s \delta k_d \s x_i} & 0 \\
	 0 & \e^{+\ii \s \delta k_d \s x_i}
	\end{pmatrix},
\end{equation}
where $\mathrm{OE}$ denotes the path-ordered exponential and 
\begin{equation} 
\Gamma \equiv \sqrt{\frac{k_r \s \lp k_r + k_\w \rp}{\om_0' \lp k_r \rp \s \om_0' \lp k_r + k_\w \rp}}.
\end{equation}
In the resonant case $\delta \om = 0$, this becomes
\begin{equation}
T^{(\om)} \lp x_1, x_2 \rp \approx 
\begin{pmatrix}
\e^{- \ii \s k_r \s \lp x_u(x_2) - x_u(x_1) \rp / 2} \cos \lp \int_{x_1}^{x_2} \delta k_d \, \dd x \rp & - \ii \, \e^{- \ii \s k_r \s \lp x_u(x_2) + x_u(x_1) \rp / 2} \sin \lp \int_{x_1}^{x_2} \delta k_d \, \dd x \rp \\
- \ii \, \e^{+ \ii \s k_r \s \lp x_u(x_2) + x_u(x_1) \rp / 2} \sin \lp \int_{x_1}^{x_2} \delta k_d \, \dd x \rp & \e^{\ii \s k_r \s \lp x_u(x_2) - x_u(x_1) \rp / 2} \cos \lp \int_{x_1}^{x_2} \delta k_d \, \dd x \rp
\end{pmatrix} .
\end{equation}
One recovers essentially the same behavior as in the case of an undulation with steplike amplitude, with $\delta k_d$ replaced by its average over the segment $[x_1, x_2]$. 

These results can be used to determine the relations between the asymptotic modes and the scattering matrix using the formulas of Appendix~\ref{app:TtoS} below. 
In particular, in the case $i_r = 3$, one sees from \eq{eq:KdV_S_31} that the $4$ coefficients of the scattering matrix relating ``resonating'' modes show the same exponential growth in $L$ as those of the transfer matrix.

\subsection{Relation between the transfer matrix and the scattering matrix} 
\label{app:TtoS}

In Appendix~\ref{app:transfer} we focused on the transfer matrix of stationary modes in the presence of an undulation, which can be straightforwardly combined with that of other sources of inhomogeneities of the flow, e.g. a localized obstacle. 
However, experiments using one-dimensional setups like the one we are considering (see for instance~\cite{Weinfurtner:2010nu,Euve:2014aga,Steinhauer:2015saa,Euve:2015vml}) usually measure the scattering matrix instead, which is more directly related to physical observables. 
To bridge the gap between these two descriptions, we here give the expression of the scattering matrix in terms of the transfer matrix. 
For definiteness we assume that, for the frequencies we are interested in, the dispersion relation is similar to that of the linear KdV equation (see \fig{fig:DR_KdV}) for $\om \in  \left] -\om_{\mathrm{max}}, \om_{\mathrm{max}} \right[$, i.e., that it has three real roots $k_\om^{(1)} < k_\om^{(2)} < k_\om^{(3)}$ at fixed $\omega$, such that 
\begin{equation}
\frac{\dd k_\om^{(1)}}{\dd \om} < 0, \; \frac{\dd k_\om^{(2)}}{\dd \om} > 0, \; \frac{\dd k_\om^{(3)}}{\dd \om} < 0. 
\end{equation}
The reasoning detailed below can be extended to more intricate dispersion relations, for instance to the one used in Appendix~\ref{app:UCP_detuned}. 

Let us first recall that the transfer matrix $T^{(\om)}$ and the scattering matrix $S^{(\om)}$ each represent a change of basis in the same (in the present case, three-dimensional) vector space, made of the modes with fixed angular frequency $\om$. 
The ``left'' and ``right'' bases are defined in \eqs{eq:L}{eq:R}. 
The transfer matrix $T^{(\om)}$ is then defined by \eq{eq:def_Tom}.  
Similarly, one defines the ``in'' and ``out'' bases by the condition that $\eta_{i,\om}^{(\mathrm{in})}$ (respectively $\eta_{i,\om}^{(\mathrm{out})}$) contains asymptotically only one plane wave with a group velocity oriented toward (respectively, away from) $x = 0$, with the wave vector $k_\om^{(i)}$ and with the  same normalization as in \eqs{eq:L}{eq:R}. 
The scattering matrix is then defined by
\begin{equation}
\begin{pmatrix}
\eta_{1,\om}^{(\mathrm{in})} \\
\eta_{2,\om}^{(\mathrm{in})} \\
\eta_{3,\om}^{(\mathrm{in})}
\end{pmatrix}
= S^{(\om)} \, 
\begin{pmatrix}
\eta_{1,\om}^{(\mathrm{out})} \\
\eta_{2,\om}^{(\mathrm{out})} \\
\eta_{3,\om}^{(\mathrm{out})}
\end{pmatrix} . 
\end{equation}

The coefficients of $S^{(\om)}$ can be related to those of $T^{(\om)}$ by expanding the ``in'' and ``out'' modes in $\eta_i^{(L)}$ and $\eta_i^{(R)}$, $i \in \lb 1, 2, 3 \rb$, then the $\eta_{i}^{(L)}$ in terms of the $\eta_{i}^{(R)}$ using \eq{eq:def_Tom}. 
We obtain:
\begin{equation}
\begin{aligned}
& S_{1,1}^{(\om)} = \frac{T_{3,3}^{(\om)}}{T_{1,1}^{(\om)} \, T_{3,3}^{(\om)} - T_{1,3}^{(\om)} \, T_{3,1}^{(\om)}}
, \; 
S_{1,3}^{(\om)} = \frac{- T_{1,3}^{(\om)}}{T_{1,1}^{(\om)} \, T_{3,3}^{(\om)} - T_{1,3}^{(\om)} \, T_{3,1}^{(\om)}}
, \;
S_{1,2}^{(\om)} = \frac{T_{1,2}^{(\om)} \, T_{3,3}^{(\om)} - T_{3,2}^{(\om)} \, T_{1,3}^{(\om)}}{T_{1,1}^{(\om)} \, T_{3,3}^{(\om)} - T_{1,3}^{(\om)} \, T_{3,1}^{(\om)}}, \\
& S_{3,3}^{(\om)} = \frac{T_{1,1}^{(\om)}}{T_{1,1}^{(\om)} \, T_{3,3}^{(\om)} - T_{1,3}^{(\om)} \, T_{3,1}^{(\om)}}
, \;  
S_{3,1}^{(\om)} = \frac{- T_{3,1}^{(\om)}}{T_{1,1}^{(\om)} \, T_{3,3}^{(\om)} - T_{1,3}^{(\om)} \, T_{3,1}^{(\om)}}
, \;
S_{3,2}^{(\om)} = \frac{T_{3,2}^{(\om)} \, T_{1,1}^{(\om)} - T_{1,2}^{(\om)} \, T_{3,1}^{(\om)}}{T_{1,1}^{(\om)} \, T_{3,3}^{(\om)} - T_{1,3}^{(\om)} \, T_{3,1}^{(\om)}}, \\
& S_{2,1}^{(\om)} = \frac{T_{2,3}^{(\om)} \, T_{3,1}^{(\om)} - T_{2,1}^{(\om)} \, T_{3,3}^{(\om)}}{T_{1,1}^{(\om)} \, T_{3,3}^{(\om)} - T_{1,3}^{(\om)} \, T_{3,1}^{(\om)}}
, \; 
S_{2,3}^{(\om)} = \frac{T_{2,1}^{(\om)} \, T_{1,3}^{(\om)} - T_{2,3}^{(\om)} \, T_{1,1}^{(\om)}}{T_{1,1}^{(\om)} \, T_{3,3}^{(\om)} - T_{1,3}^{(\om)} \, T_{3,1}^{(\om)}}, \\
& S_{2,2}^{(\om)} = T_{2,2}^{(\om)} + \frac{T_{1,2}^{(\om)} \, \lp T_{2,3}^{(\om)} \, T_{3,1}^{(\om)} - T_{2,1}^{(\om)} \, T_{3,3}^{(\om)} \rp + T_{3,2}^{(\om)} \, \lp T_{2,1}^{(\om)} \, T_{1,3}^{(\om)} - T_{2,3}^{(\om)} \, T_{1,1}^{(\om)} \rp}{T_{1,1}^{(\om)} \, T_{3,3}^{(\om)} - T_{1,3}^{(\om)} \, T_{3,1}^{(\om)}} . 
\end{aligned}
\end{equation}
In the case of a detuned undulation with steplike amplitude (see appendix~\ref{subsubsec:KdV_step}), using \eq{eq:transf_sharp_indul_KdV} (or its generalization to the case $i_r \neq 2$), we obtain three different expressions depending on $k_\w$ to leading order in $\ep$:
\begin{itemize}
\item if $i_r = 3$ and $j_r = 2$, 
\begin{equation}
S^{(\om)} \approx 
\frac{1}{T_{3,3}^{(\om)}}
\begin{pmatrix}
T_{3,3}^{(\om)} & 0 & 0 \\
0  & 1 & - T_{2,3}^{(\om)} \\
0 & T_{3,2}^{(\om)} & 1
\end{pmatrix};
\end{equation}
\item if $i_r = 3$ and $j_r = 1$, 
\begin{equation} \label{eq:KdV_S_31}
S^{(\om)} \approx 
\begin{pmatrix}
T_{3,3}^{(\om)} & 0 & - T_{1,3}^{(\om)} \\
0 & 1 & 0 \\
- T_{3,1}^{(\om)} & 0 & T_{1,1}^{(\om)}
\end{pmatrix}; 
\end{equation}
\item if $i_r = 2$ and $j_r = 1$, 
\begin{equation}
S^{(\om)} \approx 
\frac{1}{T_{1,1}^{(\om)}}
\begin{pmatrix}
1 & T_{1,2}^{(\om)} & 0 \\
- T_{2,1}^{(\om)}  & 1 & 0 \\
0 & 0 & T_{1,1}^{(\om)}
\end{pmatrix}.
\end{equation}
\end{itemize}

\section{Linearly growing modes over tuned undulations} 
\label{app:lingrowmodes}

In this appendix we provide a more general viewpoint on linearly growing modes and show their relations with nonlinear solutions. 
The nonlinear solutions of the KdV equation have been extensively studied, see for instance~\cite{Kamchatnov,Grimshaw} and references therein. A powerful tool to study their evolution is Whitham's modulation theory, which can be used to construct (approximate) scale-invariant solutions. As an example of application to a related subject, in~\cite{Michel:2015mlr} Whitham's modulation theory was used to motivate the nonlinear stability of analogue black hole flows. In the present case, however, the solutions we are interested in can be obtained by an elementary calculation, which can more directly be extended to other models such as that of Appendix~\ref{app:UCP}. It would be interesting to see if they can be recovered using Whitham's modulation theory.

\paragraph{Modes growing in space:}

$\phantom{0}$ 

\vspace*{0.25 cm}

\noindent Let us first explain on general grounds why linearly growing modes are expected in models where, as for the KdV equation, the wave vector depends on some parameters describing the solution. 
Let us consider a nonlinear, stationary equation which admits an $n$-dimensional space of periodic solutions ($n \in \mathbb{N}^*$), labeled by the parameters $\mu_1$, $\mu_2$, ..., $\mu_n$ taking values in some real intervals $I_1$, $I_2$, ..., $I_n$. 
These solutions may be written as
\begin{equation}
x \mapsto f_{\mu_1, \mu_2, ..., \mu_n} \lp k_{\mu_1, \mu_2, ..., \mu_n} \, x \rp,
\end{equation}
where, for each choice of the parameters, $f_{\mu_1, \mu_2, ..., \mu_n}$ is a periodic, differentiable function with period $2 \s \pi$, and $k_{\mu_1, \mu_2, ..., \mu_n} \in \mathbb{R}$. 
We further assume that $f_{\mu_1, \mu_2, ..., \mu_n}$ and $k_{\mu_1, \mu_2, ..., \mu_n}$ are differentiable in the parameters.~\footnote{By this, we mean that:
\begin{itemize}
\item the function $\lp \mu_1, \mu_2, ..., \mu_n \rp \mapsto k_{\mu_1, \mu_2, ..., \mu_n}$ is differentiable,
\item for all $y \in \mathbb{R}$, the function $\lp \mu_1, \mu_2, ..., \mu_n \rp \mapsto f_{\mu_1, \mu_2, ..., \mu_n}(y)$ is differentiable. 
\end{itemize}} 
Let $i \in [\![1,n]\!]$. 
Let us fix the values of $\mu_j$ for all $j \in [\![1,n]\!] \tsetminus \lb i \rb$ and define the function 
\begin{equation} \label{eq:gen_def_g}
g_i: \lp
\begin{aligned}
I_i \times \mathbb{R} &\to \mathbb{C} \\
(\mu_i, x) &\mapsto f_{\mu_1, \mu_2, ..., \mu_n} \lp k_{\mu_1, \mu_2, ..., \mu_n} \, x \rp
\end{aligned}
\rp .
\end{equation}
For all $\mu_i \in I_i$, the function $x \mapsto g_i (\mu_i, x)$ is a solution of the nonlinear equation under consideration. 
Let $\mu_i^{(0)} \in I_i$. 
Considering nearby values of $\mu_i$, we have to linear order:
\begin{equation}
\forall \, x \in \mathbb{R}, \; 
	g_i \lp \mu_i^{(0)} + \delta \mu_i, x \rp = 
		g_i \lp \mu_i^{(0)}, x \rp 
		+ \delta \mu_i \, \pd_{\mu_i} g_i \lp \mu_i^{(0)}, x \rp
		+ \ord{\delta \mu_i^2}. 
\end{equation}
This means that $x \mapsto \pd_{\mu_i} g_i \lp \mu_i^{(0)}, x \rp$ is, 
up to a constant factor, the difference between two infinitesimally close nonlinear solutions. It is thus 
a solution of the linear field equation over the background solution $x \mapsto g_i \lp \mu_i^{(0)}, x \rp$. 
Using \eq{eq:gen_def_g}, it is equal to
\begin{equation}
\pd_{\mu_i} g_i \lp \mu_i^{(0)}, x \rp = \left[
	\lp \pd_{\mu_i} f_{\mu_1, \mu_2, ..., \mu_n} \rp \lp y \rp
	+ x \s \lp \pd_{\mu_i} k_{\mu_1, \mu_2, ..., \mu_n} \rp \s \pd_y f_{\mu_1, \mu_2, ..., \mu_n} \lp y \rp
\right]_{y = k_{\mu_1, \mu_2, ..., \mu_n} \, x}.
\end{equation}
The first term in the right-hand side is periodic in $x$, and thus bounded. 
However, the second one grows linearly with $x$ 
provided $\lp \pd_{\mu_i} k_{\mu_1, \mu_2, ..., \mu_n} \rp \, \pd_y f_{\mu_1, \mu_2, ..., \mu_n}$ does not identically vanish. 
The linear equation thus has a mode with  a term whose amplitude grows linearly with $x$. 
Moreover, this mode can be obtained simply by differentiating the nonlinear solution with respect to one of its parameters. 

\begin{figure} 
\includegraphics[width = 0.49 \linewidth]{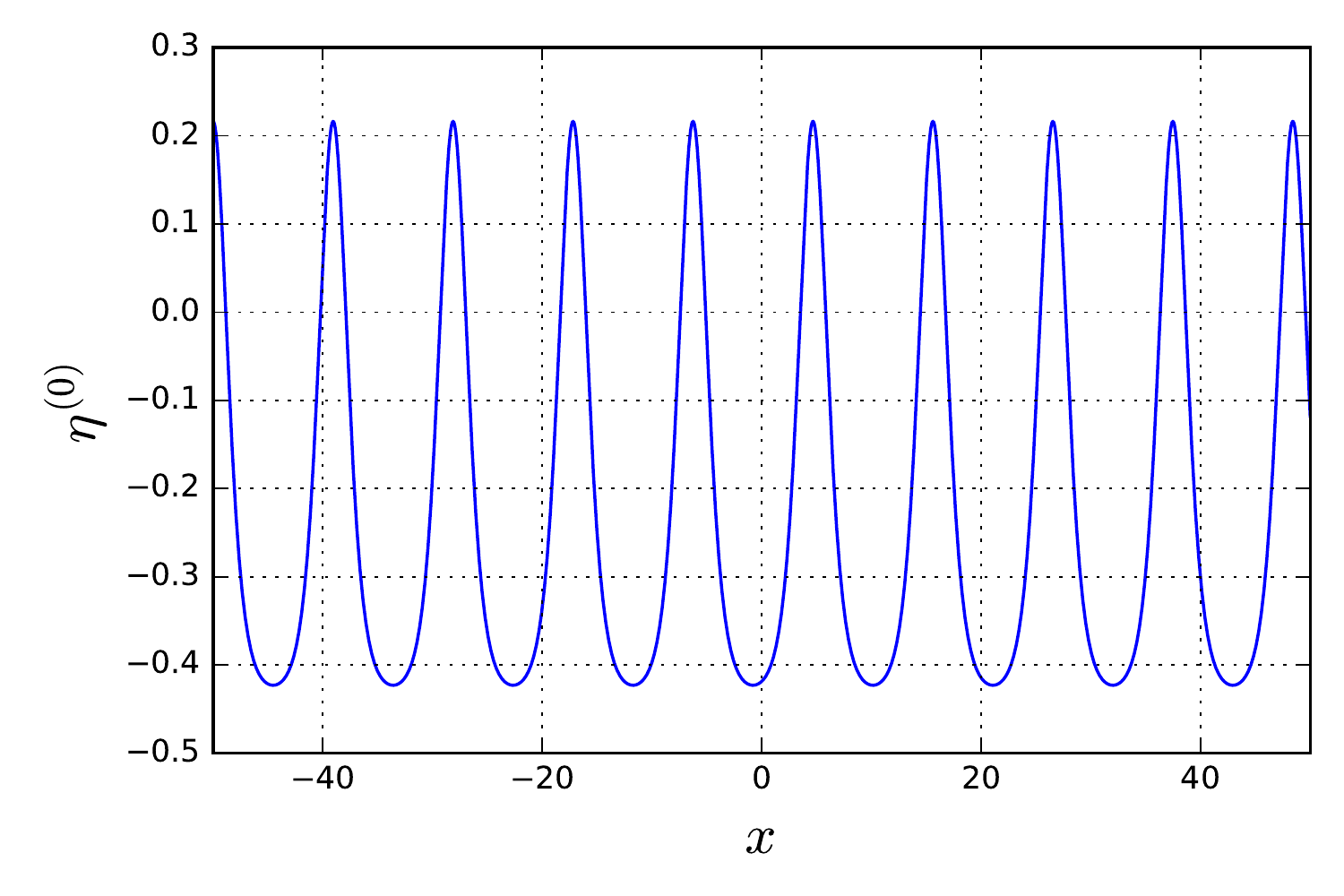} 
\includegraphics[width = 0.49 \linewidth]{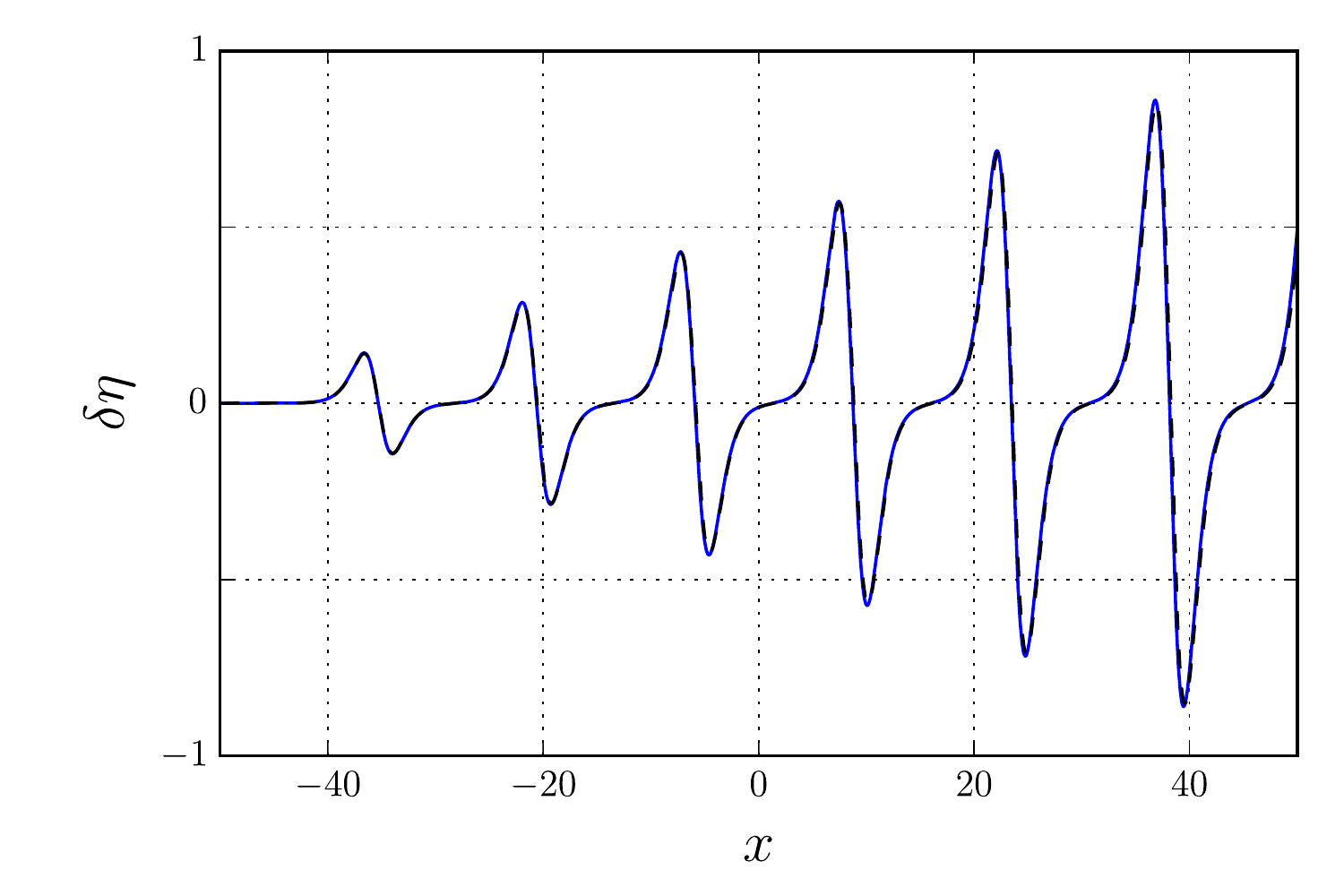}
\caption{\textit{Left panel:} A nonlinear, stationary solution of the KdV equation~\eqref{eq:KdV} for $\w = 1$ and $\eta_0 = 0$. 
\textit{Right panel:} Solution of the linearized KdV equation~\eqref{eq:KdV-linearized} over the undulation shown in the left panel, showing a linearly growing mode. 
The black, dashed line is a rescaled difference between two undulation solutions of the KdV equation with slightly different amplitudes.} \label{fig:KdV_ling}
\end{figure}

Let us now apply this argument to the KdV equation~\eqref{eq:KdV}. 
The solution \eqref{eq:KdV-undul-sols} depends on three parameters: $x_u$, $A_u$, and $\eta_0$. 
The first one does not give a linear growth since $k_u$ does not depend on $x_u$. 
One obtains
\begin{equation} \label{KdV_statmode1}
\pd_{x_u} \eta_u (x) = 
	A_u \, \sin \lp k_u \s \lp x - x_u \rp \rp
	+ \frac{A_u^2 k_u}{\w} \sin \lp 2 \s k_u \s \lp x - x_u \rp \rp
	+ \ord{A_u^3}.
\end{equation}
In the following, to simplify the expressions we choose $x_u = 0$.
Differentiation with respect to $\eta_0$ gives
\begin{equation} \label{KdV_statmode2}
\pd_{\eta_0} \eta_u (x) = 
	1
	- \frac{3 A_u}{\sqrt{\bar{\w}}} \, x \sin \lp k_u \s x \rp
	- \frac{3 A_u^2}{\bar{\w}^2} \cos \lp 2 \s k_u \s x \rp
	- \frac{3 A_u^2}{\bar{\w}^{3/2}} x \sin \lp 2 \s k_u \s x \rp
	+ \ord{A_u^3}.
\end{equation}
Finally, differentiation with respect to $A_u$ gives 
\begin{equation} \label{KdV_statmode3}
\pd_{A_u} \eta_u (x) = 
	\cos \lp k_u \s x \rp
	+ \frac{A_u}{\bar{\w}} \cos \lp k_u \s x \rp
	- \frac{3 A_u}{\bar{\w}}
	+ \frac{9 A_u^2}{16 \bar{\w}^2} \cos \lp 3 \s k_u \s x \rp
	+ \frac{15 A_u^2}{2 \bar{\w}^{3/2}} \, x \sin \lp k_u \s x \rp
	+ \ord{A_u^3}.
\end{equation}
These are the three linearly independent, zero-frequency solutions of the linearized KdV equation over the background solution $\eta_u$. 
Their most remarkable property is the presence of terms growing linearly in $x$. 
As explained in Section~\ref{subsec:tuned_mat}, these terms give the dominant contribution to the transfer matrix over long undulations. 
We show in Section~\ref{sec:WH} that they also determine the effective temperature in white-hole flows followed by a long undulation. 
The linear growth of the modes is illustrated in \fig{fig:KdV_ling}. 
The right panel displays both a solution of \eq{eq:KdV-linearized} over the undulation shown in the left panel and a function proportional to $\pd_{A_u} \eta_u$. 
The close agreement between the curves illustrates that the linear growth of the mode is due to the variation of the wave vector with the amplitude. 

Let us emphasize that the existence of these modes does not indicate the presence of unbounded solutions of the KdV equation~\eqref{eq:KdV}. 
Instead, they signal the breakdown of the linear description for the difference between two solutions with slightly different parameters at large values of $\abs{x}$. 
Indeed, the background solution $\eta^{(0)}$ and the perturbed one have phases differing by $\pi$ for values of $\abs{x}$ on the order of $1 / \abs{\delta k_u}$, where $\delta k_u$ denotes the difference between their wave vectors, so that they become very different from each other in a pointwise sense.  
Nonlinear terms must therefore be taken into account at large values of $\abs{x}$.   

However, \eq{eq:KdV-linearized} remains valid for describing small-amplitude perturbations in a finite domain with extension $L \in \mathbb{R}_+$. 
Indeed, since $\delta k_u$ is linear in the amplitude $a_p$ of the perturbation close to $x = 0$, the amplitude of the oscillations due to the linearly growing terms scales like $a_p \, L$, and can be made arbitrarily small by lowering $\abs{a_p}$. 
In the following, we assume that $\abs{a_p} \, \lp 1 + L \rp \ll A_u \ll 1$, so that Equation~\eqref{eq:KdV-undul-sols} is a good approximation of the background solution and perturbations can be described by the linear equation~\eqref{eq:KdV-linearized}. 

\paragraph{Modes growing in time:}

$\phantom{0}$

\vspace*{0.25 cm}

As explained in appendix~\ref{app:KdV_detuned}, modes unbounded in space and time have a similar origin, namely that the opposite signs of the energies or energy fluxes carried by two waves allows for an unbounded growth in their amplitudes without violation of energy conservation. 
In the present case, the waves of interest have vanishing energies. One can thus expect that there exist growing modes in time as well as in space. 
This is indeed the case: assuming $\eta^{(0)}$ has the form~\eqref{eq:KdV-undul-sols}, a straightforward calculation shows that there exists a solution of Equation~\eqref{eq:KdV-linearized} given by
\begin{equation} \label{eq:KdV-mode-lin-t}
\deta_t: (t,x) \mapsto 1 + 6 \s t \s k_u \s A_u \sin \lp k_u x \rp + \ord{A_u^2}. 
\end{equation}

One should notice, however, that this does not indicate an instability of the solution in the usual sense, but simply comes from the Galilean invariance of the KdV equation. 
Indeed, looking for solutions of the form
\begin{equation}
\eta_v: (t,x) \mapsto \xi(x - v \, t)
\end{equation}
where $v \in \mathbb{R}$, one obtains the following equation for $\xi$:
\begin{equation}
\lp \w - v \rp \, \pd_x \xi + \pd_x^3 \xi + 6 \, \xi \, \pd_x \xi = 0.
\end{equation}
This is exactly the stationary KdV equation with $\w$ replaced by $\w - v$. 
The periodic solutions in $\xi$ are thus given by \eq{eq:KdV-undul-sols}, where $k_u$ is still given by \eq{eq:KdV_ku} and $\bar{\w} \equiv \w - v + 6 \, \eta_0$. 
Working to first order in $A_u$ for simplicity, one obtains
\begin{equation}
\eta_v (t,x) = \eta_0 + A_u \, \cos \lp k_u \s (x - v \s t - x_u) \rp + \ord{A_u^2}. 
\end{equation}
Differentiating with respect to $v$ at fixed $\w$ so that $k_u$ is unchanged, one obtains
\begin{equation}
\pd_v \eta_v (t,x) = \frac{1}{6} + A_u \, t \, k_u \sin \lp k_u \, (x - v \, t - x_u) \rp + \ord{A_u^2}. 
\end{equation}
Up to a global factor, one recovers \eq{eq:KdV-mode-lin-t}. 
As in the case of a spatially growing mode the growth in time signals that the linear approximation breaks down at late times, since the ``unperturbed'' solution with vanishing velocity differs 
significantly from the ``perturbed'' one with velocity $v$ when $\abs{v \s t}$ becomes close to half the wavelength. 

\section{A more realistic model of water waves} 
\label{app:UCP} 

In the main text we used the KdV equation to approximate the dynamics of shallow-water waves. 
The aim of the present appendix is to sketch an extension of the main results to the more realistic 2D model used, e.g., in~\cite{Unruh:2013gga}. 
Since the reasoning is very similar we only give the main steps required to transpose the above calculations to this model.

\subsection{Gravity waves on a detuned undulation}
\label{app:UCP_detuned}

We consider the flow of an ideal fluid in a flume in two dimensions. 
The background flow is assumed to be stationary, and we denote by $\phi$ and $\psi$ the velocity potential and stream function, respectively. 
The latter is uniform along streamlines. 
It is defined up to a constant, which we choose so that $\psi = 0$ at the bottom of the flume. 
Let us call $\psi_s$ its value at the free surface. 
We denote by $\phi_0$ the velocity potential of the background flow. 
The unperturbed velocity is thus $\mathbf{v} = \nabla \phi_0$. 
Looking for perturbed solutions of the form $\phi = \phi_0 + \delta \phi$, to first order in $\delta \phi$, the wave equation reads~\cite{Unruh:2013gga,Coutant:2012mf}
\begin{equation} \label{eq:UCP} 
\lp \frac{1}{\mathbf{v}^2} \pd_t + \pd_{\phi_0}\rp \left[ \frac{v_x}{g+\frac{1}{2} \pd_y \mathbf{v}^2} \lp \pd_t + \mathbf{v}^2 \, \pd_{\phi_0} \rp \delta \phi \right]
+ \ii \s \pd_{\phi_0} \tanh \lp \ii \s \psi_s \s \pd_{\phi_0} \rp \delta \phi
=0,
\end{equation}
where all quantities are evaluated along the line defined by $\psi = \psi_s$, $g$ is the gravitational acceleration, and $(x,y)$ are Cartesian coordinates, $x$ in the direction of the flume and $y$ in the vertical direction. 

In this subsection we are interested in the effect of a ``detuned'' undulation, i.e., a periodic modulation of the velocity with a wavelength different from that of the zero-frequency dispersive modes. 
To simplify the notations, it is convenient to define the rescaled variables $X \equiv \phi_0 / \psi_s$ and $T \equiv t / \psi_s$, as well as the functions 
\begin{equation}
V \equiv  \mathbf{v}^2,\; G \equiv \psi_s \s \frac{v_x / \mathbf{v}^2}{g + \frac{1}{2} \s \pd_y \mathbf{v}^2}, \; \; \text{and} \; \; f(x) = x \tanh(x). 
\end{equation}
Equation~\eqref{eq:UCP} becomes
\begin{equation} 
\lp \frac{1}{V} \s \pd_T + \pd_X \rp \left[ V \s G \s \lp \pd_T + V \s \pd_X \rp \delta \phi \right]
	+ f \lp - \ii \s \pd_X \rp \s \delta \phi = 0. 
\end{equation}
Let us consider an undulation of the form 
\begin{equation} 
\begin{aligned}
& V (X) = V_0 + V_1 \s \e^{\ii \s K_u \s X} + V_1 ^*\s \e^{-\ii \s K_u \s X} + \ord{V_0 \s \ep^2}, \\
& G (X) = G_0 + G_1 \s \e^{\ii \s K_u \s X} + G_1 ^*\s \e^{-\ii \s K_u \s X} + \ord{G_0 \s \ep^2},
\end{aligned}
\end{equation}
where $\ep$ is a small parameter, $(V_0, G_0, K_u) \in \lp \mathbb{R}_+^* \rp^3$ ($K_u$ is the wave vector of the undulation in our coordinate system), and $(V_1, G_1) \in \mathbb{C}^2$ are of order $\ep \s V_0$ and $\ep \s G_0$, respectively. 
To zeroth order in $\ep$, looking for solutions of the form $\delta \phi: (T, X) \mapsto \exp \lp \ii \s \lp K \s X - \Omega \s T \rp \rp$, one obtains the dispersion relation between $K$ and $\Omega$: 
\begin{equation} \label{eq:UCP_DR}
G_0 \s \lp \Omega - V_0 \s K \rp^2 = f(K). 
\end{equation}
Let us assume there exists $(\Omega_r, K_r) \in \mathbb{R}^2$ such that $K = K_r$ and $K = K_r + K_u$ are both solutions of \eq{eq:UCP_DR} for $\Omega = \Omega_r$. 
We also assume for simplicity that for any $n \in \mathbb{Z}$, $K = K_r + n \s K_u$ is not a solution of the dispersion relation unless $n = 0$ or $n = 1$. 
The effects of the undulation on the spectrum for angular frequencies close to $\Omega_r$ and the scattering matrix can be derived using the same procedure as in appendix~\ref{app:KdV_detuned}. 
As the expressions we obtain are slightly cumbersome and do not show fundamentally new features, here we only sketch the important results:
\begin{itemize}
\item First, we find that the undulation is dynamically unstable, in that modes with real wave vectors close to $K_r$ have a complex frequency (with imaginary part of order $\ep$), if the waves of angular frequency $\Omega_r$ and wave vectors $K_r$ and $K_r + K_u$ have energies (now given by $\Omega \s \lp \Omega - V \s K \rp$) of opposite signs. 
If they have energies of the same sign, then the frequencies remain real. 
\item Second, when considering modes with real angular frequencies close to $\Omega_r$, we find that their quasimomenta are complex (again, with imaginary part of order $\ep$) if the energy fluxes of these waves have opposite signs. 
If they have the same sign, quasimomenta over the undulation remain real.
\end{itemize}
One thus obtains the same properties as for the KdV equation, namely, that the relative sign of the energies of the resonant waves determines the dynamical stability of the undulation, while that of the energy fluxes determines the existence of exponentially growing modes in space. 

\subsection{Second-order undulation}
\label{app:UCP_undul} 

To compute the resonant modes over a solution of the hydrodynamic equations, we need to know the shape of the undulation to second order in its amplitude, as well as the first variations of its wave vector. 
To this end, we look for stationary solutions for the velocity potential $\phi$ in the Cartesian coordinates $(x,y)$. 
We set $y = 0$ at the bottom of the flume and denote the vertical position of the free surface at the abscissa $x$ by $y_s(x)$. 
We then have four hydrodynamic equations to satisfy: 
\begin{itemize}
\item The continuity equation gives: $\Delta \phi = 0$;
\item The boundary condition at the bottom of the flume is: $\forall x \in \mathbb{R}, \pd_y \phi(x,0) = 0$;
\item Since the free surface must be a streamline, we have: 
$\forall x \in \mathbb{R}, y_s'(x) \, \lp \pd_x \phi(x,y) \rp_{y = y_s(x)} = \lp \pd_y \phi(x,y) \rp_{y = y_s(x)}$; 
\item The Bernoulli equation evaluated at the free surface gives 
\begin{equation}
\forall x \in \mathbb{R}, \frac{1}{2} \lp \nabla \phi (x,y_s(x))\rp^2 + g y_s(x) = C_0, 
\end{equation}
where $C_0$ is a real constant. 
\end{itemize}
We look for solutions describing small perturbations over a homogeneous flow with velocity $v_0$ in the $x$ direction. 
Writing $\phi = v_0 \s x + \delta \phi$ and expanding the four equations to second order in $\delta \phi$, one obtains after some algebraic manipulations:
\begin{equation} \label{eq:UCP_int_1}
\frac{v_0^2}{g} \lp \pd_x^2 \delta \phi \rp 
+ \lp \pd_y \delta \phi \rp
\mathop{\approx}_{y = y_{s,0}} 
\frac{v_0^3}{2g^2} \pd_x \pd_y \lp \pd_x \delta \phi \rp^2 
- \frac{v_0}{g} \lp 
\lp \pd_y \delta \phi \rp \lp \pd_y \pd_x \delta \phi \rp 
+3 \lp \pd_x \delta \phi \rp \lp \pd_x^2 \delta \phi \rp
\rp,
\end{equation}
where all quantities are evaluated along the unperturbed free surface $y = y_{s,0}$. 
Moreover, integrating the Laplace equation over $y$ gives (see for instance the appendix A of Ref.~\cite{Coutant:2012mf})
\begin{equation} \label{eq:UCP_int_2}
\pd_y \delta \phi (x,y) = \ii \s \pd_x \tanh \lp \ii \s y \s \pd_x \rp \s \delta \phi(x,y).
\end{equation}
Using Equation~\eqref{eq:UCP_int_2} to express $\pd_y \delta \phi$ in terms of $\pd_x \delta \phi$ in Equation~\eqref{eq:UCP_int_1} gives
\begin{equation}
\frac{v_0^2}{g} \pd_x^2 \delta \phi 
+ \ii \pd_x \tanh(\ii y_{s,0} \pd_x) \delta \phi
\mathop{\approx}_{y = y_{s,0}} 
\frac{v_0^3}{g^2} \pd_x \lp \lp \pd_x \delta \phi \rp \lp \ii \pd_x^2 \tanh(\ii y_{s,0} \pd_x) \delta \phi \rp \rp
- \frac{v_0}{2g} \pd_x \lp 
\lp \ii \pd_x \tanh(\ii y_{s,0} \pd_x) \delta \phi \rp^2 
+3 \lp \pd_x \delta \phi \rp^2
\rp .
\end{equation}
A straightforward calculation shows that, assuming the flow is subcritical ($v_0^2 < g \s y_{s,0}$), 
\begin{align}\label{eq:phixy} 
\delta \phi_u^{(A,C)} (x,y) = 
A \cos \lp  k_0 x \rp \frac{\cosh \lp k_0 y \rp}{\cosh \lp k_0 y_{s,0} \rp}
+ B \sin \lp 2 k_0 x \rp \frac{\cosh \lp 2 k_0 y \rp}{\cosh \lp 2 k_0 y_{s,0} \rp}
+C x
\end{align}
is a solution, where $A$ and $C$ are free real parameters, $k_0$ is the unique strictly positive solution of the equation: $g \s \tanh(k_0 \s y_{s,0}) = v_0^2 \s k_0$, and
\begin{equation}
B = \frac{3 \s \lp 1 - F^8 \s \xi^4 \rp}{8 F^4 \s \xi \s y_{s,0} \s v_0} \s A^2,
\end{equation}
where $F \equiv v_0 / \sqrt{g \s y_{s,0}}$ is the Froude number of the unperturbed solution and $\xi \equiv k_0 \s y_{s,0}$. 
The value of $C$ can be fixed by imposing that the water current be independent of $A$, as will be the case, for instance, for an undulation produced by an immersed obstacle as the current should, under the above approximations, be the same upstream and downstream. 
We obtain:
\begin{equation}
C = A^2 \frac{k_0^2 \s v_0}{4 \s g \s y_{s,0}} \frac{3 - F^4 \s \xi^2}{1 - F^2}.
\end{equation} 

The last information we will need concerning these nonlinear solutions is the variation of their wave vector (or, equivalently, their wavelength) with $A$. 
One way to obtain it would be to redo the above calculation to next order, keeping cubic terms in $\delta \phi$. 
However, there is a simpler method. 
The idea is to make use of translation invariance of the problem: translating the undulation \eq{eq:phixy} in $x$ by a constant quantity $\Delta x$ gives another solution of the fluid equations. 
Its derivative with respect to $\Delta x$ is thus a solution of the linearized equation~\eqref{eq:UCP} over the undulation. 
By definition, it has a vanishing angular frequency and the same periodicity as the undulation. 
We thus know that Equation~\eqref{eq:UCP} will have a solution with vanishing frequency and quasimomentum. 
This equation can be solved perturbatively in $A$ using the techniques of Section~\ref{sec:KdV_tuned}.  
Going to second order in $A$, we find the difference between the wave vector $k_u$ of the undulation and $k_0$ contributes to the dispersion relation at small frequencies. 
Imposing that $\Omega = K = 0$ be a solution gives 
\begin{equation}
k_u = 
	k_0  \lp 1 + 
	\frac
		{\frac{A^2 \s g^2}{8 \s v_0^6} \s \lp 9 - 6 F^4 \s \xi^2 + 5 \s F^8 \s \xi^4 \rp - 3 \frac{C}{v_0}}
		{1 - F^{-2} + F^2 \s \xi^2}
	+ \ord{\frac{A^3 \s g^3}{v_0^9}} \rp. 
\end{equation} 

\subsection{Gravity waves on a tuned undulation and effective temperature}
\label{app:UCP_tuned}

One can now solve Equation~\eqref{eq:UCP} in the background~\eqref{eq:phixy} in the zero-frequency limit.  
The calculation follows the same lines as in Section~\ref{subsec:tuned_mat}: expanding the quantities appearing in Equation~\eqref{eq:UCP} in powers of $A$, one obtains a recursion relation between the Fourier components of the perturbation, which can be solved to second order in $A$. 
As in the case of the KdV equation, one obtains two linearly growing modes in $\phi_0$. 
One of them, with a coefficient of order $A$, is related to the change in the undulation shape under variations of the mean water depth. 
The other one, with a coefficient of order $A^2$, is related to the change of wave vector of the undulation with its amplitude. 

Their precise form is not particularly enlightening. However, a useful information one can extract from them is the low-frequency effective temperature measured over a long undulation in a white-hole-like flow. 
The reasoning is similar to that of Section~\ref{sec:WH}: the condition that the mode be counterpropagating and incoming on a white-hole-like flow selects its asymptotic content, from which the scattering coefficients can be extracted. 
One difference, however, is that there are now two ``hydrodynamic'' modes whose wave vectors go to zero in the zero-frequency limit and which thus become indistinguishable. 
As we have performed the explicit calculation at a vanishing frequency only, this introduces an uncertainty about the relative amplitudes of the counterpropagating wave and that of the reflected, copropagating one. 
We thus define a parameter $\Lambda_{vu}$ equal to the fraction of the constant term due to the counterpropagating, incoming wave. 
In the absence of coupling between co- and counterpropagating modes in the rest frame of the background flow, one would have $\Lambda_{vu} = 1$. 
As it was shown (see for instance~\cite{Macher:2009tw}) that this coupling is generally small in similar setups, we expect $\Lambda_{vu}$ to be of order $1$. 
Up to this ambiguity, the effective temperature $T_\mathrm{eff}(\infty)$ determined over undulations much longer than $v_0^2 \s y_{s,0}^3 \s A^{-2}$ can be computed following the reasoning of~Section~\ref{KdV_WH:2} with minor modifications. 
We obtain 
\begin{equation} \label{eq:eq:UCP_inf}
T_\mathrm{eff}(\infty) = 
	\frac{A^2}{g^{1/2} \s y_{s,0}^{7/2}} \s 
	\frac
		{\left[ 9 - 9 \s F^2 - 12 \s F^4 \s \xi^2 - 6 \s F^6 \s \xi^2 + 13 \s F^8 \s \xi^4 - F^{10} \s \xi^4 - 2 \s F^{12} \s \xi^6 \right]^2}
		{32 \s \Lambda_{vu}^2  \s F^{10} \s \xi \s \lp 3 - F^4 \s \xi^2 \rp^2 \s \lp F^2 - 1 + F^4 \s \xi^2 \rp}
	\lp 1 + \orde{} \rp .
\end{equation}
In particular, we observe the same scaling in the amplitude $A$ of the undulation as for the KdV equation (see Equation~\eqref{Teffinftygen}). 

It is for the moment unclear to us whether \eq{eq:eq:UCP_inf} is relevant for water waves experiments. 
Indeed, in practice dissipative effects due to viscosity, for instance, may become important for a wave propagating over the undulation before reaching the lengths where \eq{eq:eq:UCP_inf} becomes valid. 
However, it shows that the scaling obtained in Section~\ref{sec:WH} is not an artifact from the specific approximations leading to the KdV equation, but also arises in a model taking the full dispersion relation of water waves (in the absence of surface tension and dissipation) into account. 
We expect that other nondissipative effects like the surface tension can be included following the same reasoning, and will lead to the same qualitative results.

\section{Low-frequency effective temperature in a subcritical flow}
\label{app:sub}

While a precise study of the scattering in a subcritical flow is beyond the scope of the present article, in this appendix we briefly comment on the main differences with respect to the white-hole-like case studied in Section~\ref{sec:WH}. 
We again focus on the zero-frequency limit of the effective temperature (see Refs.~\cite{Michel:2014zsa,Robertson:2016ocv,Coutant:2016vsf,Coutant:2017qnz} for previous analyses of the spectrum in subcritical flows). 
A more detailed study, including a calculation of the scattering coefficients for $\om \neq 0$ and their behavior for $\om \to 0$, will be presented in a future work. 

As was explained in Section~\ref{KdV_WH:2}, the zero-frequency limit of the effective temperature can be determined by first differentiating the (nonlinear) general solution of the KdV equation with given asymptotic conditions with respect to its parameters, giving a set of modes at $\om = 0$. 
One then has to select the linear combination of these modes which is incoming from the left. 
In the transcritical case of Section~\ref{KdV_WH:2}, one of the three modes is exponentially growing for $x \to \infty$, one goes to a finite constant, and one goes to zero exponentially. 
We argued that only the third one could contribute to the right-moving incoming mode. 
This condition fully determines the low-frequency effective temperature. 

In the subcritical case, where $w_+$ and $w_-$ are both positive, the dispersion relation is qualitatively the same in the two asymptotic regions (see Fig.~\ref{fig:DR_KdV}), and the exponentially growing and decaying modes are replaced by propagating waves with a negative group velocity. 
Neither of these waves can contribute to the right-moving incoming mode. 
The constant mode cannot either, for the same reason as in the transcritical case. 
This means that, in the limit $\om \to 0$, the right-moving incoming mode must correspond to $\deta = 0$, i.e., to a homogeneous $\phi = \int \deta \s \dd x$. 
(Conversely, one can check that this mode is indeed incoming and right-moving if $\w_-, \w_+ > 0$.) 
The amplitude of the oscillating part of the mode thus vanishes, i.e., $A_\mathrm{neg} = 0$. 
From Equation~\eqref{eq:effTeff}, we thus find $T_\mathrm{eff} = 0$, whether or not an undulation is present in the region $x < 0$.  
However, we expect that the presence of an undulation may strongly modify the behavior of the scattering coefficients at low but finite frequencies. 
This will be studied in a future work.

\bibliography{biblio}

\end{document}